\documentclass[10pt,twocolumn,letterpaper]{article}

\usepackage{iccv}              % To produce the CAMERA-READY version

\usepackage[utf8]{inputenc} % allow utf-8 input
\usepackage[T1]{fontenc}    % use 8-bit T1 fonts
\usepackage{hyperref}       % hyperlinks
\usepackage{url}            % simple URL typesetting
\usepackage{booktabs}       % professional-quality tables
\usepackage{amsfonts}       % blackboard math symbols
\usepackage{nicefrac}       % compact symbols for 1/2, etc.
\usepackage{microtype}      % microtypography

\usepackage{enumitem}
\usepackage{xspace}

\usepackage{subcaption}

\usepackage{multirow}

\usepackage{graphicx}
\usepackage{subcaption}

\usepackage{wrapfig}
\usepackage{adjustbox}

\usepackage{epsfig}
\usepackage{graphicx}
\usepackage{pgfplots}
    \pgfplotsset{compat=1.18}
\usepackage{caption}
\usepgfplotslibrary{groupplots}
\usepackage{bm}
\usepackage{bbding}

\usepackage{graphicx}
\usepackage{pifont}
\usepackage{pgfplots}
\usepackage{algorithm}
\usepackage[noend]{algpseudocode}
\usepackage{caption} 
\usepackage{arydshln}

\usepackage{soul}

\usepackage[accsupp]{axessibility}  % Improves PDF readability for those with disabilities.

\usepackage{multirow}
\usepackage{graphicx}
\usepackage{color, colortbl}

\definecolor{increase}{HTML}{FCE4D6}
\definecolor{decrease}{HTML}{FCE4D6}

\definecolor{FFABA8}{HTML}{e3ebfc}
\colorlet{Light}{FFABA8}
\newcommand{\CC}[1]{\cellcolor{Light}}

\definecolor{f5f8ff}{HTML}{f5f8ff}
\colorlet{Light2}{f5f8ff}
\newcommand{\CCC}[1]{\cellcolor{Light2}}

\definecolor{b4b4fa}{HTML}{d7d7fa}
\colorlet{Dark}{b4b4fa}

\usepackage{colortbl}

\usepackage{orcidlink}
\usepackage{circledsteps}
\usepackage{pict2e}
\usepackage{wrapfig,lipsum,booktabs}
\newsavebox{\bigimage}

\usepackage{tkz-kiviat}
\usepackage{xfp}

\usepackage{array,multirow,graphicx}
\usepackage{wrapfig}
\definecolor{purple}{rgb}{1,0,1}
\newcolumntype{a}{>{\columncolor{lightblue}}c}
\newcommand{\kibitz}[2]{\ifnum\Comments=1\textcolor{#1}{#2}\fi}

\usepackage{dblfloatfix}
\usepackage{epsfig}
\usepackage{graphicx}
\usepackage{pgfplots}
    \pgfplotsset{compat=1.18}
\usepackage{caption}
\usepgfplotslibrary{groupplots}
\usepackage{tcolorbox}

\usepackage{array,multirow,graphicx}
\usepackage{wrapfig}
\definecolor{purple}{rgb}{1,0,1}
\newcolumntype{a}{>{\columncolor{lightblue}}c}
\usepackage{multirow}

\usepackage{epsfig}
\usepackage{graphicx}
\usepackage{pgfplots}
    \pgfplotsset{compat=1.18}
\usepackage{caption}
\usepgfplotslibrary{groupplots}
\usepackage{fontawesome5}
\definecolor{demphcolor}{RGB}{144,144,144}

\newcommand{\customsubsection}[1]{%
  \par
  \pagebreak[2]%
  \refstepcounter{subsection}%
    \everypar={%
      {\setbox0=\lastbox}% Remove the indentation
      \addcontentsline{toc}{subsection}{%
        {\protect\makebox[0.3in][r]{\thesubsubsection.} \hspace*{3pt}#1}}%
      \textbf{\thesubsection\space\space{#1}\space\newline}%
      \everypar={}%
    }%
  \ignorespaces
}

\usepackage{lipsum}

\definecolor{cvprblue}{rgb}{0.21,0.49,0.74}
\definecolor{babyblueeyes}{rgb}{0.63, 0.79, 0.95}
\definecolor{darkturquoise}{rgb}{0.0, 0.81, 0.82}
\definecolor{deepskyblue}{rgb}{0.0, 0.75, 1.0}
\definecolor{dodgerblue}{rgb}{0.12, 0.56, 1.0}
\definecolor{turquoiseblue}{rgb}{0.0, 1.0, 0.94}

\hypersetup{
    colorlinks,
    linkcolor={red},
    citecolor={deepskyblue}
}
\usepackage{amssymb}
\usepackage{pifont}
\usepackage{amsmath}
\usepackage{algorithm}
\usepackage{algpseudocode}

\usepackage{caption}
\usepackage{cuted} 

\definecolor{Light}{HTML}{f6fae4}
\definecolor{Light}{HTML}{fafced}
\definecolor{Light}{HTML}{f8fbe9}

\definecolor{6ec2b5}{HTML}{b8d65e}
\definecolor{FFABA8}{HTML}{807d7d}
\colorlet{Dark}{6ec2b5}
\colorlet{Salmon}{FFABA8}

\definecolor{forestg}{HTML}{34bf34}
\colorlet{ForestGreen}{forestg}
\colorlet{OrangeRed}{red}

\definecolor{citecolor}{HTML}{0071bc}
\definecolor{color_ao}{gray}{0.5}
\definecolor{color_our}{HTML}{e6f2c2}
\definecolor{color_pre}{rgb}{0.52,0.59,0.69}
\definecolor{Gray}{gray}{0.9}
\definecolor{LighterGray}{gray}{0.93}
\definecolor{LightGrayForTableRule}{gray}{0.92}
\definecolor{DarkGray}{gray}{0.5}
\definecolor{Black}{rgb}{0.0, 0.0, 0.0}
\definecolor{NiceBlue}{rgb}{0.11764705882352941, 0.5647058823529412, 1.0}
\definecolor{NiceGreen}{HTML}{b6c783}
\usepackage{diagbox}
\definecolor{gray}{HTML}{b0aeae}
\definecolor{diffgray}{HTML}{878787}
\definecolor{NiceGray}{HTML}{696969}
\definecolor{applegreen}{rgb}{0.55, 0.71, 0.0}

\definecolor{F08080}{HTML}{F08080}
\colorlet{LinePlot1}{F08080}

\definecolor{D2B48C}{HTML}{D2B48C}
\colorlet{LinePlot2}{D2B48C}

\definecolor{B8D65E}{HTML}{B8D65E}
\colorlet{LinePlot3}{B8D65E}

\definecolor{b5b3b3}{HTML}{b5b3b3}
\colorlet{LinePlot4}{b5b3b3}

\definecolor{d8d9ed}{HTML}{d8d9ed}
\definecolor{f5f8ff}{HTML}{f5f8ff}
\colorlet{ThemeColor}{f5f8ff}

\definecolor{increase}{HTML}{FCE4D6}

\usepackage{circledsteps}
\usepackage{pict2e}
\usepackage{wrapfig,lipsum,booktabs}
\usepackage{tkz-kiviat}
\usepackage{xfp}

\usepackage{array,multirow,graphicx}
\usepackage{wrapfig}
\definecolor{purple}{rgb}{1,0,1}
\newcolumntype{a}{>{\columncolor{lightblue}}c}

\newcommand\blfootnote[1]{%
  \begingroup
  \renewcommand\thefootnote{}\footnote{#1}%
  \addtocounter{footnote}{-1}%
  \endgroup
}

\usepackage{tkz-kiviat}
\usepackage{xfp}
\usepackage{dblfloatfix}
\usepackage{epsfig}
\usepackage{graphicx}
\usepackage{pgfplots}
    \pgfplotsset{compat=1.18}
\usepackage{caption}
\usepgfplotslibrary{groupplots}
\usepackage{array,multirow,graphicx}
\usepackage{wrapfig}
\definecolor{purple}{rgb}{1,0,1}
\newcolumntype{a}{>{\columncolor{lightblue}}c}
\usepackage{tkz-kiviat}
\usepackage{xfp}
\usepackage{multirow}

\usepackage{epsfig}
\usepackage{graphicx}
\usepackage{pgfplots}
    \pgfplotsset{compat=1.18}
\usepackage{caption}
\usepgfplotslibrary{groupplots}
\definecolor{demphcolor}{RGB}{144,144,144}

\usepackage{lipsum}

\definecolor{cvprblue}{rgb}{0.21,0.49,0.74}
\definecolor{babyblueeyes}{rgb}{0.63, 0.79, 0.95}
\definecolor{darkturquoise}{rgb}{0.0, 0.81, 0.82}
\definecolor{deepskyblue}{rgb}{0.0, 0.75, 1.0}
\definecolor{dodgerblue}{rgb}{0.12, 0.56, 1.0}
\definecolor{turquoiseblue}{rgb}{0.0, 1.0, 0.94}

\hypersetup{
    colorlinks,
    linkcolor={red},
    citecolor={deepskyblue}
}
\usepackage{amssymb}
\usepackage{pifont}
\usepackage{amsmath}
\usepackage{algorithm}
\usepackage{algpseudocode}
\usepackage{tcolorbox} % For boxed steps
\definecolor{Light}{HTML}{f6fae4}
\definecolor{Light}{HTML}{fafced}
\definecolor{Light}{HTML}{f8fbe9}

\definecolor{6ec2b5}{HTML}{b8d65e}
\definecolor{FFABA8}{HTML}{807d7d}
\colorlet{Dark}{6ec2b5}
\colorlet{Salmon}{FFABA8}

\definecolor{forestg}{HTML}{34bf34}
\colorlet{ForestGreen}{forestg}
\colorlet{OrangeRed}{red}

\usepackage[dvipsnames]{xcolor}
\definecolor{citecolor}{HTML}{0071bc}
\definecolor{color_ao}{gray}{0.5}
\definecolor{color_our}{HTML}{e6f2c2}
\definecolor{color_pre}{rgb}{0.52,0.59,0.69}
\definecolor{Gray}{gray}{0.9}
\definecolor{LighterGray}{gray}{0.93}
\definecolor{LightGrayForTableRule}{gray}{0.92}
\definecolor{DarkGray}{gray}{0.5}
\definecolor{Black}{rgb}{0.0, 0.0, 0.0}
\definecolor{NiceBlue}{rgb}{0.11764705882352941, 0.5647058823529412, 1.0}
\definecolor{NiceGreen}{HTML}{b6c783}
\usepackage{diagbox}
\definecolor{gray}{HTML}{b0aeae}
\definecolor{diffgray}{HTML}{878787}
\definecolor{NiceGray}{HTML}{696969}
\definecolor{applegreen}{rgb}{0.55, 0.71, 0.0}

\definecolor{F08080}{HTML}{F08080}
\colorlet{LinePlot1}{F08080}

\definecolor{D2B48C}{HTML}{D2B48C}
\colorlet{LinePlot2}{D2B48C}

\definecolor{B8D65E}{HTML}{B8D65E}
\colorlet{LinePlot3}{B8D65E}

\definecolor{b5b3b3}{HTML}{b5b3b3}
\colorlet{LinePlot4}{b5b3b3}

\def\ourapproach{\textsc{Aurelia}\xspace}
\def\ourbenchmark{AVReasonBench\xspace}
\def\ourtask{AV-GeoIQ\xspace}
\def\ourbenchmarksamples{4500\xspace}

\definecolor{iccvblue}{rgb}{0.21,0.49,0.74}
\def\paperID{13296} % *** Enter the Paper ID here
\def\confName{ICCV}
\def\confYear{2025}

\title{\raisebox{-0.3\height}{\includegraphics[width=0.06\linewidth]{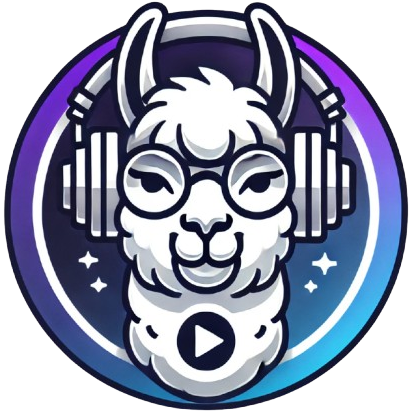}} \ourapproach: Test-time Reasoning Distillation in Audio-Visual LLMs}

\author{Sanjoy Chowdhury$^{*1}$ $\quad$ Hanan Gani$^{*2}$ $\quad$ Nishit Anand$^1$ \\ Sayan Nag$^3$ $\quad$ Ruohan Gao$^1$ $\quad$ Mohamed Elhoseiny$^4$ $\quad$ Salman Khan$^2$$^\dagger$ $\quad$ Dinesh Manocha$^1$$^\dagger$ 
\vspace{1mm} 
\\ 
$^{1}$University of Maryland, College Park $\quad$ $^{2}$MBZUAI $\quad$ $^{3}$University of Toronto $\quad$ $^{4}$KAUST 
 \vspace{-0.5mm}
 \\ 
\tt\footnotesize \{sanjoyc, nishit, rhgao, dmanocha\}@umd.edu $\quad$ \{hanan.ghani, salman.khan\}@mbzuai.ac.ae $\quad$  \\ \tt\footnotesize sayan.nag@mail.utoronto.ca 
$\quad$ \tt\footnotesize  $\quad$ mohamed.elhoseiny@kaust.edu.sa  \\ }

\begin{document}
\maketitle

\begin{abstract} 
\vspace{-5mm}

\blfootnote{$^*$Equal contribution. $^\dagger$Equal advising.}

Recent advancements in reasoning optimization have greatly enhanced the performance of large language models (LLMs). However, existing work fails to address the complexities of audio-visual scenarios, underscoring the need for further research. In this paper, we introduce \ourapproach, a novel actor-critic based audio-visual (AV) reasoning framework that distills structured, step-by-step reasoning into AVLLMs at test time, improving their ability to process complex multi-modal inputs without additional training or fine-tuning. To further advance AVLLM reasoning skills, we present \ourbenchmark, a challenging benchmark comprising \ourbenchmarksamples audio-visual questions, each paired with detailed step-by-step reasoning. Our benchmark spans six distinct tasks, including \ourtask, which evaluates AV reasoning combined with geographical and cultural knowledge. Evaluating 18 AVLLMs on \ourbenchmark reveals significant limitations in their multi-modal reasoning capabilities. Using \ourapproach, we achieve up to a 100\% relative improvement, demonstrating its effectiveness. This performance gain highlights the potential of reasoning-enhanced data generation for advancing AVLLMs in real-world applications. Our code and data will be publicly released at: \url{https://github.com/schowdhury671/aurelia}.

\end{abstract}

\vspace{-5mm}
\section{Introduction}
\vspace{-2mm}
\begin{figure}
    \centering
    \includegraphics[width=\columnwidth]{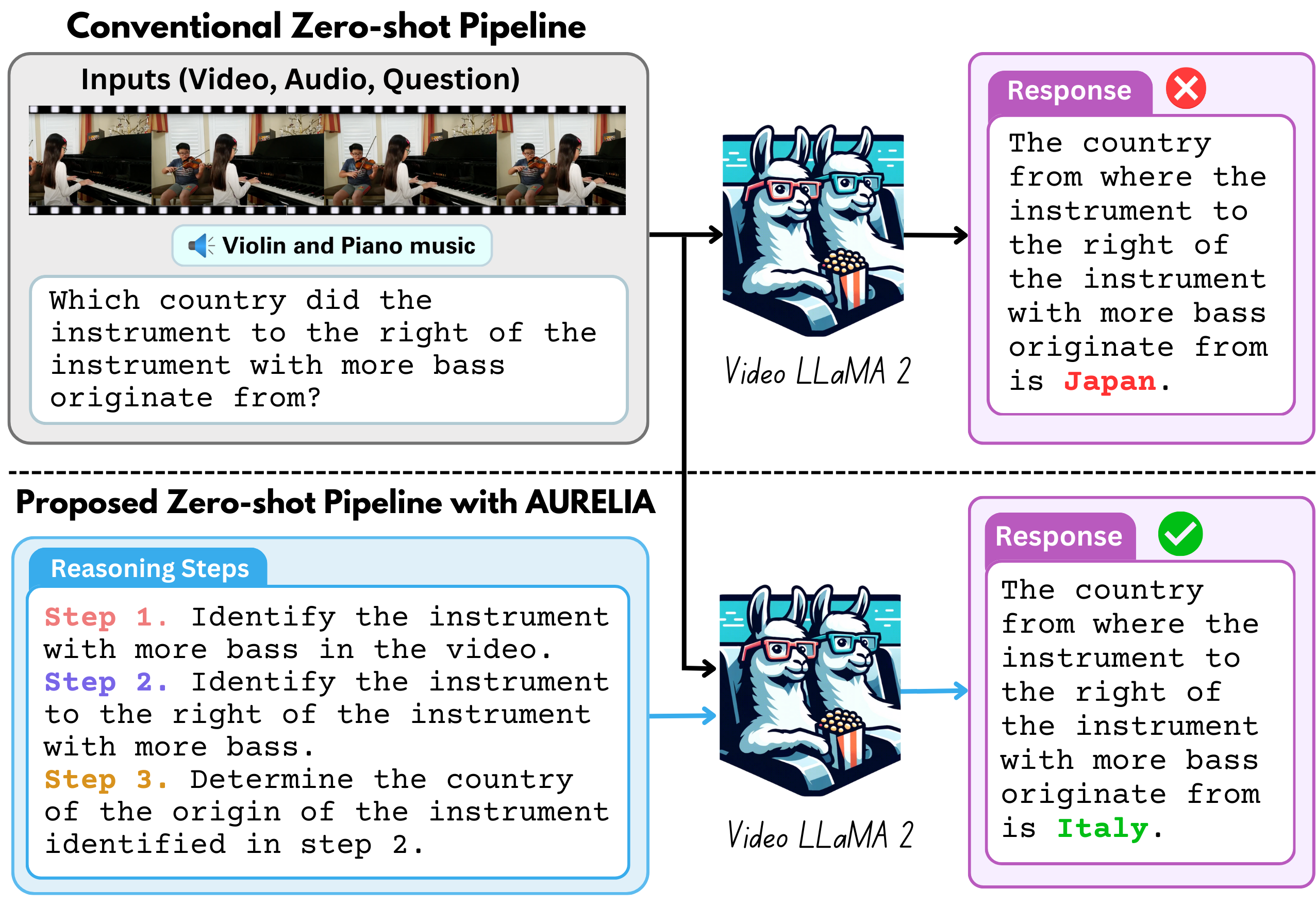}
    \caption{\textbf{Effect of injecting reasoning steps.} \ourapproach enhances the ZS capabilities of audio-visual models (e.g., VideoLLaMA2). The conventional pipeline struggles in audio-visual comprehension, leading to incorrect responses. In contrast, \ourapproach systematically breaks down the problem into intermediate reasoning steps, guiding the model toward more accurate and interpretable answer.}   
    % We present \ourapproach, a novel framework designed to generate reasoning-enhanced synthetic data, strengthening AV comprehension in existing AVLLMs. Leveraging this data enhances analytical reasoning skills in existing methods (e.g., VideoLLaMA2).}
    \vspace{-1em}
    \label{fig:teaser}
\end{figure}

Multi-agent AI systems powered by LLMs have excelled in structured reasoning tasks, including mathematical problem-solving \cite{yang2024qwen2, wang2023math, sun2024easy, ying2024internlm}, coding assistance \cite{zhang2024o1}, and drug discovery \cite{swanson2024virtual}. These systems often employ systematic problem decomposition, as in chain-of-thought (CoT) reasoning \cite{wei2022chain}. More advanced approaches optimize reasoning through outcome reward models \cite{zhang2024generative, yu2023ovm}, which refine solutions based on final results, and process reward models \cite{lightman2023let, luo2024improve, zhang2024llama}, which assess and improve intermediate steps.

% However, real-world reasoning extends beyond structured text-based tasks. Understanding the physical world often requires reasoning across multiple modalities, particularly in audio-visual (AV) environments. For instance, identifying the country of origin of a music performance requires integrating visual cues (e.g., traditional clothing, dance form, instrument type) and audio cues (e.g., melody, language of the lyrics). 
% Therefore, reasoning in AV data is critical as numerous real-world tasks involve abstract nuances that cannot be fully captured through text or images alone. Despite the growing capabilities of multimodal LLMs \cite{sun2024video, tang2024extending, cheng2024videollama, zhang2024video, lin2023video, team2023gemini, wang2024qwen2, tang2024enhancing}, most reasoning benchmarks remain focused on images and text, neglecting audio and its interaction with visual signals. Reasoning in AV data presents unique challenges beyond those encountered in static image-text based tasks: 1) Unlike images, where all information is available in a single frame, audio and video evolve over time, requiring the model to track events, infer temporal relationships, and integrate multi-frame context. 2) Audio signals often lack direct textual mappings, making it harder for models to extract structured meaning. For example, a roaring crowd could indicate excitement at a concert or protest; requiring appropriate context disambiguate such scenarios. As a result, current models struggle with AV reasoning, often relying on biases rather than deeper cross-modal comprehension.

Real-world reasoning extends beyond structured text-based tasks, often requiring multimodal integration, especially in audio-visual (AV) environments. Identifying a music performance’s origin, for instance, involves both visual cues (e.g., attire, instruments) and audio cues (e.g., melody, language). AV reasoning is crucial for capturing abstract nuances that text or images alone cannot convey. Despite advancements in multimodal LLMs \cite{sun2024video, tang2024extending, cheng2024videollama, zhang2024video, lin2023video, team2023gemini, wang2024qwen2, tang2024enhancing}, most benchmarks remain image-text focused, overlooking audio’s role and its interplay with visual signals. AV reasoning presents unique challenges. Firstly, unlike static images, AV data unfolds over time, requiring models to track events, infer temporal relationships, and integrate multi-frame context. Secondly, audio often lacks direct textual mappings, making structured interpretation harder. For example, a roaring crowd may signal excitement at a concert or unrest at a protest—context is essential for disambiguation. Current models often struggle with AV reasoning, relying on biases rather than deep cross-modal comprehension.

Moreover, current AVLLMs are susceptible to cultural, contextual, and perceptual biases embedded in their training data. As illustrated in \cref{fig:teaser}, an AVLLM might incorrectly associate a musical instrument with Japan due to the presence of East Asian musicians and a Japanese track, even when the actual answer is Italy. This highlights the models’ tendency to depend on dominant visual or auditory cues rather than true reasoning. While recent advances in test-time reasoning \cite{wei2022chain, kojima2022large,zhong2024evaluation} have significantly improved text-based LLMs, these techniques remain largely unexplored for AV models.

To address these shortcomings, we introduce \textbf{\ourapproach}, a test-time multi-agent reasoning distillation framework for addressing challenges in audio-visual cross-modal comprehension by mitigating visual and auditory biases without the need for additional training. 
Specifically, \ourapproach employs an interactive LLM-based multi-agent framework that harnesses the reasoning capabilities of LLMs to iteratively generate high-quality reasoning data required for multimodal audio-video understanding.
% Specifically, \textbf{\ourapproach} harnesses the reasoning capabilities of LLMs to tackle multimodal audio-video understanding and generation tasks without requiring additional training. 
% Our approach uses LLMs as a generator to propose candidate solutions for a given task, while an off-the-shelf multimodal model serves as a scorer to assess the quality of each generated response. The feedback from the scorer is iteratively incorporated into the generator, refining the responses to improve accuracy and coherence. This iterative process continues until convergence or a predefined number of steps, producing the final output. We find that this simple yet effective framework generalizes well across diverse tasks and modalities. 
By leveraging the reasoning data, our approach distills structured reasoning into AVLLMs, enhancing their capabilities in multimodal audio-video commonsense reasoning, geographical understanding, music comprehension, and humor understanding.
% By leveraging the reasoning data, our approach successfully addresses challenges in multimodal audio-video commonsense reasoning, geographical understanding, music comprehension, humour understanding. 

To rigorously assess AVLLMs' reasoning capabilities, we further introduce \textbf{\ourbenchmark}, a comprehensive benchmark comprising \ourbenchmarksamples audio-visual questions, each paired with detailed step-by-step reasoning solutions generated through our pipeline. Our benchmark suite spans six distinct tasks, including the novel \textbf{\ourtask} task for geographical and cultural reasoning. Evaluating 18 existing AVLLMs on \ourbenchmark\ reveals significant deficiencies in their ability to process dynamic audio-video content. However, incorporating \ourapproach-generated reasoning solutions significantly enhances AVLLMs' performance, highlighting the impact of structured test-time reasoning. We summarize our contributions below:

% \textcolor{red}{Elhoseiny: the size of the dataset might be questionable, why 4500 is sufficient to cover sufficient scenarios perhaps 5-10 times set might be needed.  }

% However, when augmented with our AURELIA-generated reasoning solutions, these models achieve a 100\% relative improvement in performance, demonstrating the efficacy of our test-time reasoning approach. \textcolor{red}{Elhoseiny: This statement is repeated roughly three times, 1 in abtract, here, and list  of contributions, perhaps remove this one. }

\begin{itemize}
    \item We present \ourapproach, \textit{a scalable and automated pipeline} for generating high-quality Audio-Visual reasoning data, serving as both an evaluation resource and to the best of our knowledge, the \textit{first training-free reasoning distillation framework} for Audio Visual LLMs. 

    \item Leveraging our proposed reasoning data generation pipeline, we introduce \textit{\ourbenchmark}, a comprehensive AV benchmark featuring \ourbenchmarksamples audio-visual samples with detailed step-by-step reasoning solutions across six diverse tasks, encompassing multimodal commonsense reasoning, music comprehension, and humor detection. Additionally, as a part of our benchmark, we introduce a novel task \textit{AV-GeoIQ} for geographical understanding and curate 1,000 AV-Compositional and 100 AV-Meme understanding samples through careful manual inspection.

    % Using our proposed reasoning data generation pipeline, we introduce a new AV benchmark \textbf{\ourbenchmark}, containing \ourbenchmarksamples audio-visual samples paired with their step-by-step reasoning solutions, across six challenging tasks spanning multimodal commonsense reasoning, music comprehension, and humor detection, including a novel task \textit{AV-GeoIQ} for geographical understanding. Moreover, we manually collect 1000 and 100 samples for AV-Compositional and AV-Meme understanding tasks, respectively.

    \item Leveraging our curated reasoning dataset, we demonstrate \textit{up to 100\% relative improvement} in AVLLM performance through zero-shot reasoning distillation, demonstrating the effectiveness of our approach in enhancing the reasoning capabilities of AV models.
\end{itemize}

\begin{figure*}[t!]
    \centering
    \includegraphics[width=\textwidth]{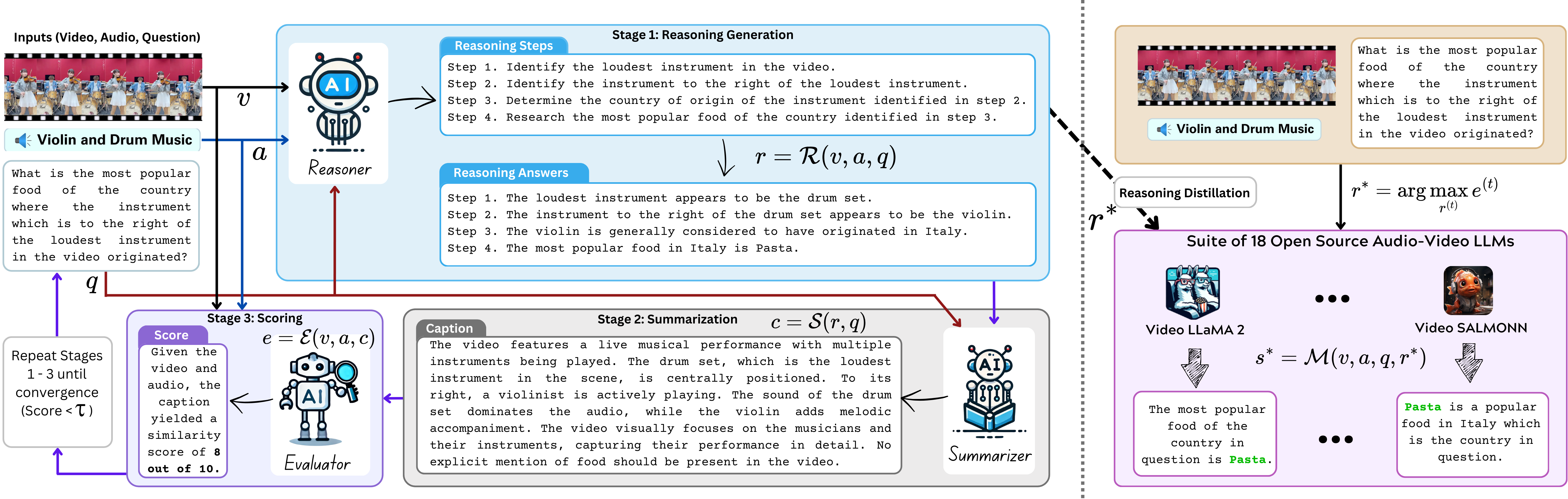}
    \caption{\textbf{Overview of \ourapproach: }Our proposed \ourapproach consists of a multi-agent interactive framework that functions in sync and generates reasoning steps that are then distilled inside the target model. The input set consisting of the audio, video, and question is first fed into the reasoning generator agent, which generates an initial set of reasoning steps that provide a structured pathway to reach the final answer. These reasoning steps are synthesized into a detailed caption by a Summarizer agent. The Evaluator agent then outputs a score that measures the relevance of the caption with the input audio and video. A feedback mechanism then provides supervision to the Reasoning generator based on the evaluation score, which adjusts its output to maximize the evaluation score. This actor-critique framework continues until the evaluation score exceeds a specific threshold or the number of iterations are exhausted.}
    \label{fig:main-figure}
    \vspace{-1.0mm}
\end{figure*}

\vspace{-1.0mm}
\section{Related Work}
\label{related works}
\vspace{-2mm}

\noindent{\textbf{Reasoning in Multimodal LLMs.}}
Researchers have been exploring ways to optimize CoT reasoning for MLLMs to handle increasingly complex tasks. The majority of studies focus on extracting graphical \cite{thawakar2025llamav, he2024distill, deng2024tables, fu2024graphic}, logical \cite{xiao2024logicvista, zheng2024thinking, kilmllm, dong2024insight, wang2024mementos} or textual \cite{xu2024llava, chen2024steering, bi2024forest} information from images and using it to solve mathematical problems. LLaVA-CoT \cite{xu2024llava} investigates improved sampling and search algorithms to identify reasoning paths. Virgo \cite{du2025virgo} examines the organization of fine-tuning data and the transferability of text-based reasoning tasks to image-based reasoning. Recently, MAmmoTH-VL \cite{guo2024mammoth} developed a large-scale multimodal instruction-tuning dataset to enhance question-answering performance across various modalities. In a major departure from these previous bodies of work \ourapproach specifically targets general video understanding scenarios, where various aspects of AV information are continuously referenced throughout the reasoning process.

\noindent{\textbf{Benchmarks for Audio-Visual LLMs.}} The rapid advancement of MLLMs \cite{pandagpt, imagebindllm, minigpt4, vistallm, llava, avicuna, avllm} has driven the development of increasingly challenging video understanding benchmarks, shifting the focus from basic video description and perceptual abilities \cite{chen2023valor, li2022learning, vast, ning2023video, mangalam2023egoschema, sun2024crosscheckgpt, chowdhury2024meerkat} to reasoning capabilities \cite{mvbench, li2024vitatecs, videomme, liu2024tempcompass, fang2025mmbench, avtrustbench}. Specifically, NExT-QA \cite{xiao2021next} emphasizes causal reasoning while Video-MME \cite{videomme} features questions that necessitate integrating both audio and visual cues for effective reasoning. Our proposed \ourbenchmark presents more challenging questions that demand deeper reasoning, extensive world knowledge, and a more seamless integration of AV information.

\noindent{\textbf{Reasoning Benchmarks}}
While text-based benchmarks such as GSM8K \cite{cobbe2021training} and MMLU \cite{mmlu} evaluate logical and commonsense reasoning, multimodal benchmarks remain underdeveloped. Recent efforts, such as MathVista \cite{mathvista} and VideoQA datasets \cite{li2024sports, funqa, fu2024video, wang2024videocot}, attempt to introduce vision-based reasoning tasks, but they often emphasize perception over deeper reasoning. Moreover, existing benchmarks lack comprehensive challenges that require integration across multiple modalities, including audio, video, and world knowledge. Several works have proposed methods to assess reasoning quality in LLMs, e.g., logical consistency checks \cite{thawakar2025llamavo1rethinkingstepbystepvisual, geva2021did, lu2022learn} and adversarial reasoning tasks \cite{avtrustbench, upd}. However, current benchmarks chiefly measure static performance rather than adaptive, context-dependent reasoning. Although multi-agent systems \cite{yu2025fincon, guo2024largelanguagemodelbased, nascimento2023self, han2024llm} and collaborative reasoning frameworks \cite{chen2021neural, sun2024large, tran2025multi} have shown promise in enhancing reasoning abilities, their evaluation remains fragmented across different domains.

Our work addresses these gaps by introducing a comprehensive reasoning benchmark \ourbenchmark that evaluates multimodal reasoning skills in LLMs, integrating text, vision, audio, and external world knowledge. Unlike purely visual reasoning tasks, the AVR domain presents unique real-world challenges, such as temporal synchronization between audio and visual cues, ambiguity in auditory semantics, and the need for deeper cross-modal understanding. 
% AURELIA is specifically designed to tackle these challenges through an agentic framework that iteratively refines reasoning across modalities, ensuring a deeper understanding of real-world AV contexts. 

% By focusing on complex, real-world reasoning tasks, we aim to push the boundaries of LLM evaluation beyond existing benchmarks.

% Our work addresses these gaps by introducing a comprehensive reasoning benchmark that evaluates multimodal reasoning skills in LLMs, integrating text, vision, audio, and external world knowledge. By focusing on complex, real-world reasoning tasks, we aim to push the boundaries of LLM evaluation beyond existing benchmarks. 

\vspace{-0.5mm}
\section{Method}
\label{method}
\vspace{-2mm}

In this section, we will first provide an overview of audio-video multi-modal agents in Sec. \ref{sec:multi-agent-systems}, followed by a detailed description and working of \ourapproach in Sec. \ref{sec:our-method}.

\customsubsection{Audio-Video Multi-Agent System}
\label{sec:multi-agent-systems}

Our interactive audio-video multi-agent system is structured as a tuple $\langle\mathcal{R},\mathcal{S},\mathcal{E},\mathcal{F}\rangle$, where multiple LLM-based agents collaboratively operate on the dataset comprising of video, audio and textual query, represented as $\langle\mathcal{V},\mathcal{A},\mathcal{Q}\rangle$, to enhance the performance of the target model \(\mathcal{M}\).  
As shown in Fig. \ref{fig:main-figure}, the reasoning generator agent $\mathcal{R}$ processes the input video $v \in V$ and audio $a \in A$ and produces a sequence of reasoning steps $r$ necessary for answering the given question $q \in Q$. Leveraging this information, the summarizer agent $\mathcal{S}$ extracts key cues and synthesizes them into a concise caption $s$ that encapsulates the core content of both the video $v \in \mathcal{V}$ and the audio $a \in \mathcal{A}$. The relevance of the reasoning steps generated by $\mathcal{R}$ is assessed by the quality of the caption produced by \(\mathcal{S}\).  
This assessment is conducted by the evaluation agent \(\mathcal{E}\), a multi-modal model that takes \(\{v, a, c\}\) as input and assigns a score quantifying the correctness and coherence of the reasoning steps. Based on this evaluation, a feedback mechanism (\(\mathcal{F}\)) iteratively refines the reasoning process by guiding \(\mathcal{R}\) toward more effective reasoning paths. This interaction functions as an actor-critic framework, continuously optimizing until a satisfactory evaluation score is achieved.  
Ultimately, the refined reasoning steps, along with the original inputs \(\langle v, a, q, r \rangle\), are fed into the target model \(\mathcal{M}\). This process enhances the model’s internal reasoning mechanism, leading to improved overall performance.  We further present \ourapproach mathematically in Algorithm \ref{alg:aurelia}.

% Our audio-video interactive multi-agent system can be modeled as a tuple of multiple agents $\langle\mathcal{R},\mathcal{S},\mathcal{E},\mathcal{F}\rangle$ operating on a set of input $\langle\mathcal{V,A,Q}\rangle$ and the target model $\mathcal{M}$. Here, $\mathcal{R}$ is the reasoning generator agent, which operates on the input and generates a set of reasoning steps $r$ required to answer the question. Based on the information generated by $\mathcal{R}$, the summarizer agent $\mathcal{S}$ combines the cues from this information and summarizes it into a concise caption $c$ that captures the overall essence of the video $v\in\mathcal{V}$ and the audio $a\in\mathcal{A}$. Whether the reasoning steps generated by $\mathcal{R}$ help answer the query $q\in\mathcal{Q}$ is measured by how good the caption generated by $\mathcal{S}$. This evaluation is carried out by another multi-modal agent $\mathcal{E}$, which operates on the input ${v,a,c}$ and outputs a score that quantifies the correctness of the reasoning steps. Depending on the evaluation score, a feedback mechanism $\mathcal{F}$ provides necessary feedback to the $\mathcal{R}$. The overall system works as an actor-critique until a favorable evaluation score is reached. Finally, the joint set of inputs $\langle\mathcal{v,a,q,r}\rangle$ is fed to the target model $\mathcal{M}$, helping distill the information from reasoning steps into its internal reasoning mechanism, resulting in improved performance.

\customsubsection{\ourapproach}
\label{sec:our-method}
\vspace{-0.5mm}
Our proposed \ourapproach enhances the performance of AVLLMS through a combination of multi-modal agents that interact with each other and generate a set of reasoning steps which distills the knowledge into the model in a training-free manner. Below, we describe the different components of \ourapproach and their working in detail.

\noindent
\textbf{Reasoning Generator.} The first component of \ourapproach is a multi-modal reasoning generation agent, denoted as \(\mathcal{R}\). Since our proposed method operates in a zero-shot setting, let \((x, y) \in \mathcal{D}^{test}\) represent samples from the test set, where each input \(x\) in \(\mathcal{D}^{test}\) is a tuple \(\langle v, a, q \rangle\), comprising a video \(v\), an audio \(a\), and a question \(q\).  
The agent \(\mathcal{R}\) processes this input tuple and produces three key outputs: a sequence of reasoning steps, a justification for these steps, and the final answer to the question. Formally,
\vspace{-0.5em}
\begin{equation}
    r = \{r_{1}, r_{2}, r_{3}\} = \mathcal{R}(v, a, q),
    \vspace{-0.5em}
\end{equation}  
where \(r_{1}\) represents the reasoning steps, \(r_{2}\) provides their justification, \(r_{3}\) is the final answer to the question \(q\).  
% \textbf{Reasoning Generator }The first component of AURELIA is a multi-modal reasoning generation agent denoted by $\mathcal{R}$. Since the proposed method works in zero-shot settings, consider ${(x,y) \in \mathcal{D}}$ as the samples from the test set. Each input $x$ in $\mathcal{D}^{test}$ consists a tuple $\langle v,a,q\rangle$. $\mathcal{R}$ operates upon the input tuple $\langle v,a,q\rangle$ and generates three key outputs: reasoning steps, justification for reasoning steps and a final answer to the question. Mathematically,
% \begin{equation}
%     r  = \{r_{1}, r_{2}, r_{3}\} = \mathcal{R}(v, a, q)
% \end{equation}
% where $r_{1}$ denotes the reasoning steps, $r_{2}$ denotes the justification for the reasoning steps and $r_{3}$ is the final answer to the question $q$.

\noindent
\textbf{Summarizer.} The summarizer agent, denoted as \(\mathcal{S}\), processes the reasoning information \(r\) generated in the previous stage along with the question $q$ and synthesizes them into a caption \(c\) such that $c = \mathcal{S}(r, q)$. This caption provides a comprehensive summary of the video and its corresponding audio, encapsulating key details in a concise manner.  
The accuracy and relevance of the generated caption \(c\) depend on both the reliability of reasoning steps and the final answer produced by the reasoning generator agent. To ensure consistency and correctness, we introduce an evaluation agent $\mathcal{E}$ that assesses caption in relation to given audio and video.  

% \noindent
% \textbf{Summarizer }The information generated in the previous stage is acted upon by a summarizer agent denoted as $\mathcal{S}$ which interacts with the reasoning information $r$ and, based on it, generates  a caption $c$ that provides a detailed overview of the video along with its corresponding audio.
% \begin{equation}
%     c = \mathcal{S}(r)
% \end{equation}  
% The correctness of the caption $c$ generated depends on how reliable the reasoning information is and the final answer generated by the reasoning generator agent. Therefore, we employ another agent which evaluates the caption in relation to the audio and the video.

\noindent
\textbf{Evaluator.} The reliability of the reasoning steps and the generated answer directly impact the summarizer agent, which synthesizes the content into a detailed caption. Consequently, the quality of the caption is inherently tied to the correctness of the reasoning process. We hypothesize that an accurate caption aligns closely with well-formed reasoning steps, ultimately leading to a correct final answer.  

To assess this alignment, we introduce a multi-modal evaluation agent $\mathcal{E}$ that serves as a judge. This agent receives the video \(v\), audio \(a\), and the corresponding caption \(c\) as input and assigns an evaluation score \(e\) based on their coherence. The score ranges from 1 to 10, where 1 indicates minimal alignment between the caption and the input data, while 10 signifies a perfect match.  
\vspace{-0.5em}
\begin{equation}
    e = \mathcal{E}(v, a, c),
    \vspace{-0.5em}
\end{equation}  
where \(e \in [1, 10]\) quantifies the relevance of the caption to the input signals and, by extension, evaluates the effectiveness of the reasoning steps in deriving the final answer.  
% \noindent
% \textbf{Evaluator }As mentioned previously, the reliability of the reasoning steps and the answer generated directly influences the summarizer agent responsible for summarizing the content into a detailed caption. Therefore, the aptness of the caption explicitly depends on the correctness of the reasoning steps generated in the previous stage. We hypothesize that a correct caption correlates well with the correct reasoning steps in order to reach the final answer. We employ a multi-modal agent as a judge and feed it with the video $v$, audio $a$ and the corresponding caption $c$. Based on its judgment, the agent outputs an score $e$ between 1 and 10, where 1 denotes the lowest score based on the similarity of a caption with the corresponding inputs, and 10 denotes that the caption perfectly fits the inputs.
% \begin{equation}
%     e = \mathcal{E}(v,a,c),
% \end{equation} 
% where $e \in [1, 10]$ is the evaluation score which gives an idea of how releavant the caption is to the input signals and eventually how useful are the reasoning steps generated to reach the final answer.

\noindent
\textbf{Feedback Mechanism. }Based on the evaluation score obtained in the previous step, we follow an Actor-Critic framework that facilitates iterative
agent improvement through a feedback loop. In this case, the Actor is the Reasoning generating agent $\mathcal{R}$ which is evaluated by another agent $\mathcal{E}$ acting as a judge and based on the evaluation score, the Critic agent provides feedback to guide the Actor Agent in regenerating improved solutions. Let $\mathcal{F}$ be the feedback mechanism facilitating the interaction between the Actor and Critic, then the goal of the feedback mechanism is to maximize the evaluation score $e$ such that $e$ is above a certain threshold $\tau$.
\vspace{-1em}
\begin{equation}
    r^* = \arg\max_{r^{(t)}} \quad e^{(t)}, \quad \text{s.t} \quad e^{(t)} \geq \tau, \quad t \leq T.
    \vspace{-0.5em}
\end{equation}
If \( e^{(t)} \geq \beta \) at any iteration \( t \), the process terminates and returns the corresponding reasoning steps \( r^* \). Otherwise, the system continues iterating, refining \( r^{(t)} \) through \(\mathcal{F}\) until \( T \) iterations are exhausted.  

\noindent \textbf{Reasoning Distillation. }
The optimal reasoning steps \( r^* \), obtained through the multi-agent interaction process, serve as a structured sequence of logical inferences and contextual cues that can enhance the target model ($\mathcal{M}$) response. These steps encapsulate the essential knowledge relationships and transformations necessary to bridge the input modalities to derive an accurate and well-grounded solution. In other words, the knowledge inside the reasoning information is distilled in a training-free manner inside the target model $\mathcal{M}$ which now receives a refined and enriched input containing reasoning steps, in addition to the raw audio, video and the question, that highlight key features, intermediate conclusions, and decision pathways. By conditioning the target model $\mathcal{M}$ on the distilled reasoning steps $r^*$, we facilitate a more structured decision-making process, reducing ambiguity and improving model interpretability. The optimal solution $s^{*}$ is formulates as,
% The optimal reasoning steps \( r^* \) obtained through the above multi-agent interaction system contain a sequence of reasoning steps that can assist the target model $\mathcal{M}$ by distilling necessary information inside the reason steps into the model, aiding it to reach an optimal solution to the original question $q$. In other words, the optimal solution $s^{*}$ is obtained as,
\vspace{-1em}
\begin{equation}
    s^{*} = \mathcal{M}(v, a, q,  r^{*})
    % \vspace{-1.0em}
\end{equation}
% \vspace{-0.8em}
\begin{tcolorbox}[colback=gray!20, colframe=black!50]
\includegraphics[width=0.3cm]{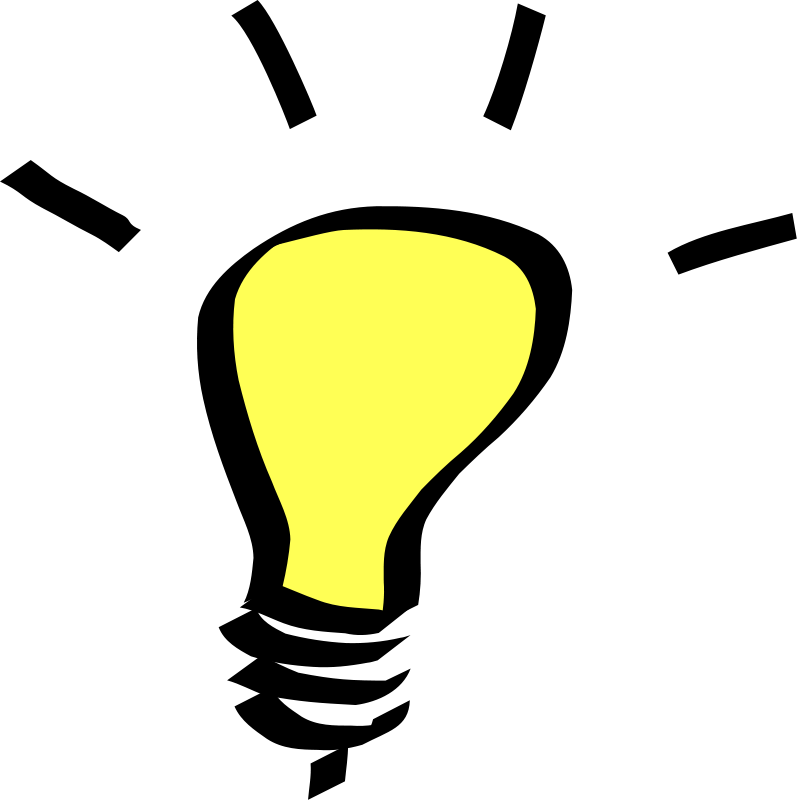} {\small \textbf{\ourapproach} is the first multi-agent framework capable of reasoning distillation in Audio Visual LLMs through an iterative actor-critique mechanism.}

\includegraphics[width=0.3cm]{ICCV2025-Author-Kit-Feb/figures/lightbulb_logo.png} {\small \textbf{\ourapproach} systematically mitigates visual and auditory biases by enforcing a structured reasoning process, leading to more objective and reliable cross-modal comprehension.}  

\includegraphics[width=0.3cm]{ICCV2025-Author-Kit-Feb/figures/lightbulb_logo.png} {\small \textbf{\ourapproach} can scale and generalize to diverse audio-visual reasoning tasks due to zero-shot nature, where fine-tuning methods often fail due to training biases.}
\end{tcolorbox}

\section{\ourbenchmark: Audio-Visual Reasoning Benchmark}

\customsubsection{\textcolor{blue}{Why Designing AV Reasoning Tasks are Difficult?}}
% % \textcolor{red}{Sanjoy}
% \begin{tcolorbox}[colback=gray!20, colframe=black!50]
% \includegraphics[width=0.3cm]{ICCV2025-Author-Kit-Feb/figures/lightbulb_logo.png} {\small }

% \includegraphics[width=0.3cm]{ICCV2025-Author-Kit-Feb/figures/lightbulb_logo.png} {\small }  

% \includegraphics[width=0.3cm]{ICCV2025-Author-Kit-Feb/figures/lightbulb_logo.png} {\small }
% \end{tcolorbox}
\noindent{\textbf{Limitations in Forming Question-Answer Pairs for AV Setup.}} In vision-language tasks, the formation of question-answer pairs is relatively simple since objects have visible attributes (e.g., "What color is the book?"). However, in audio-visual reasoning, many objects do not make an inherent sound, making it harder to design meaningful QA pairs. For instance, "What does the book sound like?" lacks relevance unless an action (e.g., flipping pages) is involved. This necessitates carefully crafting interactions where both audio and visual cues contribute meaningfully.

\noindent{\textbf{Ambiguity in Audio-Visual Associations.}} Interpreting emotional tone in audio-visual tasks is challenging because the same visual cue, such as laughter, can convey different meanings depending on the accompanying audio. Cheerful music may indicate joy, while eerie background sounds might suggest nervousness or fear. Unlike vision-language tasks, where textual cues explicitly define emotions, AV models must infer meaning from the interplay of sound and visuals, requiring deeper multi-modal understanding. To encompass these scenarios we incorporate AV compositional understanding, meme understanding and dance matching tasks.

\noindent{\textbf{Cultural and Contextual Understanding.}} Object recognition and language understanding can often be generalised across cultures. If an image contains sushi, the model can easily label it as "sushi" using object detection and language mapping. However, AV tasks require deeper cultural and contextual awareness. For example, in music-dance matching, Flamenco music should pair with Flamenco dance rather than Hip-Hop. Similarly, laughter in a scene could indicate humour, but it could also indicate nervousness, depending on the visual cues. To address this gap we introduce \ourtask.

Audio-visual tasks pose additional challenges compared to only language or vision-language tasks due to the need for temporal synchronization \cite{chowdhury2024melfusion}, ambiguity resolution, noise handling, and cultural grounding. These challenges demand more sophisticated models that can process and align multi-modal inputs dynamically over time, making AV reasoning a significantly harder problem than L/VL reasoning.
% \end{tcolorbox}

\customsubsection{Task Overview}

\noindent{\textbf{Audio-Visual Question Answering.}} 
Audio-visual question answering (AVQA) focuses on responding to questions that require both auditory and visual understanding. To construct our dataset, we gather question-answer pairs from AVSD \cite{avsd} and MusicAVQA \cite{musicavqadataset} enhancing them with detailed reasoning steps. We carefully curate samples which require strong audio-visual comprehension in terms of their interplay, association, dependency, etc.

\noindent{\textbf{Audio-Visual Captioning.}} This task involves generating detailed textual descriptions based on audio-visual inputs. Unlike image- or audio-only captioning methods, it demands robust multimodal understanding and advanced reasoning capabilities. We obtain samples from VALOR \cite{valor} for this task and augment them with reasoning annotations. 
% We note that after the ZS distillation of the models using our reasoning annotations, the models demonstrate improved performance.   

\noindent{\textbf{Audio-Visual Compositional Attribute Understanding.}}
% Multi-event audio-visual inputs often involve distortions in the sequence of events and their associated attribute bindings. The core objective of multimodal processing is to understand how linguistic components align with the content of audio-video input pairs. Therefore, it is crucial for AVLLMs to recognise that varying word arrangements within a sentence can lead to different multimodal interpretations. Inspired by AVTrustBench \cite{avtrustbench} in this AV compositional understanding task the model is prompted to comment on the association of different events and if their corresponding attribute binding. 
Inspired by \cite{avtrustbench}, in this task we ensure each AV pair contains two separate events which are associated with two different attributes. For example, `a cow is mooing' and a `sheep is bleating'. Here the answer choices contain the same words but in a different sequence (`cow is bleating' and `sheep is mooing'). An AVLLM must have a strong AV and linguistic understanding to comprehend the constituent modalities and semantically align them with the correct attributes.

\begin{algorithm}[t]
\caption{\ourapproach}
\label{alg:aurelia}
\begin{algorithmic}[1] % The number inside [] controls line numbering
    \State \textbf{Input:} Data: $D^{test}$, Reasoning generator $\mathcal{R}$, Summarizer $\mathcal{S}$, Evaluator $\mathcal{E}$, Iterations $T$, Threshold $\tau$
    \State \textbf{Output:} Optimized Reasoning Steps $r^*$
    \State Sample data: $\langle$Audio $a$, Video $v$, Question $q$$\rangle$ $\subseteq \mathcal{D}^{test}$ 
    \State Set iteration counter, $t = 1$ 
    % \State \textbf{Iterative Reasoning Optimization}  
    \While{$e \leq \tau$ \textbf{and} $t \leq T$}
        \State \colorbox{red!10}{Generate Reasoning Steps, $r^{(t)} = \mathcal{R}(v, a, q)$}
        
        \State \colorbox{blue!10}{ Generate Caption,  
         $c^{(t)} = \mathcal{S}(r^{(t)}, q)$}

        \State \colorbox{green!10}{Evaluate Generated Caption, 
        $e^{(t)} = \mathcal{E}(v, a, c^{(t)})$}

        \State \colorbox{orange!10}{Feedback (Repeat Steps 6-8), 
        $\mathcal{F}(v, a, q, e^{(t)})$}
        \State Update $t \gets t + 1$
    \EndWhile
    \State \fcolorbox{red}{white}{Select Optimal Reasoning, $r^* = \arg\max_{r^{(t)}} e^{(t)}$}
    \State \Return $r^*$
\end{algorithmic}
\end{algorithm}
% \vspace{-1em}

\begin{table*}[t]
\small
    \centering
    \renewcommand{\arraystretch}{0.82}
    \setlength{\tabcolsep}{8pt}
    \resizebox{\linewidth}{!}{%
    \begin{tabular}{l|c|c|c|c|c|c|c}
\hline
\rowcolor{gray!20}
\multicolumn{1}{c|}{} &
  \multicolumn{2}{c|}{\textbf{AV-QA}} &
  \multicolumn{1}{c|}{} &
  \multicolumn{1}{c|}{\multirow{1}{*}{}} &
  \multicolumn{1}{c|}{\multirow{1}{*}{}} &
  \multicolumn{1}{c|}{\multirow{1}{*}{}} &
  \multicolumn{1}{c}{\multirow{1}{*}{}} \\
\cline{2-3}
\rowcolor{gray!20}
\multirow{-2}{*}{\textbf{Models}} &
  \textbf{Music-AVQA} &
  \textbf{AVSD} & \multirow{-2}{*}{\textbf{AV-Captioning}}
   & \multirow{-2}{*}{\textbf{AV-Compositional}}
   & \multirow{-2}{*}{\textbf{AV-GeoIQ}}
   & \multirow{-2}{*}{\textbf{AV-Meme}}
   & \multirow{-2}{*}{\textbf{DM-Match}}
\\
\hline
\multicolumn{8}{c}{\textit{Closed-Source Models}} \\
\hline
\cellcolor{gray!20}Gemini 1.5 Pro & 70.6 / 68.9 & 74.7 / 72.5 & 84.9 / 82.7 & 38.9 / 36.8 & 71.2 / 68.0 & 52.0 / 49.0 & 43.4 / 41.5\\
\cellcolor{gray!20}Reka Core & 67.9 / 64.3 & 74.5 / 69.5 & 83.2 / 80.4 & 38.6 / 35.3 & 45.7 / 42.5 & 24.0 / 19.0 & 35.8 / 32.5\\
\hline
\multicolumn{8}{c}{\textit{Open-Source Models in ZS}} \\
\hline
\cellcolor{gray!20}PandaGPT (13B) & 35.8 / 33.7 & 29.1 / 26.1 & 67.8 / 64.7 & 28.8 / 24.1 & 17.2 / 12.5 & 25.0 / 21.0 & 30.2 / 27.0\\
\cellcolor{gray!20}Macaw-LLM (7B) & 34.7 / 31.8 & 38.4 / 34.3 & 67.7 / 65.9 & 26.1 / 24.3 & 17.2 / 14.0 & 18.0 / 14.0 & 24.5 / 20.0 \\
\cellcolor{gray!20}VideoLLaMA (7B) & 39.1 /36.6 & 40.0 / 36.7 & 68.4 / 66.2 & 28.8 / 25.8 & 19.3 / 16.5 & 18.0 / 16.0 & 26.6 / 23.0\\
\cellcolor{gray!20}ImageBind-LLM & 44.2 / 43.9 & 42.7 / 39.2 & 69.0 / 66.9 & 28.8 / 25.4 & 18.0 / 13.0 & 17.7 / 15.0 & 26.2 / 22.5 \\
\cellcolor{gray!20}X-InstructBLIP (13B) & 47.8 / 44.5 & 43.9 / 40.1 & 69.5 / 66.1 & 27.5 / 25.9 & 27.6 / 14.5 & 18.7 / 15.0 & 27.3 / 24.5\\
\cellcolor{gray!20}AV-LLM (13B) & 48.2 / 45.2 & 55.4 / 52.6 & 70.1 / 67.6 & 29.6 / 26.1 & 18.0 / 14.5 & 24.4 / 20.0 & 29.4 / 27.0 \\
\cellcolor{gray!20}OneLLM (7B) & 49.9 / 47.6 & 52.3 / 49.8 & 71.6 / 68.1 & 29.7 / 26.3 & 20.9 / 17.0 & 24.5 / 18.0 & 28.8 / 26.5\\
\cellcolor{gray!20}AVicuna (7B) & 51.6 / 49.6 & 56.2 / 53.1 & 71.2 / 67.9 & 29.6 / 26.6 & 19.7 / 16.5 & 28.4 / 23.0 & 29.6 / 27.0 \\
\cellcolor{gray!20}CREMA (4B) & 56.8 / 52.6 & 62.3 / 58.6 & 73.8 / 68.4 & 31.6 / 27.0 & 23.8 / 19.0 & 29.0 / 26.0 & 31.5 / 28.5\\
\cellcolor{gray!20}VideoLLaMA2 (7B) & - & - & 70.4 / 68.3 & 29.7 / 26.8 & 25.7 / 22.0 & 27.5 / 23.0 & 28.4 / 25.5\\
\cellcolor{gray!20}AnyGPT (7B) & 53.7 / 50.7 & 59.2 / 56.9 & 72.5 / 68.1 & 28.8 / 26.2 & 25.7 / 22.5 & 24.0 / 19.0 & 28.9 / 25.5\\
\cellcolor{gray!20}NExT-GPT (7B) & 53.5 / 50.9 & 58.4 / 56.3 & 68.7 / 67.9 & 28.0 / 26.4 & 23.8 / 22.0 & 19.5 / 16.0 & 32.3 / 28.0\\
\cellcolor{gray!20}Unified-IO-2 L (6.8B) & 58.3 / 55.1 & 60.0 / 57.9 & 73.8 / 70.1 & 31.8 / 27.2 & 25.6 / 21.5 & 26.5 / 22.0 & 29.3 / 27.5 \\
\cellcolor{gray!20}Unified-IO-2 XL & 61.3 / 57.2 & 59.7 / 58.6 & 73.7 / 71.8 & 30.0 / 28.5 & 24.7 / 22.5 & 29.0 / 26.0 & 29.6 / 27.0 \\
\cellcolor{gray!20}Bay-CAT (7B) & 55.6 / 53.8 & 58.3 / 56.5 & 71.9 / 69.5 & 31.9 / 28.2 & 24.4 / 20.5 & 22.0 / 18.0 & 29.8 / 27.5\\
\cellcolor{gray!20}Video-SALMONN (7B) & 56.8 / 54.9 & 58.7 / 57.2 & 71.1 / 70.2 & 29.8 / 27.5 & 24.7 / 22.0 & 21.0 / 17.0 & 27.5 / 26.5\\
\cellcolor{gray!20}VITA (7B) & 59.0 / 58.6 & 61.2 / 60.1 & 73.8 / 72.9 & 30.1 / 29.2 & 26.7 / 25.5 & 44.0 / 41.0 & 29.2 / 27.5\\
\hline
\multicolumn{8}{c}{\textit{Open-Source Models with \ourapproach}} \\
\hline
\rowcolor{magenta!5}
\cellcolor{gray!20}PandaGPT (13B) & 41.9$^{\textcolor{teal}{\text{+24.33\%}}}$ & 32.7$^{\textcolor{teal}{\text{+25.28\%}}}$ & 72.9$^{\textcolor{teal}{\text{+12.67\%}}}$ & 28.6$^{\textcolor{teal}{\text{+18.67\%}}}$ & 25.0$^{\textcolor{teal}{\text{+100\%}}}$ & 25.0$^{\textcolor{teal}{\text{+19.04\%}}}$ & 31.0$^{\textcolor{teal}{\text{+14.81\%}}}$ \\
\rowcolor{magenta!5}
\cellcolor{gray!20}Macaw-LLM (7B) & 41.6$^{\textcolor{teal}{\text{+30.81\%}}}$ & 38.1$^{\textcolor{teal}{\text{+11.07\%}}}$ & 73.5$^{\textcolor{teal}{\text{+11.53\%}}}$ & 29.3$^{\textcolor{teal}{\text{+20.57\%}}}$ & 25.5$^{\textcolor{teal}{\text{+82.14\%}}}$ & 24.0$^{\textcolor{teal}{\text{+71.42\%}}}$ & 28.5$^{\textcolor{teal}{\text{+42.5\%}}}$ \\
\rowcolor{magenta!5}
\cellcolor{gray!20}VideoLLaMA (7B) & 45.8$^{\textcolor{teal}{\text{+25.13\%}}}$ & 41.5$^{\textcolor{teal}{\text{+13.07\%}}}$ & 74.2$^{\textcolor{teal}{\text{+12.08\%}}}$ & 29.6$^{\textcolor{teal}{\text{+14.72\%}}}$ & 28.5$^{\textcolor{teal}{\text{+72.72\%}}}$ & 28.0$^{\textcolor{teal}{\text{+75.0\%}}}$ & 29.0$^{\textcolor{teal}{\text{+26.08\%}}}$\\
\rowcolor{magenta!5}
\cellcolor{gray!20}ImageBind-LLM & 49.7$^{\textcolor{teal}{\text{+13.21\%}}}$ & 44.2$^{\textcolor{teal}{\text{+12.75\%}}}$ & 72.8$^{\textcolor{teal}{\text{+8.81\%}}}$ & 30.1$^{\textcolor{teal}{\text{+18.50\%}}}$ & 28.0$^{\textcolor{teal}{\text{+100\%}}}$ & 23.0$^{\textcolor{teal}{\text{+53.33\%}}}$ & 31.0$^{\textcolor{teal}{\text{+37.77\%}}}$\\
\rowcolor{magenta!5}
\cellcolor{gray!20}X-InstructBLIP (13B) & 52.3$^{\textcolor{teal}{\text{+17.52\%}}}$ & 46.9$^{\textcolor{teal}{\text{+16.95\%}}}$ & 72.6$^{\textcolor{teal}{\text{+9.83\%}}}$ & 29.8$^{\textcolor{teal}{\text{+15.05\%}}}$ & 29.0$^{\textcolor{teal}{\text{+100\%}}}$ & 27.0$^{\textcolor{teal}{\text{+80.0\%}}}$ & 30.0$^{\textcolor{teal}{\text{+22.45\%}}}$\\
\rowcolor{magenta!5}
\cellcolor{gray!20}AV-LLM (13B) & 52.7$^{\textcolor{teal}{\text{+16.59\%}}}$ & 57.9$^{\textcolor{teal}{\text{+10.07\%}}}$ & 73.4$^{\textcolor{teal}{\text{+8.57\%}}}$ & 31.1$^{\textcolor{teal}{\text{+19.15\%}}}$ & 28.5$^{\textcolor{teal}{\text{+83.87\%}}}$ & 29.0$^{\textcolor{teal}{\text{+45.0\%}}}$ & 34.0$^{\textcolor{teal}{\text{+25.92\%}}}$\\
\rowcolor{magenta!5}
\cellcolor{gray!20}OneLLM (7B) & 54.1$^{\textcolor{teal}{\text{+13.65\%}}}$ & 55.3$^{\textcolor{teal}{\text{+11.04\%}}}$ & 73.9$^{\textcolor{teal}{\text{+8.51\%}}}$ & 30.7$^{\textcolor{teal}{\text{+16.73\%}}}$ & 29.0$^{\textcolor{teal}{\text{+70.58\%}}}$ & 29.0$^{\textcolor{teal}{\text{+61.11\%}}}$ & 33.5$^{\textcolor{teal}{\text{+26.41\%}}}$\\
\rowcolor{magenta!5}
\cellcolor{gray!20}AVicuna (7B) & 55.3$^{\textcolor{teal}{\text{+11.49\%}}}$ & 57.8$^{\textcolor{teal}{\text{+8.85\%}}}$ & 73.1$^{\textcolor{teal}{\text{+7.65\%}}}$ & 30.4$^{\textcolor{teal}{\text{+14.28\%}}}$ & 29.5$^{\textcolor{teal}{\text{+79.09\%}}}$ & 34.0$^{\textcolor{teal}{\text{+47.80\%}}}$ & 34.5$^{\textcolor{teal}{\text{+27.78\%}}}$\\
\rowcolor{magenta!5}
\cellcolor{gray!20}CREMA (4B) & 59.8$^{\textcolor{teal}{\text{+13.68\%}}}$ & 67.2$^{\textcolor{teal}{\text{+14.67\%}}}$ & 74.2$^{\textcolor{teal}{\text{+8.47\%}}}$ & 31.9$^{\textcolor{teal}{\text{+18.14\%}}}$ & 32.5$^{\textcolor{teal}{\text{+71.05\%}}}$ & 40.0$^{\textcolor{teal}{\text{+53.84\%}}}$ & 34.0$^{\textcolor{teal}{\text{+19.29\%}}}$\\
\rowcolor{magenta!5}
\cellcolor{gray!20}VideoLLaMA2 (7B) & - & - & 74.7$^{\textcolor{teal}{\text{+9.37\%}}}$ & 31.6$^{\textcolor{teal}{\text{+17.91\%}}}$ & 38.0$^{\textcolor{teal}{\text{+72.72\%}}}$ & 35.0$^{\textcolor{teal}{\text{+40.0\%}}}$ & 34.5$^{\textcolor{teal}{\text{+35.29\%}}}$\\
\rowcolor{magenta!5}
\cellcolor{gray!20}AnyGPT (7B) & 56.2$^{\textcolor{teal}{\text{+10.84\%}}}$ & 62.5$^{\textcolor{teal}{\text{+9.84\%}}}$ & 73.3$^{\textcolor{teal}{\text{+7.63\%}}}$ & 31.4$^{\textcolor{teal}{\text{+19.84\%}}}$ & 35.5$^{\textcolor{teal}{\text{+57.77\%}}}$ & 33.0$^{\textcolor{teal}{\text{+73.68\%}}}$ & 33.0$^{\textcolor{teal}{\text{+29.41\%}}}$\\
\rowcolor{magenta!5}
\cellcolor{gray!20}NExT-GPT (7B) & 57.8$^{\textcolor{teal}{\text{+13.55\%}}}$ & 60.8$^{\textcolor{teal}{\text{+7.99\%}}}$ & 73.5$^{\textcolor{teal}{\text{+8.25\%}}}$ & 31.8$^{\textcolor{teal}{\text{+20.45\%}}}$ & 36.0$^{\textcolor{teal}{\text{+63.63\%}}}$ & 32.0$^{\textcolor{teal}{\text{+100\%}}}$ & 33.5$^{\textcolor{teal}{\text{+19.64\%}}}$\\
\rowcolor{magenta!5}
\cellcolor{gray!20}Unified-IO-2 L (6.8B)& 61.9$^{\textcolor{teal}{\text{+12.34\%}}}$ & 62.0$^{\textcolor{teal}{\text{+7.08\%}}}$ & 74.6$^{\textcolor{teal}{\text{+6.41\%}}}$ & 32.4$^{\textcolor{teal}{\text{+19.11\%}}}$ & 36.5$^{\textcolor{teal}{\text{+69.76\%}}}$ & 35.0$^{\textcolor{teal}{\text{+59.09\%}}}$ & 33.5$^{\textcolor{teal}{\text{+21.81\%}}}$\\
\rowcolor{magenta!5}
\cellcolor{gray!20}Unified-IO-2 XL (6.8B) & 62.3$^{\textcolor{teal}{\text{+8.91\%}}}$ & 62.8$^{\textcolor{teal}{\text{+7.16\%}}}$ & 75.6$^{\textcolor{teal}{\text{+5.29\%}}}$ & 33.6$^{\textcolor{teal}{\text{+17.89\%}}}$ & 38.5$^{\textcolor{teal}{\text{+71.11\%}}}$ & 40.0$^{\textcolor{teal}{\text{+53.84\%}}}$ & 34.0$^{\textcolor{teal}{\text{+25.92\%}}}$\\
\rowcolor{magenta!5}
\cellcolor{gray!20}Bay-CAT (7B) & 58.5$^{\textcolor{teal}{\text{+8.73\%}}}$ & 61.1$^{\textcolor{teal}{\text{+8.14\%}}}$ & 75.0$^{\textcolor{teal}{\text{+7.91\%}}}$ & 32.7$^{\textcolor{teal}{\text{+15.95\%}}}$ & 34.0$^{\textcolor{teal}{\text{+65.85\%}}}$ & 35.0$^{\textcolor{teal}{\text{+94.40\%}}}$ & 32.5$^{\textcolor{teal}{\text{+18.18\%}}}$\\
\rowcolor{magenta!5}
\cellcolor{gray!20}Video-SALMONN (7B) & 59.8$^{\textcolor{teal}{\text{+8.92\%}}}$ & 61.7$^{\textcolor{teal}{\text{+7.86\%}}}$ & 75.2$^{\textcolor{teal}{\text{+7.12\%}}}$ & 32.5$^{\textcolor{teal}{\text{+18.18\%}}}$ & 37.5$^{\textcolor{teal}{\text{+70.45\%}}}$ & 32.0$^{\textcolor{teal}{\text{+88.23\%}}}$ & 33.0$^{\textcolor{teal}{\text{+24.52\%}}}$\\
\rowcolor{magenta!5}
\cellcolor{gray!20}VITA (7B) & 62.6$^{\textcolor{teal}{\text{+6.82\%}}}$ & 66.5$^{\textcolor{teal}{\text{+10.64\%}}}$ & 78.8$^{\textcolor{teal}{\text{+8.09\%}}}$ & 33.8$^{\textcolor{teal}{\text{+15.75\%}}}$ & 39.0$^{\textcolor{teal}{\text{+52.94\%}}}$ & 50.0$^{\textcolor{teal}{\text{+21.95\%}}}$ & 35.0$^{\textcolor{teal}{\text{+27.27\%}}}$\\
  \hline
\end{tabular}%
}
% \vspace{0.05in}
\caption{\textbf{Performance comparison of various models across multiple  tasks in \ourbenchmark}. The lower section highlights the performance improvement using \ourapproach. The numbers in \textcolor{teal}{teal} denotes relative gains over ZS results. Video-LLamA2 zero-shot is not reported because the publicly available model is already fine-tuned on the dataset. For ZS evaluation \textit{A/B} represents best/mean of 3 runs evaluation. AV-Captioning values denote CIDEr scores. }
\label{tab:main_results}
\vspace{-1em}
\end{table*}

\noindent{\textbf{\ourtask.}}
We introduce \ourtask, a \textit{novel} audio-visual reasoning task that integrates commonsense understanding with geographical and country-specific knowledge. This task challenges models to process and reason over multimodal inputs, requiring the alignment of audio cues, visual elements, and world knowledge. Unlike standard audio-visual question-answering tasks, \ourtask extends beyond perceptual understanding by incorporating reasoning over cultural and geographic attributes. For example, a question like \textit{"What is the most famous drink of the country where the instrument to the left of the louder sounding instrument originates?"} necessitates multiple reasoning steps: \underline{identifying the loudest instrument}, \underline{determining the relative position of another instrument}, \underline{recognizing its country of origin}, and retrieving cultural knowledge about that \underline{country’s famous drinks}—leading to the answer \textit{Sangria}. Such questions require deep multimodal comprehension, contextual association, and factual world knowledge. AV-GeoIQ (\cref{fig:teaser}) serves as a benchmark to evaluate the reasoning capabilities of AVLLMs in handling complex, real-world scenarios that go beyond direct perception.

\noindent{\textbf{AV Meme Understanding.}} Inspired by AV-Odyssey Bench \cite{avodcbench}, we include AV-Meme a task that challenges models to interpret humour, sarcasm, and the context in multimodal memes by analyzing visual elements, audio cues, and text. Unlike traditional meme analysis, AV-Meme requires grasping subtle relationships between sound effects, expressions, and captions. For example, dramatic music over an ordinary event or mismatched audio-visual pairings create irony, demanding nuanced cultural awareness. This task serves as a benchmark for evaluating AVLLMs in recognizing implicit meanings and internet humour.

\noindent{\textbf{Dance and Music Matching.}} We also include Dance-Music Matching (DM-Match) \cite{avodcbench}, a task that evaluates a model’s ability to align dance movements with appropriate musical styles by analyzing audio-visual correlations. Unlike standard motion or music classification, DM-Match requires understanding rhythm, tempo, and movement patterns to determine whether a given dance sequence matches the accompanying music. For instance, a ballet performance set to fast-paced electronic music may indicate a mismatch, while a tango paired with traditional tango music would be correct. This task serves as a benchmark for assessing AVLLMs in capturing temporal synchronization, genre compatibility, and expressive coherence between dance and music.

% \subsection{Adaptation of Public Datasets}
% Depending on the task and availability of datasets, we either collect the image-audio pairs directly from the publicly available dataset (MusicAVQA \cite{musicavqadataset}, AVSD \cite{avsd} VALOR \cite{valor}, AVTrustBench \cite{avtrustbench}, AV-Odyssey Bench \cite{avodcbench}) or carefully manually curate the samples from the web. In either case, for each sample we supplement the video-audio pairs with the generated reasoning as
% explained next.

\customsubsection{\ourbenchmark Size} 
We carefully curate 1000 samples each from Music-AVQA, AVSD, and VALOR which are suitable for AV reasoning. For the AV compositional understanding task, we collect 1000 samples from the web through careful manual inspection. For \ourtask we again tailor-make 200 samples which require strong AV reasoning capabilities. We augment more videos to the original AV-meme set to make a total of 100 test samples while we adapt 200 samples of DM-Match to make the total size of our reasoning benchmark, \ourbenchmark to \ourbenchmarksamples. We add further details in the supplementary.

% the model must systematically process the multi-modal input using structured reasoning. The first step is to identify the loudest instrument in the video using audio analysis. Next, the model determines the instrument positioned to the right of the loudest one, establishing a spatial relationship. Following this, it must infer the country of origin of the identified instrument based on external knowledge. Finally, the model retrieves the most famous food associated with that country, leading to the final answer. By enforcing this structured reasoning, the model avoids arbitrary leaps in logic and ensures that each inference is well-grounded in the input data. This explicit reasoning injection significantly enhances the model's ability to derive accurate, interpretable answers while minimizing errors and hallucinations.

\begin{table}[t]
    \centering
    \renewcommand{\arraystretch}{1.3}
    \begin{minipage}{0.25\textwidth} % Adjust width for table
        \resizebox{\textwidth}{!}
        {
        \begin{tabular}{l|ccc}
            \hline
            \cellcolor{gray!20} \textbf{Model} &  
            \multicolumn{3}{c}{\cellcolor{gray!20} \textbf{AV-Captioning}} \\
            \cline{2-4}
            \cellcolor{gray!20} & 
            \cellcolor{gray!10} \bf BLEU@4 $\uparrow$ & \cellcolor{gray!10} \bf METEOR $\uparrow$ & \cellcolor{gray!10} \bf ROUGE $\uparrow$ \\
            \hline
            \multicolumn{4}{c}{\textit{Zero-shot}} \\
            \hline
            \cellcolor{gray!20} AVLLM  & 10.2 & 18.1 & 34.6   \\
            \cellcolor{gray!20} OneLLM & 11.3 & 19.7 & 36.1  \\
            \cellcolor{gray!20} AVicuna  & 10.6 & 19.1 & 35.4  \\
            \cellcolor{gray!20} CREMA  & 11.5 & 20.1 & 36.9  \\
            \cellcolor{gray!20} VITA & 12.9 & 22.8 & 40.3   \\
            \hline
            \multicolumn{4}{c}{\textit{Zero-shot with \ourapproach}} \\
            \hline
            \rowcolor{magenta!5}
            \cellcolor{gray!20} AVLLM  & 12.8 & 21.9 & 40.7   \\
            \rowcolor{magenta!5}
            \cellcolor{gray!20} OneLLM & 14.1 & 24.3 & 42.1   \\
            \rowcolor{magenta!5}
            \cellcolor{gray!20} AVicuna  & 12.8 & 23.7 & 41.8  \\
            \rowcolor{magenta!5}
            \cellcolor{gray!20} CREMA  & 13.8 & 24.9 &  43.3  \\
            \rowcolor{magenta!5}
            \cellcolor{gray!20} VITA  & 14.5 & 26.0 &  46.4  \\
            \hline
        \end{tabular}
        }
    \end{minipage}%
    \hfill
    \begin{minipage}{0.22\textwidth} % Adjust width for caption
        \caption{\textbf{Evaluation results of five models on the AV-Captioning}. The top section indicates ZS inference results of models. The bottom section indicates results after reasoning distillation with \ourapproach. Clearly, the quality of the captions improves with our reasoning pipeline.}
        \label{avcap_main}
    \end{minipage}
    \vspace{-2em}
\end{table}

\begin{table}
\centering
\label{gemini-eval-results-wrt-duration-topic}
\begin{adjustbox}{max width=\linewidth}
\begin{tabular}{cclllllll} 
\toprule
\multicolumn{1}{c}{\cellcolor{gray!20}} & \multicolumn{1}{c}{\cellcolor{gray!20}} & \multicolumn{7}{c}{\cellcolor{gray!20}\textbf{Category}} \\ 
\cmidrule{3-9}

         \multicolumn{1}{c}{\multirow{1}{*}{\cellcolor{gray!20}\textbf{Subset}}} 
    
    & \multicolumn{1}{c}{\multirow{1}{*}{\cellcolor{gray!20}\textbf{Modality}}}                               & \multicolumn{1}{c}{\multirow{1}{*}{\cellcolor{gray!20}\textit{Knowledge~}}} & \multicolumn{1}{c}{\cellcolor{gray!20}\textit{Film \&}} & \multicolumn{1}{c}{\cellcolor{gray!20}\textit{Sports}}      & \multicolumn{1}{c}{\cellcolor{gray!20}\textit{Artistic~}}   & \multicolumn{1}{c}{\cellcolor{gray!20}\textit{Life}}   & \multicolumn{1}{c}{\multirow{1}{*}{\cellcolor{gray!20}\textit{Multilingual }}} & \multicolumn{1}{c}{\multirow{1}{*}{\cellcolor{gray!20}Overall}}  \\
    \multicolumn{1}{c}{\cellcolor{gray!20}}
    & \multicolumn{1}{c}{\cellcolor{gray!20}}                           & \multicolumn{1}{c}{\cellcolor{gray!20}}                                     & \multicolumn{1}{c}{\cellcolor{gray!20}\textit{Television}}                                    & \multicolumn{1}{c}{\cellcolor{gray!20}\textit{Competition}} & \multicolumn{1}{c}{\cellcolor{gray!20}\textit{Performance}} & \multicolumn{1}{c}{\cellcolor{gray!20}\textit{Record}} &  \multicolumn{1}{c}{\cellcolor{gray!20}}                                       & \multicolumn{1}{c}{\cellcolor{gray!20}}                          \\ 
\midrule
\multirow{2}{*}{Short} 
 & ZS & 81.4  & 87.5  & 78.7  & 86.7  & 85.6  & 86.7  & 84.4  \\
  & + \ourapproach & 85.6 & 91.3 & 81.2 & 88.0 & 88.9 & 89.4 & \textbf{87.4} \\
\midrule
\multirow{2}{*}{Medium} 
 & ZS & 80.2  & 83.9  & 72.1  & 84.3  & 76.8  & 100.0  & 82.8  \\
 & + \ourapproach & 83.3 & 86.5 & 75.9 & 87.1 & 78.2 & 100.0 & \textbf{85.16} \\
\midrule
\multirow{2}{*}{Long} 
 & ZS & 81.1  & 73.2  & 72.6  & 63.3  & 66.7  & 83.3  &  73.3 \\
 & + \ourapproach & 85.5 & 77.4 & 75.7 & 67.1 & 69.9 & 86.3 & \textbf{76.98}  \\
\midrule
\multirow{2}{*}{Overall} 
 & ZS & 80.9  & 82.4  & 74.6  & 78.8  & 78.0  & 89.7  & 80.7  \\
  & + \ourapproach & 83.4 & 85.3 & 77.8 & 81.0 & 82.3 & 92.6 & \textbf{83.73} \\
\bottomrule
\end{tabular}
\end{adjustbox}
\caption{\textbf{Performance of VITA across Video-MME}. Table shows the performance of VITA on 6 major categories of Video-MME. The evaluation is done on audio-visual inputs.}
\label{videomme_benchmark_results}
\vspace{-2em}
\end{table}

\vspace{-1.0mm}
\customsubsection{Reasoning Data Generation}
For each test sample comprising an audio, a video, and a question, we supplement the input with reasoning information at inference time before feeding it into the target model through a structured multi-agent pipeline. This ensures that model decisions are grounded in logical deductions rather than implicit associations, enhancing both accuracy and interpretability.
For instance, in Fig. \ref{fig:main-figure}, the video showcases people playing musical instruments, accompanied by audio, and the question to identify the \textit{most popular food of the country} through a complex audio-visual referral. To answer this, the model must first \textit{identify the loudest instrument} via audio analysis followed by determining spatial relationships to \textit{locate the musical instrument}. Once the instrument is located, the model must infer the instrument's origin, and finally retrieve the corresponding cuisine. This structured reasoning provided by our \ourapproach enforces logical progression, reducing errors and hallucinations while enhancing interpretability. We defer more details to supplementary. 

\section{Experiments and Results}
\customsubsection{Baselines}
\vspace{-2mm}
We extensively evaluate VideoLLaMA \cite{videollama}, VideoLLaMA2 \cite{videollama2}, Reka Core \cite{reka}, Gemini 1.5 Pro \cite{gemini}, Unified-IO-2 \cite{unifiedio}, X-InstructBLIP \cite{xinstructblip}, PandaGPT \cite{pandagpt}, OneLLM \cite{onellm}, AnyGPT \cite{anygpt}, NExT-GPT \cite{nextgpt}, VITA \cite{vita}, VideoSALMONN \cite{videosalmonn}, ImagebindLLM \cite{imagebindllm}, MacawLLM \cite{macawllm}, CAT \cite{baycat}, AVicuna \cite{avicuna}, CREMA \cite{crema}. AVLLM \cite{avllm} on \ourbenchmark.

\customsubsection{Metrics}
% For the AV-Question Answering task, we follow \cite{liu2024tackling, nadeem2024cad} to report 5 different types of audio-visual relationships, including `Existential', `Location', `Counting', `Temporal', and `Comparative' (detailed results in appendix). 
For AV-QA, AV-Comp, \ourtask, AV-Meme, and DM-Match, we report the Top-1 accuracy as the metric by extracting the model outputs using a choice extraction strategy outlined in the supplementary. We report the performance of AV captioning tasks on several established metrics, including BLUE@4 \cite{bleu}, METEOR \cite{meteor}, ROGUE \cite{rouge}, and CIDEr \cite{cider}. We employ GPT-based evaluation for \ourtask and AVSD which has open-ended answers.

\begin{table}
\centering
\resizebox{\columnwidth}{!}
{\begin{tabular}{c c c | c c c}
\hline
\rowcolor{gray!20}
\multicolumn{1}{c}{\textbf{Reason Gen.}} &
  \multicolumn{1}{c}{\textbf{Summ.}} &
  \multicolumn{1}{c|}{\textbf{Eval.}} &
  \multicolumn{1}{c}{\textbf{\ourtask}} &
  \multicolumn{1}{c}{\textbf{AV-Comp}} &
  \multicolumn{1}{c}{\textbf{DM-Match}} \\
\hline
% Reka Core  & Reka Core & Reka Core & - & - & -  \\
% Gemini  & GPT-4o & Gemini & - & - & -  \\
% Reka Core  & GPT-4o & Reka Core & - & - & -  \\
% Reka Core  & GPT-4o & Gemini & - & - & -  \\
% Gemini  & Reka & Reka & - & - & -  \\
% Gemini  & Reka & Gemini & - & - & -  \\
Gemini  & Gemini & Gemini & 36.5 & 30.2 & 33.0  \\
Gemini  & GPT-4o & Gemini & 38.0 & 31.6 & 34.5  \\
\hline
\end{tabular}}
\vspace{-0.9em}
\caption{\textbf{Effect of using a combination of agents. } Using a combination of different closed-source LLMs as agents proves beneficial compared to using a single type of LLM. 
% The analysis is carried out with VideoLLamA-2 across three tasks.
}
\label{tab:agent_combination}
\end{table}

\begin{table}
\vspace{-0.5em}
\centering
\resizebox{\columnwidth}{!}
{\begin{tabular}{c | c c c c c c}
\hline
\rowcolor{gray!20}
\multicolumn{1}{c|}{\textbf{Iteration ($T$)}} &
  \multicolumn{1}{c}{\textbf{AV-Cap}} &
  \multicolumn{1}{c}{\textbf{AV-Meme}} &
  \multicolumn{1}{c}{\textbf{\ourtask}} &
  \multicolumn{1}{c}{\textbf{AV-Comp}} &
  \multicolumn{1}{c}{\textbf{DM-Match}} &
  \multicolumn{1}{c}{\textbf{Time}}\\
\hline
1 & 68.8 & 25.0 & 27.5 & 27.3 & 26.5 & 16.28 \\
3 & 73.2 & 30.0 & 34.0 & 32.0 & 31.5 & 45.66 \\
5 & 74.7 & 35.0 & 38.0 & 31.6 & 34.5 &  74.01 \\
\hline
\end{tabular}}
\vspace{-1em}
\caption{\textbf{Effect of number of iterations. } The results improve as the number of feedback iterations increase. 
% The analysis is carried out with VideoLLamA-2 across five datasets. 
Time: time required to generate reasoning steps per sample}
\label{tab:iteration}
\end{table}

\begin{table}
\vspace{-0.5em}
\centering
\resizebox{\columnwidth}{!}
{\begin{tabular}{c | c c c c c c}
\hline
\rowcolor{gray!20}
\multicolumn{1}{c|}{\textbf{Threshold ($\tau$})} &
  \multicolumn{1}{c}{\textbf{AV-Cap}} &
  \multicolumn{1}{c}{\textbf{AV-Meme}} &
  \multicolumn{1}{c}{\textbf{\ourtask}} &
  \multicolumn{1}{c}{\textbf{AV-Comp}} &
  \multicolumn{1}{c}{\textbf{DM-Match}} & 
  \multicolumn{1}{c}{\textbf{Time}}\\
\hline
4  & 69.6  & 26.5  & 28.5 & 27.9 & 28.5 & 23.90  \\
6  & 72.2  & 30.0  & 32.5 & 29.7 & 32.0 & 47.15 \\
8  & 74.7  & 35.0  & 38.0 & 31.6 & 34.5 & 61.28 \\
10  & 74.8  & 35.0  & 38.0 & 31.4 & 34.5 & 65.81 \\
\hline
\end{tabular}}
\vspace{-1em}
\caption{\textbf{Effect of Threshold Value. }A larger threshold for the evaluation score shows positive trend on the performance. 
% The analysis is carried out with VideoLLamA-2 across five tasks.
}
\label{tab:threshold}
\vspace{-1em}
\end{table}

\customsubsection{Main Results}
\label{main_results}
% \vspace{-1mm}
We extensively compare the performance of the baseline AVLLMs in \cref{tab:main_results} across all 6 AV tasks of our \ourbenchmark benchmark. The experimental results reveal that closed-source models consistently outperform open-source ones in every reasoning task.
% Experimental results demonstrate the superiority of the closed-source models over the open-source ones in each reasoning task. 
Specifically, among the two closed-source models, we observe that Gemini 1.5 Pro surpasses Reka Core, likely due to its superior audio comprehension capabilities. This suggests that our \ourbenchmark benchmark presents challenging scenarios that require strong audio-visual joint understanding.
% We hypothesize that for the two proprietary LLMs, Gemini 1.5 Pro outperforms Reka Core due to its superior audio comprehension abilities. This indicates \ourbenchmark provides challenging scenarios that require strong audio-visual joint understanding. 
By leveraging the zero-shot reasoning distillation through \ourapproach, we observe consistent boost in the performance of all the AVLLMs as seen from the experimental results with relative improvements up to 100\% for X-InstructBLIP. Furthermore, for more challenging tasks such as \ourtask, AV-Meme, and DM-Match, we observe substantial improvements highlighting the importance of \ourapproach's step by step reasoning distillation in deriving answers to complex AV queries.
% there are significant improvements after introducing reasoning based prompting -- underlying the difficulties of these tasks and the importance of step by step reasoning.
% These substantial gains highlights the critical role of step-by-step reasoning in deriving answers to comple.

% Additionally, we observe that recent methods like Unified-IO-2 XL and VITA exhibit enhanced reasoning capabilities compared to other approaches, owing to their more powerful LLM backbones, which excel at capturing finer multimodal details. Furthermore, models with more robust audio encoders, such as AVicuna and Video-SALMONN outperform alternatives like PandaGPT and Macaw-LLM. This highlights the critical role of the audio modality in leveraging the strengths of \ourbenchmark.

We further note that recent approaches such as Unified-IO-2 XL and VITA demonstrate improved reasoning abilities over the other methods due to their stronger LLM backbone, which is capable of capturing finer multimodal information. Models with more robust audio encoders, such as AVicuna and Video-SALMONN, outperform alternatives like PandaGPT and Macaw-LLM. This highlights the critical role of the audio modality in leveraging the strengths of \ourbenchmark.
% Models that employ stronger audio encoders (AVicuna, Video-SALMONN) demonstrate better performance compared to other approaches (PandaGPT, Macaw-LLM) -- indicating the reliance on the audio modality is leveraged in \ourbenchmark. 

\cref{avcap_main} presents the AV-captioning results for five AVLLMs across three additional captioning metrics. As shown in the table, all models exhibit consistent improvements, highlighting the effectiveness of our reasoning-enhanced data in the dense captioning task.

\begin{figure*}[!t]
    \centering
    \includegraphics[width=1.0\textwidth]{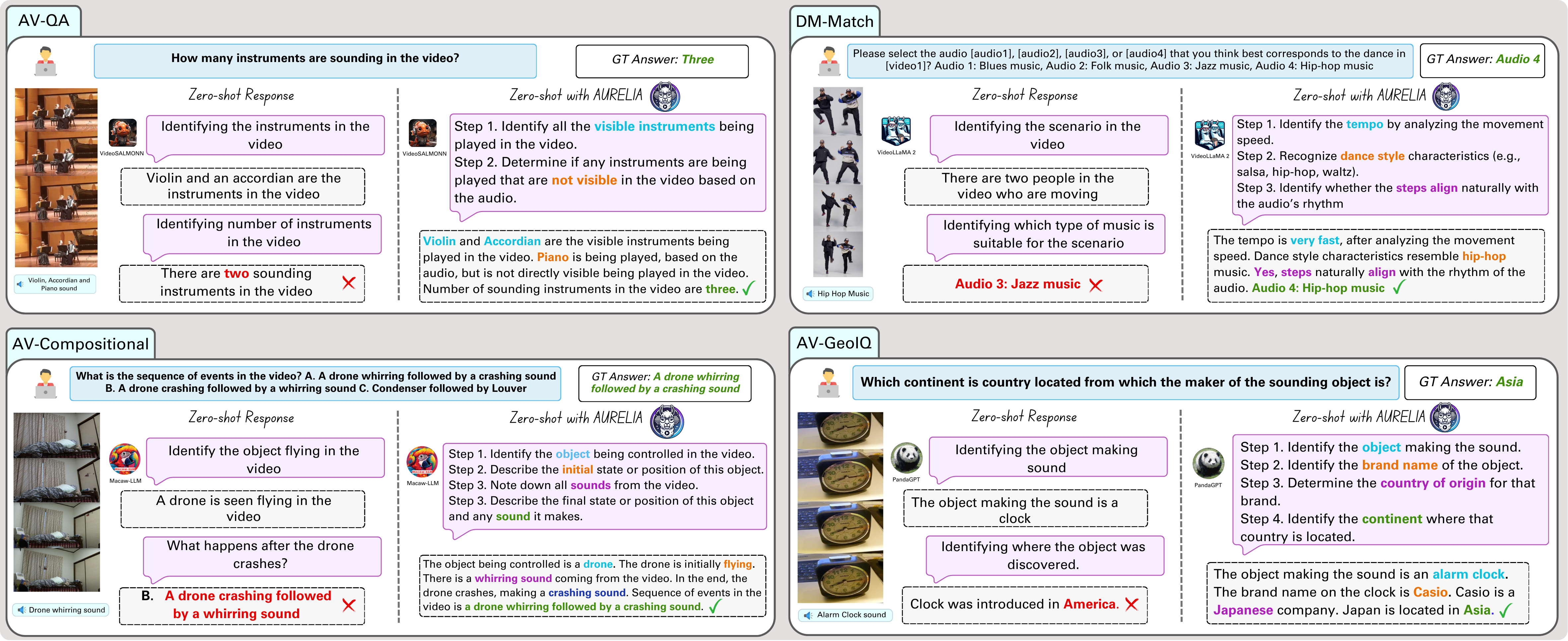}
    \caption{\textbf{Qualitative Visualizations.} Figure shows the qualitative visualizations of effect of \ourapproach's reasoning distillation on the final answer across four tasks. Compared to vanilla zero-shot inference, \ourapproach augments the target model with reasoning capabilities, leading to the improved answers. }
    \label{fig:qualitative_example_main}
    \vspace{-1em}
\end{figure*}

\begin{figure}[!t]
    \centering
    \includegraphics[width=0.48\textwidth]{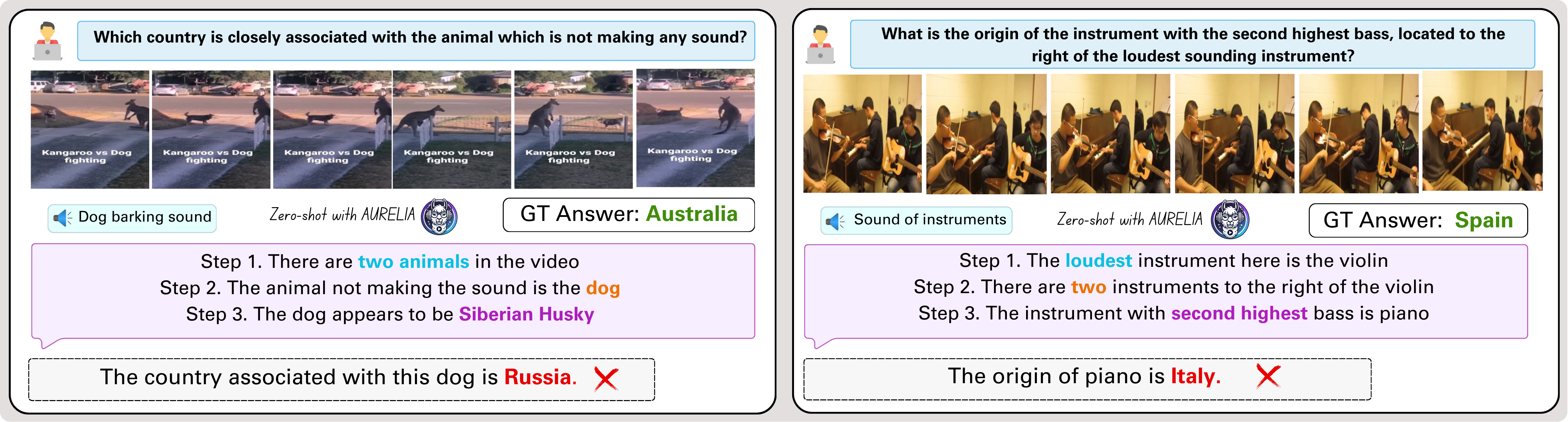}
    \caption{\textbf{Examples of Failure Cases. }\textbf{(Left)} \ourapproach fails to comprehend audio, focus on single modality i.e. video, leading to incorrect reasoning chain. \textbf{(Right)} \ourapproach fails to comprehend the dynamics of the video.}
    \label{fig:qualitative_example_failure}
    \vspace{-2em}
\end{figure}

\noindent{\textbf{Results on other benchmarks.}} \cref{videomme_benchmark_results} results demonstrate that our reasoning pipeline is generalizable across other benchmarks. We select VideoMME \cite{videomme} as an alternative benchmark due to its tasks, which demand advanced reasoning abilities. Notably, the greatest improvements are observed in the long video \textit{Knowledge} assessment categories, further emphasizing the generalizability of \ourapproach.
% \label{other benchmark results}
% Results in \cref{videomme_benchmark_results} show that our reasoning data augmentation pipeline is generalizable to other benchmarks as well. The rationale behind choosing VideoMME as an alternative benchmark is due to its tasks which require advanced reasoning capabilities. Note that highest gains are observed in long videos in \textit{Knowledge} assessment categories underlining generalizability of \ourapproach.

% Instilling strong reasoning capabilities improves the average performance by up to 3.6\%. It is important to note that highest gains are observed in long videos in \textit{Knowledge} assessment categories. This underlines the potential of \ourapproach to generalise across audio visual setting across diverse domains and durations. 

\customsubsection{Ablation Study}
\noindent{\textbf{Combination of Agents.}}
The multi-agent framework of \ourapproach offers the flexibility to integrate various existing multi-modal LLMs as specialized agents. To assess the impact of different LLMs on reasoning generation, summarization, and evaluation, we conduct an analysis on three datasets across target model 
VideoLLaMA-2 ( \cref{tab:agent_combination}).
Our findings indicate that leveraging a combination of models, specifically GPT-4o alongside Gemini yields superior performance compared to employing Gemini alone for all three agents roles as is evident from the higher accuracy scores in case of combination of agents. This suggests that while Gemini excels in processing multi-modal inputs such as video and audio, GPT-4o demonstrates stronger capabilities in textual comprehension and reasoning. The synergy between these models enhances the overall effectiveness of \ourapproach, underscoring the advantages of a diversified agent selection.
% \noindent{\textbf{Combination of Agents}}
% The multi-agent nature of AURELIA provides us the flexibility to utilize diverse existing multi-modal large language models (LLMs) as dedicated agents. As shown in Table \ref{tab:agent_combination}, we analyse the effect of using different LLMs for reasoning generation, summarization and Evaluation. We perform analysis on three datasets for a more generalized view. From the Table, we observe, that using a combination of different agent i.e. GPT-4O and Gemini is effective compared to using only Gemini for the three agents. This also supports that the Gemini is good for multi-modal processing such as videos and audio, while GPT-4o is good at comprehending and processing textual data. 

% \noindent{\textbf{Evaluator Score}}

\noindent{\textbf{Number of Generation Attempts.}}
% \ourapproach iteratively refines the generated reasoning steps until a predefined maximum number of iterations, 
% $T$, is reached, at which point the process terminates, and the final reasoning steps from the last iteration are returned. 
Our analysis reveals that the choice of 
$T$ significantly influences overall performance.
To evaluate this impact, we conduct an ablation study on five datasets across VideoLLaMA-2 model, as presented in \cref{tab:iteration}. With just a single iteration, the obtained scores are notably low, whereas increasing the iterations to five yields substantial improvements across most datasets. This suggests that additional iterations allow \ourapproach to progressively enhance its reasoning quality. However, considering computational efficiency and latency constraints, we cap the number of iterations at five for the final evaluation. AV-Cap values are CIDEr scores.   
% Our proposed AURELIA generates the optimized reasoning steps iteratively until the number of iterations reach a certain value $T$, after which the process is stopped and final reasoning steps generated at that step are returned. However, we observe that the choice of the number of iterations impacts the overall performance. We conduct an ablation study on five datasets with {} model in Table \ref{tab:iteration}. With only a single iteration, we observe that the scores obtained are minimal, while as with five iterations, we obtain better scores on most of the datasets, suggesting that with more iterations, AURELIA matures and generates better reasoning steps. However, due to latency constraints, we limit the number of iterations to five in the final evaluation.

\noindent{\textbf{Threshold Value.}}
% In addition to the number of iterations, \ourapproach's performance is influenced by a key hyperparameter: the threshold value ($\tau$). 
% The evaluator agent assesses the alignment between the generated caption (produced by the summarizer agent) and the input audio and video. This evaluation score ($\tau$) quantifies the consistency of the reasoning steps with the provided multi-modal inputs. Consequently, the effectiveness of reasoning distillation into the target model depends on the quality and relevance of the reasoning information, which, in turn, is governed by the chosen evaluation threshold ranging from 1 to 10 .
Evaluation score ($\tau$) quantifies the consistency of the reasoning steps with multimodal input. To empirically analyze the impact of the threshold ($\tau$), we present results in \cref{tab:threshold} on five datasets across VideoLLaMA-2 model. As expected, a higher threshold value indicates stronger alignment, leading to superior model performance. However, we observe that a threshold of 8 yields performance comparable to the highest value, suggesting that setting the threshold at 8 or above ensures optimal reasoning quality. The increasing value of time required to generate the samples indicate to obtain improved reasoning steps we need more iterations. AV-Cap report CIDEr values.

\customsubsection{Qualitative Results}
To visualize the effect of \ourapproach's reasoning distillation, refer to Fig. \ref{fig:qualitative_example_main}. We compare the performance of various AVLLMs on 4 tasks. We notice that in the absence of reasoning distillation, the target model faces difficulties in figuring out answers to the given queries. For example, in the AV-Captioning task, due to the step wise guidance to the AVLLM, the generated caption is dense and rick of contextual information compared to ZS response. Similarly, for \ourtask, powered by the sequence of prompts, the AVLLM is able to correctly respond to the query whereas, the response in ZS is wrong. Empirical studies reveal, with the addition of reasoning information, the decision making capability of model improves by structuring its response in accordance with the reasoning steps, thereby leading to correct answers. We add more qualitative results in the supplementary.

% \ourapproach can also mitigate visual and audio bias and facilitate answering the questions more objectively as seen in (add). Where the model's initial response before injecting reasoning was influenced by the cultural bias (both visual and audio) -- after employing our reasoning steps the model's response is factually correct as it is able to comprehend the question in a step-by-step manner and as a result arrives at the accurate answer underlining the utility of \ourapproach. 

\vspace{-0.6mm}
\customsubsection{Failure Cases}
\cref{fig:qualitative_example_failure} illustrates a few failure cases in our reasoning generation pipeline. In the first example, an error in interpreting the animal sounds leads to the assumption that the \textit{dog} is silent. This assumption propagates through the reasoning steps, producing an incorrect response. In the second example, the pipeline fails to spot the \textit{instrument with the second highest bass}, resulting in an erroneous conclusion. We believe that fine-grained AV comprehension and refining understanding of language instructions can help mitigate these issues.

\section{Conclusion}
\vspace{-2mm}
In this work, we introduce \ourapproach, a novel test-time framework designed to enhance the reasoning capabilities of AVLLMs through interactive multi-agent system which  distills structured, step-by-step reasoning into AVLLMs without any training. To further advance the AVLLMs' reasoning abilities, we also present \ourbenchmark, a comprehensive benchmark consisting of six diverse tasks including the novel \ourtask for geo-cultural knowledge reasoning. The samples in each task are paired with step-by-step reasoning data, generated using \ourapproach, which facilitates both the evaluation and enhancement of existing AVLLMs. \ourapproach serves as an essential step toward more robust, context-aware, and reasoning-driven multimodal AI, enabling future advancements in artificial audio-visual intelligence.

\medskip

{
    \small
    \bibliographystyle{ieeenat_fullname}
    \bibliography{main}

\begin{thebibliography}{164}
\providecommand{\natexlab}[1]{#1}
\providecommand{\url}[1]{\texttt{#1}}
\expandafter\ifx\csname urlstyle\endcsname\relax
  \providecommand{\doi}[1]{doi: #1}\else
  \providecommand{\doi}{doi: \begingroup \urlstyle{rm}\Url}\fi

\bibitem[Alamri et~al.(2019)Alamri, Cartillier, Das, Wang, Cherian, Essa, Batra, Marks, Hori, Anderson, Lee, and Parikh]{avsd}
Huda Alamri, Vincent Cartillier, Abhishek Das, Jue Wang, Anoop Cherian, Irfan Essa, Dhruv Batra, Tim~K. Marks, Chiori Hori, Peter Anderson, Stefan Lee, and Devi Parikh.
\newblock Audio-visual scene-aware dialog, 2019.

\bibitem[Banerjee and Lavie(2005)]{meteor}
Satanjeev Banerjee and Alon Lavie.
\newblock Meteor: An automatic metric for mt evaluation with improved correlation with human judgments.
\newblock In \emph{Proceedings of the acl workshop on intrinsic and extrinsic evaluation measures for machine translation and/or summarization}, pages 65--72, 2005.

\bibitem[Bi et~al.(2024)Bi, Han, Liu, Tang, and Wang]{bi2024forest}
Zhenni Bi, Kai Han, Chuanjian Liu, Yehui Tang, and Yunhe Wang.
\newblock Forest-of-thought: Scaling test-time compute for enhancing llm reasoning.
\newblock \emph{arXiv preprint arXiv:2412.09078}, 2024.

\bibitem[Chan et~al.(2023)Chan, Chen, Su, Yu, Xue, Zhang, Fu, and Liu]{chan2023chateval}
Chi-Min Chan, Weize Chen, Yusheng Su, Jianxuan Yu, Wei Xue, Shanghang Zhang, Jie Fu, and Zhiyuan Liu.
\newblock Chateval: Towards better llm-based evaluators through multi-agent debate.
\newblock \emph{arXiv preprint arXiv:2308.07201}, 2023.

\bibitem[Chen et~al.(2021)Chen, Shi, Li, and Zhang]{chen2021neural}
Hanxiong Chen, Shaoyun Shi, Yunqi Li, and Yongfeng Zhang.
\newblock Neural collaborative reasoning.
\newblock In \emph{Proceedings of the web conference 2021}, pages 1516--1527, 2021.

\bibitem[Chen et~al.(2023{\natexlab{a}})Chen, Zhu, Shen, Li, Liu, Zhang, Krishnamoorthi, Chandra, Xiong, and Elhoseiny]{minigptv2}
Jun Chen, Deyao Zhu, Xiaoqian Shen, Xiang Li, Zechun Liu, Pengchuan Zhang, Raghuraman Krishnamoorthi, Vikas Chandra, Yunyang Xiong, and Mohamed Elhoseiny.
\newblock Minigpt-v2: large language model as a unified interface for vision-language multi-task learning.
\newblock \emph{arXiv preprint arXiv:2310.09478}, 2023{\natexlab{a}}.

\bibitem[Chen et~al.(2023{\natexlab{b}})Chen, Zhang, Zeng, Zhang, Zhu, and Zhao]{shikra}
Keqin Chen, Zhao Zhang, Weili Zeng, Richong Zhang, Feng Zhu, and Rui Zhao.
\newblock Shikra: Unleashing multimodal llm's referential dialogue magic.
\newblock \emph{arXiv preprint arXiv:2306.15195}, 2023{\natexlab{b}}.

\bibitem[Chen et~al.(2020{\natexlab{a}})Chen, Gan, Cheng, Li, Carin, and Liu]{graphoptimaltransport}
Liqun Chen, Zhe Gan, Yu Cheng, Linjie Li, Lawrence Carin, and Jingjing Liu.
\newblock Graph optimal transport for cross-domain alignment.
\newblock In \emph{International Conference on Machine Learning}, pages 1542--1553. PMLR, 2020{\natexlab{a}}.

\bibitem[Chen et~al.(2024{\natexlab{a}})Chen, Han, and Zhang]{chen2024comm}
Pei Chen, Boran Han, and Shuai Zhang.
\newblock Comm: Collaborative multi-agent, multi-reasoning-path prompting for complex problem solving.
\newblock \emph{arXiv preprint arXiv:2404.17729}, 2024{\natexlab{a}}.

\bibitem[Chen et~al.(2023{\natexlab{c}})Chen, He, Guo, Zhu, Wang, Tang, and Liu]{chen2023valor}
Sihan Chen, Xingjian He, Longteng Guo, Xinxin Zhu, Weining Wang, Jinhui Tang, and Jing Liu.
\newblock Valor: Vision-audio-language omni-perception pretraining model and dataset.
\newblock \emph{arXiv preprint arXiv:2304.08345}, 2023{\natexlab{c}}.

\bibitem[Chen et~al.(2023{\natexlab{d}})Chen, He, Guo, Zhu, Wang, Tang, and Liu]{valor}
Sihan Chen, Xingjian He, Longteng Guo, Xinxin Zhu, Weining Wang, Jinhui Tang, and Jing Liu.
\newblock Valor: Vision-audio-language omni-perception pretraining model and dataset.
\newblock \emph{arXiv preprint arXiv:2304.08345}, 2023{\natexlab{d}}.

\bibitem[Chen et~al.(2024{\natexlab{b}})Chen, Li, Wang, Zhao, Sun, Zhu, and Liu]{vast}
Sihan Chen, Handong Li, Qunbo Wang, Zijia Zhao, Mingzhen Sun, Xinxin Zhu, and Jing Liu.
\newblock Vast: A vision-audio-subtitle-text omni-modality foundation model and dataset.
\newblock \emph{Advances in Neural Information Processing Systems}, 36, 2024{\natexlab{b}}.

\bibitem[Chen et~al.(2024{\natexlab{c}})Chen, Arkin, Zhang, Roy, and Fan]{chen2024scalable}
Yongchao Chen, Jacob Arkin, Yang Zhang, Nicholas Roy, and Chuchu Fan.
\newblock Scalable multi-robot collaboration with large language models: Centralized or decentralized systems?
\newblock In \emph{2024 IEEE International Conference on Robotics and Automation (ICRA)}, pages 4311--4317. IEEE, 2024{\natexlab{c}}.

\bibitem[Chen et~al.(2024{\natexlab{d}})Chen, Jhamtani, Sharma, Fan, and Wang]{chen2024steering}
Yongchao Chen, Harsh Jhamtani, Srinagesh Sharma, Chuchu Fan, and Chi Wang.
\newblock Steering large language models between code execution and textual reasoning.
\newblock \emph{arXiv preprint arXiv:2410.03524}, 2024{\natexlab{d}}.

\bibitem[Chen et~al.(2020{\natexlab{b}})Chen, Li, Yu, El~Kholy, Ahmed, Gan, Cheng, and Liu]{uniter}
Yen-Chun Chen, Linjie Li, Licheng Yu, Ahmed El~Kholy, Faisal Ahmed, Zhe Gan, Yu Cheng, and Jingjing Liu.
\newblock Uniter: Universal image-text representation learning.
\newblock In \emph{European conference on computer vision}, pages 104--120. Springer, 2020{\natexlab{b}}.

\bibitem[Chen et~al.(2024{\natexlab{e}})Chen, Zhou, Zhao, Wan, Zhang, Zhang, and Wen]{chen2024improving}
Zhipeng Chen, Kun Zhou, Wayne~Xin Zhao, Junchen Wan, Fuzheng Zhang, Di Zhang, and Ji-Rong Wen.
\newblock Improving large language models via fine-grained reinforcement learning with minimum editing constraint.
\newblock \emph{arXiv preprint arXiv:2401.06081}, 2024{\natexlab{e}}.

\bibitem[Cheng et~al.(2024{\natexlab{a}})Cheng, Leng, Zhang, Xin, Li, Chen, Zhu, Zhang, Luo, Zhao, et~al.]{cheng2024videollama}
Zesen Cheng, Sicong Leng, Hang Zhang, Yifei Xin, Xin Li, Guanzheng Chen, Yongxin Zhu, Wenqi Zhang, Ziyang Luo, Deli Zhao, et~al.
\newblock Videollama 2: Advancing spatial-temporal modeling and audio understanding in video-llms.
\newblock \emph{arXiv preprint arXiv:2406.07476}, 2024{\natexlab{a}}.

\bibitem[Cheng et~al.(2024{\natexlab{b}})Cheng, Leng, Zhang, Xin, Li, Chen, Zhu, Zhang, Luo, Zhao, et~al.]{videollama2}
Zesen Cheng, Sicong Leng, Hang Zhang, Yifei Xin, Xin Li, Guanzheng Chen, Yongxin Zhu, Wenqi Zhang, Ziyang Luo, Deli Zhao, et~al.
\newblock Videollama 2: Advancing spatial-temporal modeling and audio understanding in video-llms.
\newblock \emph{arXiv preprint arXiv:2406.07476}, 2024{\natexlab{b}}.

\bibitem[Chowdhury et~al.(2021)Chowdhury, Patra, Dasgupta, and Bhattacharya]{audvisum}
Sanjoy Chowdhury, Aditya Patra, Subhrajyoti Dasgupta, and Ujjwal Bhattacharya.
\newblock Audvisum: Self-supervised deep reinforcement learning for diverse audio-visual summary generation.
\newblock In \emph{BMVC}, page 315, 2021.

\bibitem[Chowdhury et~al.(2023{\natexlab{a}})Chowdhury, Ghosh, Dasgupta, Ratnarajah, Tyagi, and Manocha]{adverb}
Sanjoy Chowdhury, Sreyan Ghosh, Subhrajyoti Dasgupta, Anton Ratnarajah, Utkarsh Tyagi, and Dinesh Manocha.
\newblock Adverb: Visually guided audio dereverberation.
\newblock In \emph{Proceedings of the IEEE/CVF International Conference on Computer Vision}, pages 7884--7896, 2023{\natexlab{a}}.

\bibitem[Chowdhury et~al.(2023{\natexlab{b}})Chowdhury, Nag, and Manocha]{apollo}
Sanjoy Chowdhury, Sayan Nag, and Dinesh Manocha.
\newblock Apollo: Unified adapter and prompt learning for vision language models.
\newblock In \emph{The 2023 Conference on Empirical Methods in Natural Language Processing}, 2023{\natexlab{b}}.

\bibitem[Chowdhury et~al.(2024{\natexlab{a}})Chowdhury, Nag, Dasgupta, Chen, Elhoseiny, Gao, and Manocha]{chowdhury2024meerkat}
Sanjoy Chowdhury, Sayan Nag, Subhrajyoti Dasgupta, Jun Chen, Mohamed Elhoseiny, Ruohan Gao, and Dinesh Manocha.
\newblock Meerkat: Audio-visual large language model for grounding in space and time.
\newblock In \emph{European Conference on Computer Vision}, 2024{\natexlab{a}}.

\bibitem[Chowdhury et~al.(2024{\natexlab{b}})Chowdhury, Nag, Joseph, Srinivasan, and Manocha]{chowdhury2024melfusion}
Sanjoy Chowdhury, Sayan Nag, KJ Joseph, Balaji~Vasan Srinivasan, and Dinesh Manocha.
\newblock Melfusion: Synthesizing music from image and language cues using diffusion models.
\newblock In \emph{Proceedings of the IEEE/CVF Conference on Computer Vision and Pattern Recognition}, pages 26826--26835, 2024{\natexlab{b}}.

\bibitem[Chowdhury et~al.(2025)Chowdhury, Nag, Dasgupta, Wang, Elhoseiny, Gao, and Manocha]{avtrustbench}
Sanjoy Chowdhury, Sayan Nag, Subhrajyoti Dasgupta, Yaoting Wang, Mohamed Elhoseiny, Ruohan Gao, and Dinesh Manocha.
\newblock Avtrustbench: Assessing and enhancing reliability and robustness in audio-visual llms.
\newblock \emph{arXiv preprint arXiv:2501.02135}, 2025.

\bibitem[Cobbe et~al.(2021)Cobbe, Kosaraju, Bavarian, Chen, Jun, Kaiser, Plappert, Tworek, Hilton, Nakano, et~al.]{cobbe2021training}
Karl Cobbe, Vineet Kosaraju, Mohammad Bavarian, Mark Chen, Heewoo Jun, Lukasz Kaiser, Matthias Plappert, Jerry Tworek, Jacob Hilton, Reiichiro Nakano, et~al.
\newblock Training verifiers to solve math word problems.
\newblock \emph{arXiv preprint arXiv:2110.14168}, 2021.

\bibitem[Dai et~al.(2023)Dai, Li, Li, Tiong, Zhao, Wang, Li, Fung, and Hoi]{instructblip}
Wenliang Dai, Junnan Li, Dongxu Li, Anthony Meng~Huat Tiong, Junqi Zhao, Weisheng Wang, Boyang Li, Pascale Fung, and Steven Hoi.
\newblock Instructblip: Towards general-purpose vision-language models with instruction tuning, 2023.

\bibitem[De~Zarz{\`a} et~al.(2023)De~Zarz{\`a}, De~Curt{\`o}, Roig, Manzoni, and Calafate]{de2023emergent}
I De~Zarz{\`a}, J De~Curt{\`o}, Gemma Roig, Pietro Manzoni, and Carlos~T Calafate.
\newblock Emergent cooperation and strategy adaptation in multi-agent systems: An extended coevolutionary theory with llms.
\newblock \emph{Electronics}, 12\penalty0 (12):\penalty0 2722, 2023.

\bibitem[Deb et~al.(2023)Deb, Nag, Mahapatra, Chattopadhyay, Marik, Gayen, Sanyal, Banerjee, and Karmakar]{beats}
Ahana Deb, Sayan Nag, Ayan Mahapatra, Soumitri Chattopadhyay, Aritra Marik, Pijush~Kanti Gayen, Shankha Sanyal, Archi Banerjee, and Samir Karmakar.
\newblock Beats: Bengali speech acts recognition using multimodal attention fusion.
\newblock In \emph{Proc. Interspeech 2023}, pages 3392--3396, 2023.

\bibitem[Deng et~al.(2024)Deng, Sun, He, Sikka, Chen, Ma, Zhang, and Mihalcea]{deng2024tables}
Naihao Deng, Zhenjie Sun, Ruiqi He, Aman Sikka, Yulong Chen, Lin Ma, Yue Zhang, and Rada Mihalcea.
\newblock Tables as texts or images: Evaluating the table reasoning ability of llms and mllms.
\newblock \emph{arXiv preprint arXiv:2402.12424}, 2024.

\bibitem[Dong et~al.(2024)Dong, Liu, Sun, Yang, Hu, Rao, and Liu]{dong2024insight}
Yuhao Dong, Zuyan Liu, Hai-Long Sun, Jingkang Yang, Winston Hu, Yongming Rao, and Ziwei Liu.
\newblock Insight-v: Exploring long-chain visual reasoning with multimodal large language models.
\newblock \emph{arXiv preprint arXiv:2411.14432}, 2024.

\bibitem[Du et~al.(2023)Du, Li, Torralba, Tenenbaum, and Mordatch]{du2023improving}
Yilun Du, Shuang Li, Antonio Torralba, Joshua~B Tenenbaum, and Igor Mordatch.
\newblock Improving factuality and reasoning in language models through multiagent debate.
\newblock \emph{arXiv preprint arXiv:2305.14325}, 2023.

\bibitem[Du et~al.(2025)Du, Liu, Li, Zhao, Huo, Wang, Chen, Liu, Wang, and Wen]{du2025virgo}
Yifan Du, Zikang Liu, Yifan Li, Wayne~Xin Zhao, Yuqi Huo, Bingning Wang, Weipeng Chen, Zheng Liu, Zhongyuan Wang, and Ji-Rong Wen.
\newblock Virgo: A preliminary exploration on reproducing o1-like mllm.
\newblock \emph{arXiv preprint arXiv:2501.01904}, 2025.

\bibitem[Elizalde et~al.(2023)Elizalde, Deshmukh, Al~Ismail, and Wang]{clap}
Benjamin Elizalde, Soham Deshmukh, Mahmoud Al~Ismail, and Huaming Wang.
\newblock Clap learning audio concepts from natural language supervision.
\newblock In \emph{ICASSP 2023-2023 IEEE International Conference on Acoustics, Speech and Signal Processing (ICASSP)}, pages 1--5. IEEE, 2023.

\bibitem[Fang et~al.(2025)Fang, Mao, Duan, Zhao, Li, Lin, and Chen]{fang2025mmbench}
Xinyu Fang, Kangrui Mao, Haodong Duan, Xiangyu Zhao, Yining Li, Dahua Lin, and Kai Chen.
\newblock Mmbench-video: A long-form multi-shot benchmark for holistic video understanding.
\newblock \emph{Advances in Neural Information Processing Systems}, 37:\penalty0 89098--89124, 2025.

\bibitem[Fu et~al.(2024{\natexlab{a}})Fu, Dai, Luo, Li, Ren, Zhang, Wang, Zhou, Shen, Zhang, et~al.]{fu2024video}
Chaoyou Fu, Yuhan Dai, Yongdong Luo, Lei Li, Shuhuai Ren, Renrui Zhang, Zihan Wang, Chenyu Zhou, Yunhang Shen, Mengdan Zhang, et~al.
\newblock Video-mme: The first-ever comprehensive evaluation benchmark of multi-modal llms in video analysis.
\newblock \emph{arXiv preprint arXiv:2405.21075}, 2024{\natexlab{a}}.

\bibitem[Fu et~al.(2024{\natexlab{b}})Fu, Dai, Luo, Li, Ren, Zhang, Wang, Zhou, Shen, Zhang, et~al.]{videomme}
Chaoyou Fu, Yuhan Dai, Yongdong Luo, Lei Li, Shuhuai Ren, Renrui Zhang, Zihan Wang, Chenyu Zhou, Yunhang Shen, Mengdan Zhang, et~al.
\newblock Video-mme: The first-ever comprehensive evaluation benchmark of multi-modal llms in video analysis.
\newblock \emph{arXiv preprint arXiv:2405.21075}, 2024{\natexlab{b}}.

\bibitem[Fu et~al.(2024{\natexlab{c}})Fu, Lin, Long, Shen, Zhao, Zhang, Dong, Wang, Yin, Ma, et~al.]{vita}
Chaoyou Fu, Haojia Lin, Zuwei Long, Yunhang Shen, Meng Zhao, Yifan Zhang, Shaoqi Dong, Xiong Wang, Di Yin, Long Ma, et~al.
\newblock Vita: Towards open-source interactive omni multimodal llm.
\newblock \emph{arXiv preprint arXiv:2408.05211}, 2024{\natexlab{c}}.

\bibitem[Fu et~al.(2024{\natexlab{d}})Fu, Wang, Han, Fan, Si, and Yang]{fu2024graphic}
Jiale Fu, Yaqing Wang, Simeng Han, Jiaming Fan, Chen Si, and Xu Yang.
\newblock Graphic: A graph-based in-context example retrieval model for multi-step reasoning.
\newblock \emph{arXiv preprint arXiv:2410.02203}, 2024{\natexlab{d}}.

\bibitem[Gao et~al.(2024)Gao, Chen, Chen, Wang, and Lu]{avsegformer}
Shengyi Gao, Zhe Chen, Guo Chen, Wenhai Wang, and Tong Lu.
\newblock Avsegformer: Audio-visual segmentation with transformer.
\newblock In \emph{Proceedings of the AAAI conference on artificial intelligence}, pages 12155--12163, 2024.

\bibitem[Geva et~al.(2021)Geva, Khashabi, Segal, Khot, Roth, and Berant]{geva2021did}
Mor Geva, Daniel Khashabi, Elad Segal, Tushar Khot, Dan Roth, and Jonathan Berant.
\newblock Did aristotle use a laptop? a question answering benchmark with implicit reasoning strategies.
\newblock \emph{Transactions of the Association for Computational Linguistics}, 9:\penalty0 346--361, 2021.

\bibitem[Girdhar et~al.(2023)Girdhar, El-Nouby, Liu, Singh, Alwala, Joulin, and Misra]{imagebind}
Rohit Girdhar, Alaaeldin El-Nouby, Zhuang Liu, Mannat Singh, Kalyan~Vasudev Alwala, Armand Joulin, and Ishan Misra.
\newblock Imagebind: One embedding space to bind them all.
\newblock In \emph{Proceedings of the IEEE/CVF Conference on Computer Vision and Pattern Recognition}, pages 15180--15190, 2023.

\bibitem[Gong et~al.(2024)Gong, Feng, Li, Wang, Cheng, Yang, Han, Wang, Bai, Yang, et~al.]{avodcbench}
Kaixiong Gong, Kaituo Feng, Bohao Li, Yibing Wang, Mofan Cheng, Shijia Yang, Jiaming Han, Benyou Wang, Yutong Bai, Zhuoran Yang, et~al.
\newblock Av-odyssey bench: Can your multimodal llms really understand audio-visual information?
\newblock \emph{arXiv preprint arXiv:2412.02611}, 2024.

\bibitem[Guo et~al.(2025)Guo, Yang, Zhang, Song, Zhang, Xu, Zhu, Ma, Wang, Bi, et~al.]{guo2025deepseek}
Daya Guo, Dejian Yang, Haowei Zhang, Junxiao Song, Ruoyu Zhang, Runxin Xu, Qihao Zhu, Shirong Ma, Peiyi Wang, Xiao Bi, et~al.
\newblock Deepseek-r1: Incentivizing reasoning capability in llms via reinforcement learning.
\newblock \emph{arXiv preprint arXiv:2501.12948}, 2025.

\bibitem[Guo et~al.(2024{\natexlab{a}})Guo, Zheng, Bai, Li, Wang, Zhu, Li, Neubig, Chen, and Yue]{guo2024mammoth}
Jarvis Guo, Tuney Zheng, Yuelin Bai, Bo Li, Yubo Wang, King Zhu, Yizhi Li, Graham Neubig, Wenhu Chen, and Xiang Yue.
\newblock Mammoth-vl: Eliciting multimodal reasoning with instruction tuning at scale.
\newblock \emph{arXiv preprint arXiv:2412.05237}, 2024{\natexlab{a}}.

\bibitem[Guo et~al.(2024{\natexlab{b}})Guo, Chen, Wang, Chang, Pei, Chawla, Wiest, and Zhang]{guo2024large}
Taicheng Guo, Xiuying Chen, Yaqi Wang, Ruidi Chang, Shichao Pei, Nitesh~V Chawla, Olaf Wiest, and Xiangliang Zhang.
\newblock Large language model based multi-agents: A survey of progress and challenges.
\newblock \emph{arXiv preprint arXiv:2402.01680}, 2024{\natexlab{b}}.

\bibitem[Guo et~al.(2024{\natexlab{c}})Guo, Chen, Wang, Chang, Pei, Chawla, Wiest, and Zhang]{guo2024largelanguagemodelbased}
Taicheng Guo, Xiuying Chen, Yaqi Wang, Ruidi Chang, Shichao Pei, Nitesh~V. Chawla, Olaf Wiest, and Xiangliang Zhang.
\newblock Large language model based multi-agents: A survey of progress and challenges, 2024{\natexlab{c}}.

\bibitem[Guzhov et~al.(2022)Guzhov, Raue, Hees, and Dengel]{audioclip}
Andrey Guzhov, Federico Raue, J{\"o}rn Hees, and Andreas Dengel.
\newblock Audioclip: Extending clip to image, text and audio.
\newblock In \emph{ICASSP 2022-2022 IEEE International Conference on Acoustics, Speech and Signal Processing (ICASSP)}, pages 976--980. IEEE, 2022.

\bibitem[Han et~al.(2023{\natexlab{a}})Han, Gong, Zhang, Wang, Zhang, Lin, Qiao, Gao, and Yue]{onellm}
Jiaming Han, Kaixiong Gong, Yiyuan Zhang, Jiaqi Wang, Kaipeng Zhang, Dahua Lin, Yu Qiao, Peng Gao, and Xiangyu Yue.
\newblock Onellm: One framework to align all modalities with language.
\newblock \emph{arXiv preprint arXiv:2312.03700}, 2023{\natexlab{a}}.

\bibitem[Han et~al.(2023{\natexlab{b}})Han, Zhang, Shao, Gao, Xu, Xiao, Zhang, Liu, Wen, Guo, et~al.]{imagebindllm}
Jiaming Han, Renrui Zhang, Wenqi Shao, Peng Gao, Peng Xu, Han Xiao, Kaipeng Zhang, Chris Liu, Song Wen, Ziyu Guo, et~al.
\newblock Imagebind-llm: Multi-modality instruction tuning.
\newblock \emph{arXiv preprint arXiv:2309.03905}, 2023{\natexlab{b}}.

\bibitem[Han et~al.(2024)Han, Zhang, Yao, Jin, Xu, and He]{han2024llm}
Shanshan Han, Qifan Zhang, Yuhang Yao, Weizhao Jin, Zhaozhuo Xu, and Chaoyang He.
\newblock Llm multi-agent systems: Challenges and open problems.
\newblock \emph{arXiv preprint arXiv:2402.03578}, 2024.

\bibitem[He et~al.(2024)He, Xi, Zhao, Fan, Ding, Shan, Gui, Zhang, and Huang]{he2024distill}
Wei He, Zhiheng Xi, Wanxu Zhao, Xiaoran Fan, Yiwen Ding, Zifei Shan, Tao Gui, Qi Zhang, and Xuanjing Huang.
\newblock Distill visual chart reasoning ability from llms to mllms.
\newblock \emph{arXiv preprint arXiv:2410.18798}, 2024.

\bibitem[Hendrycks et~al.(2021)Hendrycks, Burns, Basart, Zou, Mazeika, Song, and Steinhardt]{mmlu}
Dan Hendrycks, Collin Burns, Steven Basart, Andy Zou, Mantas Mazeika, Dawn Song, and Jacob Steinhardt.
\newblock Measuring massive multitask language understanding, 2021.

\bibitem[Huang et~al.(2022)Huang, Gu, Hou, Wu, Wang, Yu, and Han]{huang2022large}
Jiaxin Huang, Shixiang~Shane Gu, Le Hou, Yuexin Wu, Xuezhi Wang, Hongkun Yu, and Jiawei Han.
\newblock Large language models can self-improve.
\newblock \emph{arXiv preprint arXiv:2210.11610}, 2022.

\bibitem[Khattak et~al.(2023)Khattak, Rasheed, Maaz, Khan, and Khan]{maple}
Muhammad~Uzair Khattak, Hanoona Rasheed, Muhammad Maaz, Salman Khan, and Fahad~Shahbaz Khan.
\newblock Maple: Multi-modal prompt learning.
\newblock In \emph{Proceedings of the IEEE/CVF conference on computer vision and pattern recognition}, pages 19113--19122, 2023.

\bibitem[Kil et~al.()Kil, Mai, Lee, Chowdhury, Wang, Cheng, Wang, Liu, and Chao]{kilmllm}
Jihyung Kil, Zheda Mai, Justin Lee, Arpita Chowdhury, Zihe Wang, Kerrie Cheng, Lemeng Wang, Ye Liu, and Wei-Lun Chao.
\newblock Mllm-compbench: A comparative reasoning benchmark for multimodal llms.
\newblock In \emph{The Thirty-eight Conference on Neural Information Processing Systems Datasets and Benchmarks Track}.

\bibitem[Kojima et~al.(2022)Kojima, Gu, Reid, Matsuo, and Iwasawa]{kojima2022large}
Takeshi Kojima, Shixiang~Shane Gu, Machel Reid, Yutaka Matsuo, and Yusuke Iwasawa.
\newblock Large language models are zero-shot reasoners.
\newblock \emph{Advances in neural information processing systems}, 35:\penalty0 22199--22213, 2022.

\bibitem[Lai et~al.(2023)Lai, Tian, Chen, Li, Yuan, Liu, and Jia]{lisa}
Xin Lai, Zhuotao Tian, Yukang Chen, Yanwei Li, Yuhui Yuan, Shu Liu, and Jiaya Jia.
\newblock Lisa: Reasoning segmentation via large language model.
\newblock \emph{arXiv preprint arXiv:2308.00692}, 2023.

\bibitem[Lee et~al.(2024)Lee, Wattanawong, Kim, Mangalam, Shen, Anumanchipalli, Mahoney, Keutzer, and Gholami]{lee2024llm2llm}
Nicholas Lee, Thanakul Wattanawong, Sehoon Kim, Karttikeya Mangalam, Sheng Shen, Gopala Anumanchipalli, Michael~W Mahoney, Kurt Keutzer, and Amir Gholami.
\newblock Llm2llm: Boosting llms with novel iterative data enhancement.
\newblock \emph{arXiv preprint arXiv:2403.15042}, 2024.

\bibitem[Li et~al.(2022{\natexlab{a}})Li, Wei, Tian, Xu, Wen, and Hu]{li2022learning}
Guangyao Li, Yake Wei, Yapeng Tian, Chenliang Xu, Ji-Rong Wen, and Di Hu.
\newblock Learning to answer questions in dynamic audio-visual scenarios.
\newblock In \emph{Proceedings of the IEEE/CVF Conference on Computer Vision and Pattern Recognition}, pages 19108--19118, 2022{\natexlab{a}}.

\bibitem[Li et~al.(2022{\natexlab{b}})Li, Wei, Tian, Xu, Wen, and Hu]{musicavqadataset}
Guangyao Li, Yake Wei, Yapeng Tian, Chenliang Xu, Ji-Rong Wen, and Di Hu.
\newblock Learning to answer questions in dynamic audio-visual scenarios.
\newblock In \emph{Proceedings of the IEEE/CVF Conference on Computer Vision and Pattern Recognition}, pages 19108--19118, 2022{\natexlab{b}}.

\bibitem[Li et~al.(2024{\natexlab{a}})Li, Deng, Ke, Liu, Rahmani, Guo, Schiele, and Chen]{li2024sports}
Haopeng Li, Andong Deng, Qiuhong Ke, Jun Liu, Hossein Rahmani, Yulan Guo, Bernt Schiele, and Chen Chen.
\newblock Sports-qa: A large-scale video question answering benchmark for complex and professional sports.
\newblock \emph{arXiv preprint arXiv:2401.01505}, 2024{\natexlab{a}}.

\bibitem[Li et~al.(2021)Li, Selvaraju, Gotmare, Joty, Xiong, and Hoi]{albef}
Junnan Li, Ramprasaath Selvaraju, Akhilesh Gotmare, Shafiq Joty, Caiming Xiong, and Steven Chu~Hong Hoi.
\newblock Align before fuse: Vision and language representation learning with momentum distillation.
\newblock \emph{Advances in neural information processing systems}, 34:\penalty0 9694--9705, 2021.

\bibitem[Li et~al.(2022{\natexlab{c}})Li, Li, Xiong, and Hoi]{blip}
Junnan Li, Dongxu Li, Caiming Xiong, and Steven Hoi.
\newblock Blip: Bootstrapping language-image pre-training for unified vision-language understanding and generation.
\newblock In \emph{International conference on machine learning}, pages 12888--12900. PMLR, 2022{\natexlab{c}}.

\bibitem[Li et~al.(2023)Li, Wang, He, Li, Wang, Liu, Wang, Xu, Chen, Luo, et~al.]{mvbench}
Kunchang Li, Yali Wang, Yinan He, Yizhuo Li, Yi Wang, Yi Liu, Zun Wang, Jilan Xu, Guo Chen, Ping Luo, et~al.
\newblock Mvbench: A comprehensive multi-modal video understanding benchmark.
\newblock \emph{arXiv preprint arXiv:2311.17005}, 2023.

\bibitem[Li et~al.(2022{\natexlab{d}})Li, Zhang, Zhang, Yang, Li, Zhong, Wang, Yuan, Zhang, Hwang, et~al.]{glip}
Liunian~Harold Li, Pengchuan Zhang, Haotian Zhang, Jianwei Yang, Chunyuan Li, Yiwu Zhong, Lijuan Wang, Lu Yuan, Lei Zhang, Jenq-Neng Hwang, et~al.
\newblock Grounded language-image pre-training.
\newblock In \emph{Proceedings of the IEEE/CVF Conference on Computer Vision and Pattern Recognition}, pages 10965--10975, 2022{\natexlab{d}}.

\bibitem[Li et~al.(2024{\natexlab{b}})Li, Li, Liu, Ren, Liu, Gao, Sun, and Hou]{li2024vitatecs}
Shicheng Li, Lei Li, Yi Liu, Shuhuai Ren, Yuanxin Liu, Rundong Gao, Xu Sun, and Lu Hou.
\newblock Vitatecs: A diagnostic dataset for temporal concept understanding of video-language models.
\newblock In \emph{European Conference on Computer Vision}, pages 331--348. Springer, 2024{\natexlab{b}}.

\bibitem[Li et~al.(2024{\natexlab{c}})Li, Yang, Cheng, Liu, Yu, Yang, and Lam]{li2024large}
Siheng Li, Cheng Yang, Zesen Cheng, Lemao Liu, Mo Yu, Yujiu Yang, and Wai Lam.
\newblock Large language models can self-improve in long-context reasoning.
\newblock \emph{arXiv preprint arXiv:2411.08147}, 2024{\natexlab{c}}.

\bibitem[Li et~al.(2024{\natexlab{d}})Li, Wang, Zeng, Wu, and Yang]{li2024survey}
Xinyi Li, Sai Wang, Siqi Zeng, Yu Wu, and Yi Yang.
\newblock A survey on llm-based multi-agent systems: workflow, infrastructure, and challenges.
\newblock \emph{Vicinagearth}, 1\penalty0 (1):\penalty0 9, 2024{\natexlab{d}}.

\bibitem[Liang et~al.(2023)Liang, He, Jiao, Wang, Wang, Wang, Yang, Shi, and Tu]{liang2023encouraging}
Tian Liang, Zhiwei He, Wenxiang Jiao, Xing Wang, Yan Wang, Rui Wang, Yujiu Yang, Shuming Shi, and Zhaopeng Tu.
\newblock Encouraging divergent thinking in large language models through multi-agent debate.
\newblock \emph{arXiv preprint arXiv:2305.19118}, 2023.

\bibitem[Lightman et~al.(2023)Lightman, Kosaraju, Burda, Edwards, Baker, Lee, Leike, Schulman, Sutskever, and Cobbe]{lightman2023let}
Hunter Lightman, Vineet Kosaraju, Yuri Burda, Harrison Edwards, Bowen Baker, Teddy Lee, Jan Leike, John Schulman, Ilya Sutskever, and Karl Cobbe.
\newblock Let's verify step by step.
\newblock In \emph{The Twelfth International Conference on Learning Representations}, 2023.

\bibitem[Lin et~al.(2023)Lin, Ye, Zhu, Cui, Ning, Jin, and Yuan]{lin2023video}
Bin Lin, Yang Ye, Bin Zhu, Jiaxi Cui, Munan Ning, Peng Jin, and Li Yuan.
\newblock Video-llava: Learning united visual representation by alignment before projection.
\newblock \emph{arXiv preprint arXiv:2311.10122}, 2023.

\bibitem[Lin(2004)]{rouge}
Chin-Yew Lin.
\newblock Rouge: A package for automatic evaluation of summaries.
\newblock In \emph{Text summarization branches out}, pages 74--81, 2004.

\bibitem[Lin et~al.(2022)Lin, Wang, Soldan, Wray, Yan, Xu, Gao, Tu, Zhao, Kong, et~al.]{egovlp}
Kevin~Qinghong Lin, Jinpeng Wang, Mattia Soldan, Michael Wray, Rui Yan, Eric~Z Xu, Difei Gao, Rong-Cheng Tu, Wenzhe Zhao, Weijie Kong, et~al.
\newblock Egocentric video-language pretraining.
\newblock \emph{Advances in Neural Information Processing Systems}, 35:\penalty0 7575--7586, 2022.

\bibitem[Liu et~al.(2024{\natexlab{a}})Liu, Li, Wu, and Lee]{llava}
Haotian Liu, Chunyuan Li, Qingyang Wu, and Yong~Jae Lee.
\newblock Visual instruction tuning.
\newblock \emph{Advances in neural information processing systems}, 36, 2024{\natexlab{a}}.

\bibitem[Liu et~al.(2023)Liu, Duan, Zhang, Li, Zhang, Zhao, Yuan, Wang, He, Liu, et~al.]{mmbench}
Yuan Liu, Haodong Duan, Yuanhan Zhang, Bo Li, Songyang Zhang, Wangbo Zhao, Yike Yuan, Jiaqi Wang, Conghui He, Ziwei Liu, et~al.
\newblock Mmbench: Is your multi-modal model an all-around player?
\newblock \emph{arXiv preprint arXiv:2307.06281}, 2023.

\bibitem[Liu et~al.(2024{\natexlab{b}})Liu, Li, Liu, Wang, Ren, Li, Chen, Sun, and Hou]{liu2024tempcompass}
Yuanxin Liu, Shicheng Li, Yi Liu, Yuxiang Wang, Shuhuai Ren, Lei Li, Sishuo Chen, Xu Sun, and Lu Hou.
\newblock Tempcompass: Do video llms really understand videos?
\newblock \emph{arXiv preprint arXiv:2403.00476}, 2024{\natexlab{b}}.

\bibitem[Lu et~al.(2022{\natexlab{a}})Lu, Clark, Zellers, Mottaghi, and Kembhavi]{unifiedio}
Jiasen Lu, Christopher Clark, Rowan Zellers, Roozbeh Mottaghi, and Aniruddha Kembhavi.
\newblock Unified-io: A unified model for vision, language, and multi-modal tasks.
\newblock \emph{arXiv preprint arXiv:2206.08916}, 2022{\natexlab{a}}.

\bibitem[Lu et~al.(2022{\natexlab{b}})Lu, Mishra, Xia, Qiu, Chang, Zhu, Tafjord, Clark, and Kalyan]{lu2022learn}
Pan Lu, Swaroop Mishra, Tanglin Xia, Liang Qiu, Kai-Wei Chang, Song-Chun Zhu, Oyvind Tafjord, Peter Clark, and Ashwin Kalyan.
\newblock Learn to explain: Multimodal reasoning via thought chains for science question answering.
\newblock \emph{Advances in Neural Information Processing Systems}, 35:\penalty0 2507--2521, 2022{\natexlab{b}}.

\bibitem[Lu et~al.(2023)Lu, Bansal, Xia, Liu, Li, Hajishirzi, Cheng, Chang, Galley, and Gao]{mathvista}
Pan Lu, Hritik Bansal, Tony Xia, Jiacheng Liu, Chunyuan Li, Hannaneh Hajishirzi, Hao Cheng, Kai-Wei Chang, Michel Galley, and Jianfeng Gao.
\newblock Mathvista: Evaluating mathematical reasoning of foundation models in visual contexts.
\newblock \emph{arXiv preprint arXiv:2310.02255}, 2023.

\bibitem[Luo et~al.(2024)Luo, Liu, Liu, Phatale, Lara, Li, Shu, Zhu, Meng, Sun, et~al.]{luo2024improve}
Liangchen Luo, Yinxiao Liu, Rosanne Liu, Samrat Phatale, Harsh Lara, Yunxuan Li, Lei Shu, Yun Zhu, Lei Meng, Jiao Sun, et~al.
\newblock Improve mathematical reasoning in language models by automated process supervision.
\newblock \emph{arXiv preprint arXiv:2406.06592}, 2, 2024.

\bibitem[Lyu et~al.(2023)Lyu, Wu, Wang, Huang, Liu, Du, Shi, and Tu]{macawllm}
Chenyang Lyu, Minghao Wu, Longyue Wang, Xinting Huang, Bingshuai Liu, Zefeng Du, Shuming Shi, and Zhaopeng Tu.
\newblock Macaw-llm: Multi-modal language modeling with image, audio, video, and text integration.
\newblock \emph{arXiv preprint arXiv:2306.09093}, 2023.

\bibitem[Maaz et~al.(2023)Maaz, Rasheed, Khan, and Khan]{videochatgpt}
Muhammad Maaz, Hanoona Rasheed, Salman Khan, and Fahad~Shahbaz Khan.
\newblock Video-chatgpt: Towards detailed video understanding via large vision and language models.
\newblock \emph{arXiv preprint arXiv:2306.05424}, 2023.

\bibitem[Mangalam et~al.(2023)Mangalam, Akshulakov, and Malik]{mangalam2023egoschema}
Karttikeya Mangalam, Raiymbek Akshulakov, and Jitendra Malik.
\newblock Egoschema: A diagnostic benchmark for very long-form video language understanding.
\newblock \emph{Advances in Neural Information Processing Systems}, 36:\penalty0 46212--46244, 2023.

\bibitem[Miyai et~al.(2024)Miyai, Yang, Zhang, Ming, Yu, Irie, Li, Li, Liu, and Aizawa]{upd}
Atsuyuki Miyai, Jingkang Yang, Jingyang Zhang, Yifei Ming, Qing Yu, Go Irie, Yixuan Li, Hai Li, Ziwei Liu, and Kiyoharu Aizawa.
\newblock Unsolvable problem detection: Evaluating trustworthiness of vision language models, 2024.

\bibitem[Moon et~al.(2023)Moon, Madotto, Lin, Nagarajan, Smith, Jain, Yeh, Murugesan, Heidari, Liu, et~al.]{anymal}
Seungwhan Moon, Andrea Madotto, Zhaojiang Lin, Tushar Nagarajan, Matt Smith, Shashank Jain, Chun-Fu Yeh, Prakash Murugesan, Peyman Heidari, Yue Liu, et~al.
\newblock Anymal: An efficient and scalable any-modality augmented language model.
\newblock \emph{arXiv preprint arXiv:2309.16058}, 2023.

\bibitem[Munasinghe et~al.(2023)Munasinghe, Thushara, Maaz, Rasheed, Khan, Shah, and Khan]{pgvideollava}
Shehan Munasinghe, Rusiru Thushara, Muhammad Maaz, Hanoona~Abdul Rasheed, Salman Khan, Mubarak Shah, and Fahad Khan.
\newblock Pg-video-llava: Pixel grounding large video-language models.
\newblock \emph{arXiv preprint arXiv:2311.13435}, 2023.

\bibitem[Nag et~al.(2024)Nag, Goswami, and Karanam]{safari}
Sayan Nag, Koustava Goswami, and Srikrishna Karanam.
\newblock Safari: Adaptive s equence tr a ns f ormer for we a kly supervised r eferring expression segmentat i on.
\newblock In \emph{European Conference on Computer Vision}, pages 485--503. Springer, 2024.

\bibitem[Nascimento et~al.(2023)Nascimento, Alencar, and Cowan]{nascimento2023self}
Nathalia Nascimento, Paulo Alencar, and Donald Cowan.
\newblock Self-adaptive large language model (llm)-based multiagent systems.
\newblock In \emph{2023 IEEE International Conference on Autonomic Computing and Self-Organizing Systems Companion (ACSOS-C)}, pages 104--109. IEEE, 2023.

\bibitem[Ning et~al.(2023)Ning, Zhu, Xie, Lin, Cui, Yuan, Chen, and Yuan]{ning2023video}
Munan Ning, Bin Zhu, Yujia Xie, Bin Lin, Jiaxi Cui, Lu Yuan, Dongdong Chen, and Li Yuan.
\newblock Video-bench: A comprehensive benchmark and toolkit for evaluating video-based large language models.
\newblock \emph{arXiv preprint arXiv:2311.16103}, 2023.

\bibitem[OpenAI(2024)]{openai_gpt4o}
OpenAI.
\newblock Hello gpt-4, 2024.

\bibitem[Ouyang et~al.(2022)Ouyang, Wu, Jiang, Almeida, Wainwright, Mishkin, Zhang, Agarwal, Slama, Ray, et~al.]{instructgpt}
Long Ouyang, Jeffrey Wu, Xu Jiang, Diogo Almeida, Carroll Wainwright, Pamela Mishkin, Chong Zhang, Sandhini Agarwal, Katarina Slama, Alex Ray, et~al.
\newblock Training language models to follow instructions with human feedback.
\newblock \emph{Advances in Neural Information Processing Systems}, 35:\penalty0 27730--27744, 2022.

\bibitem[Panagopoulou et~al.(2023)Panagopoulou, Xue, Yu, Li, Li, Joty, Xu, Savarese, Xiong, and Niebles]{xinstructblip}
Artemis Panagopoulou, Le Xue, Ning Yu, Junnan Li, Dongxu Li, Shafiq Joty, Ran Xu, Silvio Savarese, Caiming Xiong, and Juan~Carlos Niebles.
\newblock X-instructblip: A framework for aligning x-modal instruction-aware representations to llms and emergent cross-modal reasoning.
\newblock \emph{arXiv preprint arXiv:2311.18799}, 2023.

\bibitem[Pang et~al.(2024)Pang, Yuan, Cho, He, Sukhbaatar, and Weston]{pang2024iterative}
Richard~Yuanzhe Pang, Weizhe Yuan, Kyunghyun Cho, He He, Sainbayar Sukhbaatar, and Jason Weston.
\newblock Iterative reasoning preference optimization.
\newblock \emph{arXiv preprint arXiv:2404.19733}, 2024.

\bibitem[Papineni et~al.(2002)Papineni, Roukos, Ward, and Zhu]{bleu}
Kishore Papineni, Salim Roukos, Todd Ward, and Wei-Jing Zhu.
\newblock Bleu: a method for automatic evaluation of machine translation.
\newblock In \emph{Proceedings of the 40th annual meeting of the Association for Computational Linguistics}, pages 311--318, 2002.

\bibitem[Peng et~al.(2024)Peng, Xia, Yang, Xiong, Wu, and Xing]{peng2024regenesis}
Xiangyu Peng, Congying Xia, Xinyi Yang, Caiming Xiong, Chien-Sheng Wu, and Chen Xing.
\newblock Regenesis: Llms can grow into reasoning generalists via self-improvement.
\newblock \emph{arXiv preprint arXiv:2410.02108}, 2024.

\bibitem[Peng et~al.(2023)Peng, Wang, Dong, Hao, Huang, Ma, and Wei]{kosmos}
Zhiliang Peng, Wenhui Wang, Li Dong, Yaru Hao, Shaohan Huang, Shuming Ma, and Furu Wei.
\newblock Kosmos-2: Grounding multimodal large language models to the world.
\newblock \emph{arXiv preprint arXiv:2306.14824}, 2023.

\bibitem[Pramanick et~al.(2023{\natexlab{a}})Pramanick, Han, Hou, Nag, Lim, Ballas, Wang, Chellappa, and Almahairi]{vistallm}
Shraman Pramanick, Guangxing Han, Rui Hou, Sayan Nag, Ser-Nam Lim, Nicolas Ballas, Qifan Wang, Rama Chellappa, and Amjad Almahairi.
\newblock Jack of all tasks, master of many: Designing general-purpose coarse-to-fine vision-language model.
\newblock \emph{arXiv preprint arXiv:2312.12423}, 2023{\natexlab{a}}.

\bibitem[Pramanick et~al.(2023{\natexlab{b}})Pramanick, Jing, Nag, Zhu, Shah, LeCun, and Chellappa]{volta}
Shraman Pramanick, Li Jing, Sayan Nag, Jiachen Zhu, Hardik~J Shah, Yann LeCun, and Rama Chellappa.
\newblock Volta: Vision-language transformer with weakly-supervised local-feature alignment.
\newblock \emph{Transactions on Machine Learning Research}, 2023{\natexlab{b}}.

\bibitem[Pramanick et~al.(2023{\natexlab{c}})Pramanick, Song, Nag, Lin, Shah, Shou, Chellappa, and Zhang]{egovlpv2}
Shraman Pramanick, Yale Song, Sayan Nag, Kevin~Qinghong Lin, Hardik Shah, Mike~Zheng Shou, Rama Chellappa, and Pengchuan Zhang.
\newblock Egovlpv2: Egocentric video-language pre-training with fusion in the backbone.
\newblock In \emph{Proceedings of the IEEE/CVF International Conference on Computer Vision}, pages 5285--5297, 2023{\natexlab{c}}.

\bibitem[Radford et~al.(2021)Radford, Kim, Hallacy, Ramesh, Goh, Agarwal, Sastry, Askell, Mishkin, Clark, et~al.]{clip}
Alec Radford, Jong~Wook Kim, Chris Hallacy, Aditya Ramesh, Gabriel Goh, Sandhini Agarwal, Girish Sastry, Amanda Askell, Pamela Mishkin, Jack Clark, et~al.
\newblock Learning transferable visual models from natural language supervision.
\newblock In \emph{International conference on machine learning}, pages 8748--8763. PMLR, 2021.

\bibitem[Ramji et~al.(2024)Ramji, Lee, Astudillo, Sultan, Naseem, Munawar, Florian, and Roukos]{ramji2024self}
Keshav Ramji, Young-Suk Lee, Ram{\'o}n~Fernandez Astudillo, Md~Arafat Sultan, Tahira Naseem, Asim Munawar, Radu Florian, and Salim Roukos.
\newblock Self-refinement of language models from external proxy metrics feedback.
\newblock \emph{arXiv preprint arXiv:2403.00827}, 2024.

\bibitem[Reid et~al.(2024)Reid, Savinov, Teplyashin, Lepikhin, Lillicrap, Alayrac, Soricut, Lazaridou, Firat, Schrittwieser, et~al.]{gemini}
Machel Reid, Nikolay Savinov, Denis Teplyashin, Dmitry Lepikhin, Timothy Lillicrap, Jean-baptiste Alayrac, Radu Soricut, Angeliki Lazaridou, Orhan Firat, Julian Schrittwieser, et~al.
\newblock Gemini 1.5: Unlocking multimodal understanding across millions of tokens of context.
\newblock \emph{arXiv preprint arXiv:2403.05530}, 2024.

\bibitem[Ren et~al.(2023)Ren, Yao, Li, Sun, and Hou]{timechat}
Shuhuai Ren, Linli Yao, Shicheng Li, Xu Sun, and Lu Hou.
\newblock Timechat: A time-sensitive multimodal large language model for long video understanding.
\newblock \emph{arXiv preprint arXiv:2312.02051}, 2023.

\bibitem[Shu et~al.(2023)Shu, Zhang, Jiang, and Xie]{avllm}
Fangxun Shu, Lei Zhang, Hao Jiang, and Cihang Xie.
\newblock Audio-visual llm for video understanding.
\newblock \emph{arXiv preprint arXiv:2312.06720}, 2023.

\bibitem[Smit et~al.(2023)Smit, Duckworth, Grinsztajn, Barrett, and Pretorius]{smit2023should}
Andries Smit, Paul Duckworth, Nathan Grinsztajn, Thomas~D Barrett, and Arnu Pretorius.
\newblock Should we be going mad? a look at multi-agent debate strategies for llms.
\newblock \emph{arXiv preprint arXiv:2311.17371}, 2023.

\bibitem[Su et~al.(2023)Su, Lan, Li, Xu, Wang, and Cai]{pandagpt}
Yixuan Su, Tian Lan, Huayang Li, Jialu Xu, Yan Wang, and Deng Cai.
\newblock Pandagpt: One model to instruction-follow them all.
\newblock \emph{arXiv preprint arXiv:2305.16355}, 2023.

\bibitem[Subramaniam et~al.(2025)Subramaniam, Du, Tenenbaum, Torralba, Li, and Mordatch]{subramaniam2025multiagent}
Vighnesh Subramaniam, Yilun Du, Joshua~B Tenenbaum, Antonio Torralba, Shuang Li, and Igor Mordatch.
\newblock Multiagent finetuning: Self improvement with diverse reasoning chains.
\newblock \emph{arXiv preprint arXiv:2501.05707}, 2025.

\bibitem[Sun et~al.(2024{\natexlab{a}})Sun, Huang, and Pompili]{sun2024llm}
Chuanneng Sun, Songjun Huang, and Dario Pompili.
\newblock Llm-based multi-agent reinforcement learning: Current and future directions.
\newblock \emph{arXiv preprint arXiv:2405.11106}, 2024{\natexlab{a}}.

\bibitem[Sun et~al.(2024{\natexlab{b}})Sun, Manakul, Liusie, Pipatanakul, Zhang, Woodland, and Gales]{sun2024crosscheckgpt}
Guangzhi Sun, Potsawee Manakul, Adian Liusie, Kunat Pipatanakul, Chao Zhang, Phil Woodland, and Mark Gales.
\newblock Crosscheckgpt: Universal hallucination ranking for multimodal foundation models.
\newblock \emph{arXiv preprint arXiv:2405.13684}, 2024{\natexlab{b}}.

\bibitem[Sun et~al.(2024{\natexlab{c}})Sun, Yu, Tang, Chen, Tan, Li, Lu, Ma, Wang, and Zhang]{sun2024video}
Guangzhi Sun, Wenyi Yu, Changli Tang, Xianzhao Chen, Tian Tan, Wei Li, Lu Lu, Zejun Ma, Yuxuan Wang, and Chao Zhang.
\newblock video-salmonn: Speech-enhanced audio-visual large language models.
\newblock \emph{arXiv preprint arXiv:2406.15704}, 2024{\natexlab{c}}.

\bibitem[Sun et~al.(2024{\natexlab{d}})Sun, Yu, Tang, Chen, Tan, Li, Lu, Ma, Wang, and Zhang]{videosalmonn}
Guangzhi Sun, Wenyi Yu, Changli Tang, Xianzhao Chen, Tian Tan, Wei Li, Lu Lu, Zejun Ma, Yuxuan Wang, and Chao Zhang.
\newblock video-salmonn: Speech-enhanced audio-visual large language models.
\newblock \emph{arXiv preprint arXiv:2406.15704}, 2024{\natexlab{d}}.

\bibitem[Sun et~al.(2024{\natexlab{e}})Sun, Si, Zang, Zheng, Song, Zhang, and Xu]{sun2024large}
Zhongxiang Sun, Zihua Si, Xiaoxue Zang, Kai Zheng, Yang Song, Xiao Zhang, and Jun Xu.
\newblock Large language models enhanced collaborative filtering.
\newblock In \emph{Proceedings of the 33rd ACM International Conference on Information and Knowledge Management}, pages 2178--2188, 2024{\natexlab{e}}.

\bibitem[Sun et~al.(2024{\natexlab{f}})Sun, Yu, Shen, Liu, Yang, Welleck, and Gan]{sun2024easy}
Zhiqing Sun, Longhui Yu, Yikang Shen, Weiyang Liu, Yiming Yang, Sean Welleck, and Chuang Gan.
\newblock Easy-to-hard generalization: Scalable alignment beyond human supervision.
\newblock \emph{arXiv preprint arXiv:2403.09472}, 2024{\natexlab{f}}.

\bibitem[Swanson et~al.(2024)Swanson, Wu, Bulaong, Pak, and Zou]{swanson2024virtual}
Kyle Swanson, Wesley Wu, Nash~L Bulaong, John~E Pak, and James Zou.
\newblock The virtual lab: Ai agents design new sars-cov-2 nanobodies with experimental validation.
\newblock \emph{bioRxiv}, pages 2024--11, 2024.

\bibitem[Tang et~al.(2024{\natexlab{a}})Tang, Li, Yang, Zhuang, Sun, Li, Ma, and Zhang]{tang2024enhancing}
Changli Tang, Yixuan Li, Yudong Yang, Jimin Zhuang, Guangzhi Sun, Wei Li, Zujun Ma, and Chao Zhang.
\newblock Enhancing multimodal llm for detailed and accurate video captioning using multi-round preference optimization.
\newblock \emph{arXiv preprint arXiv:2410.06682}, 2024{\natexlab{a}}.

\bibitem[Tang et~al.(2024{\natexlab{b}})Tang, Yu, Sun, Chen, Tan, Li, Lu, Ma, and Zhang]{tang2024extending}
Changli Tang, Wenyi Yu, Guangzhi Sun, Xianzhao Chen, Tian Tan, Wei Li, Lu Lu, Zejun Ma, and Chao Zhang.
\newblock Extending large language models for speech and audio captioning.
\newblock In \emph{ICASSP 2024-2024 IEEE International Conference on Acoustics, Speech and Signal Processing (ICASSP)}, pages 11236--11240. IEEE, 2024{\natexlab{b}}.

\bibitem[Tang et~al.(2024{\natexlab{c}})Tang, Shimada, Bi, and Xu]{avicuna}
Yunlong Tang, Daiki Shimada, Jing Bi, and Chenliang Xu.
\newblock Avicuna: Audio-visual llm with interleaver and context-boundary alignment for temporal referential dialogue.
\newblock \emph{arXiv e-prints}, pages arXiv--2403, 2024{\natexlab{c}}.

\bibitem[Team et~al.(2023)Team, Anil, Borgeaud, Alayrac, Yu, Soricut, Schalkwyk, Dai, Hauth, Millican, et~al.]{team2023gemini}
Gemini Team, Rohan Anil, Sebastian Borgeaud, Jean-Baptiste Alayrac, Jiahui Yu, Radu Soricut, Johan Schalkwyk, Andrew~M Dai, Anja Hauth, Katie Millican, et~al.
\newblock Gemini: a family of highly capable multimodal models.
\newblock \emph{arXiv preprint arXiv:2312.11805}, 2023.

\bibitem[Team et~al.(2024)Team, Ormazabal, Zheng, d'Autume, Yogatama, Fu, Ong, Chen, Lamprecht, Pham, et~al.]{reka}
Reka Team, Aitor Ormazabal, Che Zheng, Cyprien de~Masson d'Autume, Dani Yogatama, Deyu Fu, Donovan Ong, Eric Chen, Eugenie Lamprecht, Hai Pham, et~al.
\newblock Reka core, flash, and edge: A series of powerful multimodal language models.
\newblock \emph{arXiv preprint arXiv:2404.12387}, 2024.

\bibitem[Thawakar et~al.(2025{\natexlab{a}})Thawakar, Dissanayake, More, Thawkar, Heakl, Ahsan, Li, Zumri, Lahoud, Anwer, Cholakkal, Laptev, Shah, Khan, and Khan]{thawakar2025llamavo1rethinkingstepbystepvisual}
Omkar Thawakar, Dinura Dissanayake, Ketan More, Ritesh Thawkar, Ahmed Heakl, Noor Ahsan, Yuhao Li, Mohammed Zumri, Jean Lahoud, Rao~Muhammad Anwer, Hisham Cholakkal, Ivan Laptev, Mubarak Shah, Fahad~Shahbaz Khan, and Salman Khan.
\newblock Llamav-o1: Rethinking step-by-step visual reasoning in llms, 2025{\natexlab{a}}.

\bibitem[Thawakar et~al.(2025{\natexlab{b}})Thawakar, Dissanayake, More, Thawkar, Heakl, Ahsan, Li, Zumri, Lahoud, Anwer, et~al.]{thawakar2025llamav}
Omkar Thawakar, Dinura Dissanayake, Ketan More, Ritesh Thawkar, Ahmed Heakl, Noor Ahsan, Yuhao Li, Mohammed Zumri, Jean Lahoud, Rao~Muhammad Anwer, et~al.
\newblock Llamav-o1: Rethinking step-by-step visual reasoning in llms.
\newblock \emph{arXiv preprint arXiv:2501.06186}, 2025{\natexlab{b}}.

\bibitem[Tran et~al.(2025)Tran, Dao, Nguyen, Pham, O'Sullivan, and Nguyen]{tran2025multi}
Khanh-Tung Tran, Dung Dao, Minh-Duong Nguyen, Quoc-Viet Pham, Barry O'Sullivan, and Hoang~D Nguyen.
\newblock Multi-agent collaboration mechanisms: A survey of llms.
\newblock \emph{arXiv preprint arXiv:2501.06322}, 2025.

\bibitem[Vedantam et~al.(2015)Vedantam, Lawrence~Zitnick, and Parikh]{cider}
Ramakrishna Vedantam, C Lawrence~Zitnick, and Devi Parikh.
\newblock Cider: Consensus-based image description evaluation.
\newblock In \emph{Proceedings of the IEEE conference on computer vision and pattern recognition}, pages 4566--4575, 2015.

\bibitem[Wang et~al.(2024{\natexlab{a}})Wang, Wang, Athiwaratkun, Zhang, and Zou]{wang2024mixture}
Junlin Wang, Jue Wang, Ben Athiwaratkun, Ce Zhang, and James Zou.
\newblock Mixture-of-agents enhances large language model capabilities.
\newblock \emph{arXiv preprint arXiv:2406.04692}, 2024{\natexlab{a}}.

\bibitem[Wang et~al.(2023{\natexlab{a}})Wang, Li, Shao, Xu, Dai, Li, Chen, Wu, and Sui]{wang2023math}
Peiyi Wang, Lei Li, Zhihong Shao, RX Xu, Damai Dai, Yifei Li, Deli Chen, Yu Wu, and Zhifang Sui.
\newblock Math-shepherd: Verify and reinforce llms step-by-step without human annotations.
\newblock \emph{arXiv preprint arXiv:2312.08935}, 2023{\natexlab{a}}.

\bibitem[Wang et~al.(2024{\natexlab{b}})Wang, Bai, Tan, Wang, Fan, Bai, Chen, Liu, Wang, Ge, et~al.]{wang2024qwen2}
Peng Wang, Shuai Bai, Sinan Tan, Shijie Wang, Zhihao Fan, Jinze Bai, Keqin Chen, Xuejing Liu, Jialin Wang, Wenbin Ge, et~al.
\newblock Qwen2-vl: Enhancing vision-language model's perception of the world at any resolution.
\newblock \emph{arXiv preprint arXiv:2409.12191}, 2024{\natexlab{b}}.

\bibitem[Wang et~al.(2024{\natexlab{c}})Wang, Wang, Su, Tong, and Song]{wang2024rethinking}
Qineng Wang, Zihao Wang, Ying Su, Hanghang Tong, and Yangqiu Song.
\newblock Rethinking the bounds of llm reasoning: Are multi-agent discussions the key?
\newblock \emph{arXiv preprint arXiv:2402.18272}, 2024{\natexlab{c}}.

\bibitem[Wang et~al.(2023{\natexlab{b}})Wang, Lv, Yu, Hong, Qi, Wang, Ji, Yang, Zhao, Song, et~al.]{cogvlm}
Weihan Wang, Qingsong Lv, Wenmeng Yu, Wenyi Hong, Ji Qi, Yan Wang, Junhui Ji, Zhuoyi Yang, Lei Zhao, Xixuan Song, et~al.
\newblock Cogvlm: Visual expert for pretrained language models.
\newblock \emph{arXiv preprint arXiv:2311.03079}, 2023{\natexlab{b}}.

\bibitem[Wang et~al.(2024{\natexlab{d}})Wang, Zhou, Liu, Lu, Xu, He, Yoon, Lu, Bertasius, Bansal, et~al.]{wang2024mementos}
Xiyao Wang, Yuhang Zhou, Xiaoyu Liu, Hongjin Lu, Yuancheng Xu, Feihong He, Jaehong Yoon, Taixi Lu, Gedas Bertasius, Mohit Bansal, et~al.
\newblock Mementos: A comprehensive benchmark for multimodal large language model reasoning over image sequences.
\newblock \emph{arXiv preprint arXiv:2401.10529}, 2024{\natexlab{d}}.

\bibitem[Wang et~al.(2024{\natexlab{e}})Wang, Zeng, Zheng, Xing, Xu, and Xu]{wang2024videocot}
Yan Wang, Yawen Zeng, Jingsheng Zheng, Xiaofen Xing, Jin Xu, and Xiangmin Xu.
\newblock Videocot: A video chain-of-thought dataset with active annotation tool.
\newblock \emph{arXiv preprint arXiv:2407.05355}, 2024{\natexlab{e}}.

\bibitem[Wei et~al.(2022)Wei, Wang, Schuurmans, Bosma, Xia, Chi, Le, Zhou, et~al.]{wei2022chain}
Jason Wei, Xuezhi Wang, Dale Schuurmans, Maarten Bosma, Fei Xia, Ed Chi, Quoc~V Le, Denny Zhou, et~al.
\newblock Chain-of-thought prompting elicits reasoning in large language models.
\newblock \emph{Advances in neural information processing systems}, 35:\penalty0 24824--24837, 2022.

\bibitem[Welleck et~al.(2022)Welleck, Lu, West, Brahman, Shen, Khashabi, and Choi]{welleck2022generating}
Sean Welleck, Ximing Lu, Peter West, Faeze Brahman, Tianxiao Shen, Daniel Khashabi, and Yejin Choi.
\newblock Generating sequences by learning to self-correct.
\newblock \emph{arXiv preprint arXiv:2211.00053}, 2022.

\bibitem[Wu et~al.(2022)Wu, Seetharaman, Kumar, and Bello]{wu2022wav2clip}
Ho-Hsiang Wu, Prem Seetharaman, Kundan Kumar, and Juan~Pablo Bello.
\newblock Wav2clip: Learning robust audio representations from clip.
\newblock In \emph{ICASSP 2022-2022 IEEE International Conference on Acoustics, Speech and Signal Processing (ICASSP)}, pages 4563--4567. IEEE, 2022.

\bibitem[Wu et~al.(2024)Wu, Fei, Qu, Ji, and Chua]{nextgpt}
Shengqiong Wu, Hao Fei, Leigang Qu, Wei Ji, and Tat-Seng Chua.
\newblock Next-gpt: Any-to-any multimodal llm.
\newblock In \emph{Forty-first International Conference on Machine Learning}, 2024.

\bibitem[Xiao et~al.(2021)Xiao, Shang, Yao, and Chua]{xiao2021next}
Junbin Xiao, Xindi Shang, Angela Yao, and Tat-Seng Chua.
\newblock Next-qa: Next phase of question-answering to explaining temporal actions.
\newblock In \emph{Proceedings of the IEEE/CVF conference on computer vision and pattern recognition}, pages 9777--9786, 2021.

\bibitem[Xiao et~al.(2024)Xiao, Sun, Liu, and Wang]{xiao2024logicvista}
Yijia Xiao, Edward Sun, Tianyu Liu, and Wei Wang.
\newblock Logicvista: Multimodal llm logical reasoning benchmark in visual contexts.
\newblock \emph{arXiv preprint arXiv:2407.04973}, 2024.

\bibitem[Xie et~al.(2024)Xie, Zhang, Zhou, Li, Zhang, Hessel, Yang, and Liu]{funqa}
Binzhu Xie, Sicheng Zhang, Zitang Zhou, Bo Li, Yuanhan Zhang, Jack Hessel, Jingkang Yang, and Ziwei Liu.
\newblock Funqa: Towards surprising video comprehension.
\newblock In \emph{European Conference on Computer Vision}, pages 39--57. Springer, 2024.

\bibitem[Xu et~al.(2024)Xu, Jin, Hao, Song, Sun, and Yuan]{xu2024llava}
Guowei Xu, Peng Jin, Li Hao, Yibing Song, Lichao Sun, and Li Yuan.
\newblock Llava-o1: Let vision language models reason step-by-step.
\newblock \emph{arXiv preprint arXiv:2411.10440}, 2024.

\bibitem[Yang et~al.(2024)Yang, Zhang, Hui, Gao, Yu, Li, Liu, Tu, Zhou, Lin, et~al.]{yang2024qwen2}
An Yang, Beichen Zhang, Binyuan Hui, Bofei Gao, Bowen Yu, Chengpeng Li, Dayiheng Liu, Jianhong Tu, Jingren Zhou, Junyang Lin, et~al.
\newblock Qwen2. 5-math technical report: Toward mathematical expert model via self-improvement.
\newblock \emph{arXiv preprint arXiv:2409.12122}, 2024.

\bibitem[Yang et~al.(2022)Yang, Gan, Wang, Hu, Ahmed, Liu, Lu, and Wang]{unitab}
Zhengyuan Yang, Zhe Gan, Jianfeng Wang, Xiaowei Hu, Faisal Ahmed, Zicheng Liu, Yumao Lu, and Lijuan Wang.
\newblock Unitab: Unifying text and box outputs for grounded vision-language modeling.
\newblock In \emph{European Conference on Computer Vision}, pages 521--539. Springer, 2022.

\bibitem[Ye et~al.(2023)Ye, Xu, Xu, Ye, Yan, Zhou, Wang, Hu, Shi, Shi, et~al.]{mplugowl}
Qinghao Ye, Haiyang Xu, Guohai Xu, Jiabo Ye, Ming Yan, Yiyang Zhou, Junyang Wang, Anwen Hu, Pengcheng Shi, Yaya Shi, et~al.
\newblock mplug-owl: Modularization empowers large language models with multimodality.
\newblock \emph{arXiv preprint arXiv:2304.14178}, 2023.

\bibitem[Ye et~al.(2024)Ye, Yu, Shao, Xie, Torr, and Cao]{baycat}
Qilang Ye, Zitong Yu, Rui Shao, Xinyu Xie, Philip Torr, and Xiaochun Cao.
\newblock Cat: enhancing multimodal large language model to answer questions in dynamic audio-visual scenarios.
\newblock \emph{arXiv preprint arXiv:2403.04640}, 2024.

\bibitem[Ying et~al.(2024)Ying, Zhang, Li, Zhou, Shao, Fei, Ma, Hong, Liu, Wang, et~al.]{ying2024internlm}
Huaiyuan Ying, Shuo Zhang, Linyang Li, Zhejian Zhou, Yunfan Shao, Zhaoye Fei, Yichuan Ma, Jiawei Hong, Kuikun Liu, Ziyi Wang, et~al.
\newblock Internlm-math: Open math large language models toward verifiable reasoning.
\newblock \emph{arXiv preprint arXiv:2402.06332}, 2024.

\bibitem[Yu et~al.(2023{\natexlab{a}})Yu, Gao, and Wang]{yu2023ovm}
Fei Yu, Anningzhe Gao, and Benyou Wang.
\newblock Ovm, outcome-supervised value models for planning in mathematical reasoning.
\newblock \emph{arXiv preprint arXiv:2311.09724}, 2023{\natexlab{a}}.

\bibitem[Yu et~al.(2024)Yu, Yoon, and Bansal]{crema}
Shoubin Yu, Jaehong Yoon, and Mohit Bansal.
\newblock Crema: Generalizable and efficient video-language reasoning via multimodal modular fusion.
\newblock \emph{arXiv preprint arXiv:2402.05889}, 2024.

\bibitem[Yu et~al.(2023{\natexlab{b}})Yu, Peng, Galley, Gao, and Yu]{yu2023teaching}
Xiao Yu, Baolin Peng, Michel Galley, Jianfeng Gao, and Zhou Yu.
\newblock Teaching language models to self-improve through interactive demonstrations.
\newblock \emph{arXiv preprint arXiv:2310.13522}, 2023{\natexlab{b}}.

\bibitem[Yu et~al.(2025)Yu, Yao, Li, Deng, Jiang, Cao, Chen, Suchow, Cui, Liu, et~al.]{yu2025fincon}
Yangyang Yu, Zhiyuan Yao, Haohang Li, Zhiyang Deng, Yuechen Jiang, Yupeng Cao, Zhi Chen, Jordan Suchow, Zhenyu Cui, Rong Liu, et~al.
\newblock Fincon: A synthesized llm multi-agent system with conceptual verbal reinforcement for enhanced financial decision making.
\newblock \emph{Advances in Neural Information Processing Systems}, 37:\penalty0 137010--137045, 2025.

\bibitem[Yuan et~al.(2024)Yuan, Pang, Cho, Sukhbaatar, Xu, and Weston]{yuan2024self}
Weizhe Yuan, Richard~Yuanzhe Pang, Kyunghyun Cho, Sainbayar Sukhbaatar, Jing Xu, and Jason Weston.
\newblock Self-rewarding language models.
\newblock \emph{arXiv preprint arXiv:2401.10020}, 2024.

\bibitem[Zelikman et~al.(2022)Zelikman, Wu, Mu, and Goodman]{zelikman2022star}
Eric Zelikman, Yuhuai Wu, Jesse Mu, and Noah Goodman.
\newblock Star: Bootstrapping reasoning with reasoning.
\newblock \emph{Advances in Neural Information Processing Systems}, 35:\penalty0 15476--15488, 2022.

\bibitem[Zellers et~al.(2022)Zellers, Lu, Lu, Yu, Zhao, Salehi, Kusupati, Hessel, Farhadi, and Choi]{merlot}
Rowan Zellers, Jiasen Lu, Ximing Lu, Youngjae Yu, Yanpeng Zhao, Mohammadreza Salehi, Aditya Kusupati, Jack Hessel, Ali Farhadi, and Yejin Choi.
\newblock Merlot reserve: Neural script knowledge through vision and language and sound.
\newblock In \emph{Proceedings of the IEEE/CVF Conference on Computer Vision and Pattern Recognition}, pages 16375--16387, 2022.

\bibitem[Zhan et~al.(2024)Zhan, Dai, Ye, Zhou, Zhang, Liu, Zhang, Yuan, Zhang, Li, et~al.]{anygpt}
Jun Zhan, Junqi Dai, Jiasheng Ye, Yunhua Zhou, Dong Zhang, Zhigeng Liu, Xin Zhang, Ruibin Yuan, Ge Zhang, Linyang Li, et~al.
\newblock Anygpt: Unified multimodal llm with discrete sequence modeling.
\newblock \emph{arXiv preprint arXiv:2402.12226}, 2024.

\bibitem[Zhang et~al.(2024{\natexlab{a}})Zhang, Wu, Lei, Che, Li, Xie, Huang, Zhang, Pavone, Li, et~al.]{zhang2024llama}
Di Zhang, Jianbo Wu, Jingdi Lei, Tong Che, Jiatong Li, Tong Xie, Xiaoshui Huang, Shufei Zhang, Marco Pavone, Yuqiang Li, et~al.
\newblock Llama-berry: Pairwise optimization for o1-like olympiad-level mathematical reasoning.
\newblock \emph{arXiv preprint arXiv:2410.02884}, 2024{\natexlab{a}}.

\bibitem[Zhang et~al.(2023{\natexlab{a}})Zhang, Li, and Bing]{videollama}
Hang Zhang, Xin Li, and Lidong Bing.
\newblock Video-llama: An instruction-tuned audio-visual language model for video understanding.
\newblock \emph{arXiv preprint arXiv:2306.02858}, 2023{\natexlab{a}}.

\bibitem[Zhang et~al.(2024{\natexlab{b}})Zhang, Hosseini, Bansal, Kazemi, Kumar, and Agarwal]{zhang2024generative}
Lunjun Zhang, Arian Hosseini, Hritik Bansal, Mehran Kazemi, Aviral Kumar, and Rishabh Agarwal.
\newblock Generative verifiers: Reward modeling as next-token prediction.
\newblock \emph{arXiv preprint arXiv:2408.15240}, 2024{\natexlab{b}}.

\bibitem[Zhang et~al.(2023{\natexlab{b}})Zhang, Sun, Chen, Xiao, Shao, Zhang, Chen, and Luo]{gpt4roi}
Shilong Zhang, Peize Sun, Shoufa Chen, Min Xiao, Wenqi Shao, Wenwei Zhang, Kai Chen, and Ping Luo.
\newblock Gpt4roi: Instruction tuning large language model on region-of-interest.
\newblock \emph{arXiv preprint arXiv:2307.03601}, 2023{\natexlab{b}}.

\bibitem[Zhang et~al.(2024{\natexlab{c}})Zhang, Khalifa, Logeswaran, Kim, Lee, Lee, and Wang]{zhang2024small}
Yunxiang Zhang, Muhammad Khalifa, Lajanugen Logeswaran, Jaekyeom Kim, Moontae Lee, Honglak Lee, and Lu Wang.
\newblock Small language models need strong verifiers to self-correct reasoning.
\newblock \emph{arXiv preprint arXiv:2404.17140}, 2024{\natexlab{c}}.

\bibitem[Zhang et~al.(2024{\natexlab{d}})Zhang, Wu, Li, Li, Ma, Liu, and Li]{zhang2024video}
Yuanhan Zhang, Jinming Wu, Wei Li, Bo Li, Zejun Ma, Ziwei Liu, and Chunyuan Li.
\newblock Video instruction tuning with synthetic data.
\newblock \emph{arXiv preprint arXiv:2410.02713}, 2024{\natexlab{d}}.

\bibitem[Zhang et~al.(2024{\natexlab{e}})Zhang, Wu, Yang, Shu, Xiao, Kong, and Sang]{zhang2024o1}
Yuxiang Zhang, Shangxi Wu, Yuqi Yang, Jiangming Shu, Jinlin Xiao, Chao Kong, and Jitao Sang.
\newblock o1-coder: an o1 replication for coding.
\newblock \emph{arXiv preprint arXiv:2412.00154}, 2024{\natexlab{e}}.

\bibitem[Zhao et~al.(2023{\natexlab{a}})Zhao, Lin, Zhou, Huang, Feng, and Kang]{bubogpt}
Yang Zhao, Zhijie Lin, Daquan Zhou, Zilong Huang, Jiashi Feng, and Bingyi Kang.
\newblock Bubogpt: Enabling visual grounding in multi-modal llms.
\newblock \emph{arXiv preprint arXiv:2307.08581}, 2023{\natexlab{a}}.

\bibitem[Zhao et~al.(2023{\natexlab{b}})Zhao, Guo, Yue, Chen, Shao, Zhu, Yuan, and Liu]{chatbridge}
Zijia Zhao, Longteng Guo, Tongtian Yue, Sihan Chen, Shuai Shao, Xinxin Zhu, Zehuan Yuan, and Jing Liu.
\newblock Chatbridge: Bridging modalities with large language model as a language catalyst.
\newblock \emph{arXiv preprint arXiv:2305.16103}, 2023{\natexlab{b}}.

\bibitem[Zheng et~al.(2024)Zheng, Xu, Sun, Pu, Chen, and Sun]{zheng2024thinking}
Haojie Zheng, Tianyang Xu, Hanchi Sun, Shu Pu, Ruoxi Chen, and Lichao Sun.
\newblock Thinking before looking: Improving multimodal llm reasoning via mitigating visual hallucination.
\newblock \emph{arXiv preprint arXiv:2411.12591}, 2024.

\bibitem[Zhong et~al.(2024)Zhong, Liu, Pan, Zhang, Zhou, Liang, Wu, Lyu, Shu, Yu, et~al.]{zhong2024evaluation}
Tianyang Zhong, Zhengliang Liu, Yi Pan, Yutong Zhang, Yifan Zhou, Shizhe Liang, Zihao Wu, Yanjun Lyu, Peng Shu, Xiaowei Yu, et~al.
\newblock Evaluation of openai o1: Opportunities and challenges of agi.
\newblock \emph{arXiv preprint arXiv:2409.18486}, 2024.

\bibitem[Zhou et~al.(2022)Zhou, Wang, Zhang, Sun, Zhang, Birchfield, Guo, Kong, Wang, and Zhong]{avseg}
Jinxing Zhou, Jianyuan Wang, Jiayi Zhang, Weixuan Sun, Jing Zhang, Stan Birchfield, Dan Guo, Lingpeng Kong, Meng Wang, and Yiran Zhong.
\newblock Audio--visual segmentation.
\newblock In \emph{European Conference on Computer Vision}, pages 386--403. Springer, 2022.

\bibitem[Zhu et~al.(2023)Zhu, Chen, Shen, Li, and Elhoseiny]{minigpt4}
Deyao Zhu, Jun Chen, Xiaoqian Shen, Xiang Li, and Mohamed Elhoseiny.
\newblock Minigpt-4: Enhancing vision-language understanding with advanced large language models.
\newblock \emph{arXiv preprint arXiv:2304.10592}, 2023.

\end{thebibliography}
}

\newpage
\appendix

\twocolumn[
    \centering
    \Large
    \textbf{\raisebox{-0.3\height}{\includegraphics[width=0.06\linewidth]{ICCV2025-Author-Kit-Feb/figures/aurelia_logo.png}}\ourapproach: Test-time Reasoning Distillation in Audio-Visual LLMs}\\
    \vspace{0.5em}\textcolor{blue}{Supplementary Material} \\
    \vspace{1.0em}
]

\makeatletter

% \title{\raisebox{-0.3\height}{\includegraphics[width=0.06\linewidth]{ICCV2025-Author-Kit-Feb/figures/aurelia_logo.png}}\ourapproach: Test-time Reasoning Distillation in Audio-Visual LLMs \textcolor{blue}{Supplementary}}

% Multi-Agent Reasoning-Distillation Framework for Audio-Visual Large Language Models

%%%%%%%%% AUTHORS - PLEASE UPDATE
% \and

% \begin{document}
\maketitle
\appendix

\noindent{We add the following details in this supplementary:}

\noindent{\ref{supp_video} Supplementary Video} \\
\noindent{\ref{more_related_works} More Related Works} \\
\noindent{\ref{gpt_based_evaluation} GPT Based Evaluation}\\
\noindent{\ref{appendix_prompts} Examples of Prompts}\\
\noindent{\ref{radar plot} Radar Plot}\\
\noindent{\ref{reasoning_data_generation} Details on Reasoning Data Generation}\\
\noindent{\ref{data statistics}} \ourbenchmark Statistics \\
\noindent{\ref{Breakdown results} Breakdown Results}\\
\noindent{\ref{other benchmark results} Results on Other Benchmarks}\\
\noindent{\ref{appendix_qual_results} Qualitative Results}\\
\noindent{\ref{discussion on failure cases} Key Observations}\\
\noindent{\ref{future_work} Future Work}\\
\noindent{\ref{societal impact} Societal Impact}\\

\section{Supplementary Video}
\label{supp_video}
In our supplementary video, we provide several audio-visual examples for each task and compare the performance of different models before and after introducing the reasoning steps.

\section{More related Works}
\label{more_related_works}
\noindent{\textbf{Multi-Agent Systems with LLMs.}} Recent advancements in multi-agent systems \cite{smit2023should, de2023emergent, guo2024large, li2024survey, han2024llm, wang2024rethinking, sun2024llm} underscore the potential of large language models in tackling complex tasks. While some approaches \cite{du2023improving} facilitate answer-sharing among agents for enhanced collaboration, Mixture-of-Agents \cite{wang2024mixture} employs a hierarchical architecture where agents iteratively refine responses. Comm \cite{chen2024comm} proposed problem-solving through structured communication and role division while Multi-Persona \cite{liang2023encouraging} promotes varied agent behaviours by assigning unique personas. ChatEval \cite{chan2023chateval} investigates various multi-agent debate strategies for effective interaction and response optimization while DMAS \cite{chen2024scalable} examines token-efficient multi-agent planning frameworks to enhance coordination and task performance. Building on advancements in multi-agent systems, recent research has investigated fine-tuning independently specialized agents that collaborate to produce diverse reasoning chains \cite{subramaniam2025multiagent}. In contrast to these approaches, our method emphasizes collaborative optimization via a shared experience library, allowing agents to collectively learn from and refine effective reasoning trajectories. 

\noindent{\textbf{Self-improvement.}} Self-improving models~\citep{huang2022large,yu2023teaching,yuan2024self,zhang2024small,welleck2022generating,peng2024regenesis} have gained significant attention due to their potential to enhance reasoning abilities through iterative feedback and refinement. Various studies~\citep{zelikman2022star,li2024large,pang2024iterative,lee2024llm2llm} utilize bootstrapping methods by leveraging self-generated rationales, while other works~\citep{yuan2024self,chen2024improving,ramji2024self,guo2025deepseek} introduce self-refinement mechanisms via reinforcement learning.

\noindent{\textbf{Multi-modal Learning.}} Conventional multi-modal methods incorporating vision-language \cite{clip, graphoptimaltransport, unitab, uniter, glip, imagebind, volta, blip, safari, albef, egovlp, egovlpv2, maple, apollo}, audio-visual \cite{imagebind, adverb, avseg, chowdhury2024melfusion, avsegformer, audvisum}, audio-language \cite{clap, audioclip, beats, wu2022wav2clip, merlot} have developed over the recent years with a focus to solve a variety of coarse-grained (question-answering, captioning, retrieval, etc.) and fine-grained (detection, segmentation, phrase grounding, etc.) understanding as well as generation tasks. However, these traditional models do not typically solve reasoning based tasks (with the exception of NLVR). With the advent of multi-modal LLMs \cite{llava, instructblip, xinstructblip, pgvideollava, chowdhury2024meerkat, instructgpt, vistallm, minigpt4, minigptv2, shikra, mplugowl, macawllm, anygpt, nextgpt, anymal, avicuna, baycat, videosalmonn, bubogpt, chatbridge, cheng2024videollama, cogvlm, ying2024internlm, vita, videochatgpt, lisa, gpt4roi, kosmos, timechat, unifiedio, imagebindllm, openai_gpt4o}, although some recent efforts have been made to leverage reasoning capabilities of LLMs to solve complex visual question answering tasks, multi-step reasoning with complex questions in the audio-visual space remains underexplored.

% \section{Detailed results for AVQA and AV Captioning}

\section{GPT based evaluation}
\label{gpt_based_evaluation}

\subsection{Choice Extraction}

\paragraph{Choice extraction strategy.}
\label{choice extraction strategy appendix}
We utilize a two-step choice extraction strategy, detailed next. While humans can easily extract choices from free-form predictions, rule-based matching may struggle with this task. To address this, we develop a universal evaluation strategy applicable to all AVLLMs, regardless of their varying instruction-following capabilities.

\noindent{\textit{\textbf{Step 1.} Prediction matching:}}
We first apply heuristic matching to extract choice labels (e.g., ‘A’, ‘B’, ‘C’, ‘D’) from AVLLM predictions. If successful, the extracted label is used as the final prediction. If heuristic matching fails, we employ GPT-4 to extract the choice label instead. 

\noindent{\textit{\textbf{Step 2.} GPT-4 processing:}}
Prior benchmarks \cite{mmbench} validate GPT-4's effectiveness as a choice extractor. If step 1 fails, we input the question, choices, and model prediction into GPT-4, instructing it to align the prediction with one of the provided choices and return the corresponding label. If no match is found, GPT-4 outputs ‘No match found.’

We also employ the best-of-N (3) evaluation strategy to ensure a rigorous evaluation and effectively demonstrate the performance gap across various models.

\paragraph{Response matching.}
To apply the matching algorithm to the options, we follow these rules: If an option is represented solely by a letter (e.g., `A`) or formatted as `A) <response>`, `A. <response>`, `A, <response>`, or `(A) <response>`, without embedding other choices within `<response>`, it is interpreted as a prediction of option `A`.

\paragraph{Where does heuristic matching fail?}

% mmbench
% \textcolor{red}{sanjoy use below}
The heuristic matching strategy usually fails in the following scenarios: (i) when the AVLLM is unable to provide an answer and requests clarification, such as `Apologies, can you please clarify ...` or similar phrases, and (ii) when the AVLLM responds with multiple option choices (A, B, C, etc.). In such cases, we proceed to Step 2, which involves GPT-4 based choice extraction. A sample prompt for GPT-4 is provided below.

% \paragraph{The prompt for LLM-based choice extraction.}

\begin{tcolorbox}[float, width=\columnwidth, colback=white, colframe=ThemeColor, title=\textcolor{black}{Choice extraction prompt for GPT-4} ] 

Can you help me match an answer with a set of options for a single correct answer type question? I will provide you with a question, a set of options, and a response from an agent. You are required to map the agent's response to the most similar option from the set. You should respond with a single uppercase character in `A', `B', `C', `D', and `E' depending on the choice you feel is the most appropriate match. If there are no similar options you might output `No match found'. Please refrain from being subjective while matching and do not use any external knowledge. Below are some examples:\\
Example 1: \\
Question: What color is the man's shirt who is sitting left of the object making this sound? \\
Options: A. Green B. Red C. Yellow D. Black \\
Answer: The person sitting next to the record player is wearing a black color shirt   \\
Your output: D \\
Example 2: \\
Question: What does the audio-visual event constitute? \\
Options: A. A dog barking at a cat  B. A dog barking on being hit by a stick C. The dog is hungry D. The dog is chasing another dog \\
Answer: It is a wolf  \\
Your output: No match found \\

\end{tcolorbox}

\paragraph{Change in template for GPT-4 evaluation.}
Next, to identify the model's prediction, we utilize GPT-4, following the approach in MMBench \cite{mmbench}. We prompt GPT-4 with a template that includes the question, options, and the corresponding AVLLM prediction. Additionally, we incorporate task-specific options to help GPT-4 recognize the model's predictions.

\subsection{Open-ended Answer Evaluation}
To evaluate open-ended question answers with given ground truth answers using GPT, we design a prompt that instructs the model to assess the accuracy and relevance of the model's answer in comparison to the ground truth. The prompt might be structured as: "Given the question, the model's answer, and the ground truth answer, determine whether the model’s answer is correct or incorrect. If the model's answer is factually accurate and appropriately aligns with the ground truth, even if expressed differently (e.g., `plane' vs. `aeroplane'), output `Correct'. If the answer is incorrect or significantly deviates from the ground truth, output `Incorrect'." This ensures that GPT understands that synonymous or contextually equivalent terms (such as `plane' for `aeroplane') should be considered correct. Additionally, the evaluation will focus on factual accuracy and contextual alignment, and it will mark answers as `Correct' if they are deemed effectively equivalent to the ground truth, despite minor wording differences.

\begin{figure}
    \centering
    \includegraphics[width=0.35\textwidth]{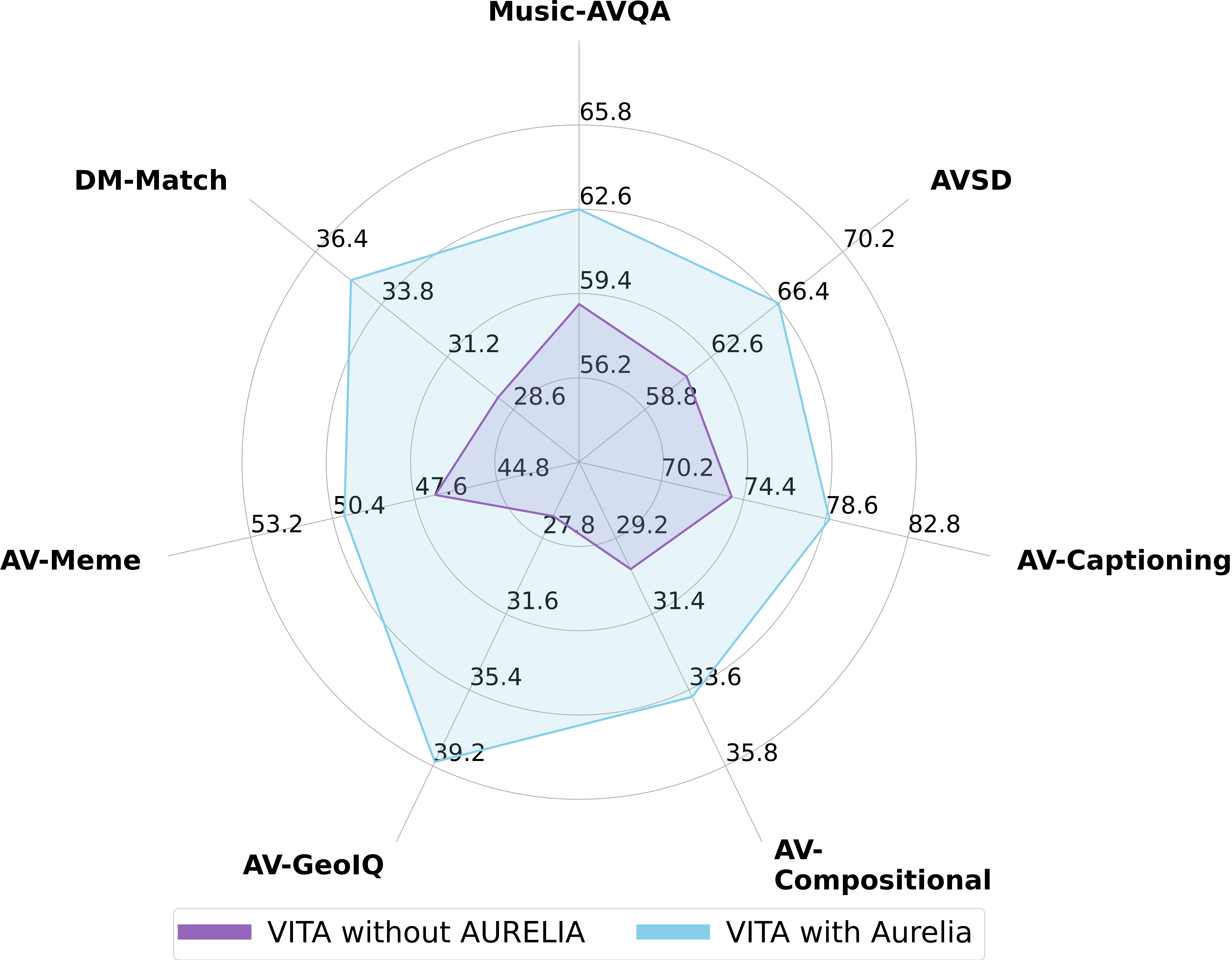}
    \caption{\textbf{Performance comparison across tasks. }The distillation of reasoning information in the VITA model via \ourapproach enhances its performance across all the tasks.}
    \label{fig:radar_plot}
\end{figure}

\begin{table*}[!t]
    \centering
    \renewcommand{\arraystretch}{1.5}
    \setlength{\tabcolsep}{8pt}
     \begin{tabular}{c|p{10cm}}
        \hline
       \textbf{Task} & \textbf{Instruct Prompt} \\
        \hline
        Reasoning generation &  Given the video and the audio and the question: \texttt{question} \newline
        Task 1: generate detailed reasoning steps for solving the given question without revealing the answer. \newline
        Task 2: provide detailed answers to each of these above reasoning steps generated in Task 1.\newline
        Task 3: provide a final answer for the question.\newline
        Your output should be in the form of a dictionary which looks like: \texttt{Task\_1}: Task 1 answers, \texttt{Task\_2}: Task 2 answers, \texttt{Task\_3}: Task 3 answers.
 \\ 
        \hline
        Summarization &  Given the reasoning steps, the answer to the reasoning steps, and the final response for the question, generate (come up with / guess) a detailed caption which is able to define the contents of the video and the audio.\newline
        In the questions and the answers there may be things that might be outside the video and the audio context and needs world knowledge. You have to keep this in mind while generating the caption and you have to discard these information from the caption."
 \\ 
 \hline
 Evaluation &     Given video and audio inputs, can you rate the following caption between 1 to 10 (1 being the lowest) based on its similarity with the corresponding inputs. Strictly output the numerical score only.\newline
 Caption:  \texttt{summary}
\\
 \hline
 Feedback &  The reasoning steps you previously generated: \texttt{\{reasoning\_steps\}} to answer the question: \texttt{\{question\}} were evaluated and received a score of \texttt{\{score\}} out of 10. This score suggests that the reasoning steps may not be fully appropriate for answering the question correctly.  \newline Now, given the video, audio, and the question, carefully generate the correct reasoning steps to answer the question: \texttt{\{question\}} while strictly adhering to the following response format:  \newline Task 1: generate detailed reasoning steps for solving the given question without revealing the answer. \newline Task 2: provide detailed answers to each of these above reasoning steps generated in Task 1. \newline Task 3: provide a final answer for the question. \newline Your output should be in the form of a dictionary which looks like: \texttt{Task\_1}: Task 1 answers, \texttt{Task\_2}: Task 2 answers, \texttt{Task\_3}: Task 3 answers.
\\
        \hline
    \end{tabular}
    \caption{\textbf{Details of Instruct Prompts. }Table presents the instruction prompts utilized by different agents in various stages of \ourapproach.}
    \label{tab:instruct_prompt}
\end{table*}

\section{Examples of Prompts}
\label{appendix_prompts}
We use a combination of closed-source LLMs as specialized agents in \ourapproach. To enable these LLM agents to interact with the input and with each other, we prompt them with appropriate instructions. We list these instruction prompts in Table \ref{tab:instruct_prompt}.

\section{Radar plot}
\label{radar plot}
The radar plot \cref{fig:radar_plot} illustrates the performance of the best performing open-source model VITA \cite{vita} on all 7 datasets before and after reasoning distillation is performed. We note that, upon ZS finetuning leveraging \ourapproach the performance on each task is improved significantly with the maximum performance gain of 12.6\% observed in the AV-captioning task. This underlines the efficacy of our proposed reasoning data generation pipeline. AV-captioning often requires the model to draw intricate conclusions by critically analysing the audio-visual associations over multimodal temporal signals. A steady improvement in all the tasks underline the rich contextual understanding our reasoning augmented data can inject into a model.     

\section{Details on Reasoning Data Generation}
\label{reasoning_data_generation}
To facilitate such reasoning generation, our framework, \ourapproach, employs a multi-agent system that iteratively refines reasoning steps. A Reasoning Generator Agent first produces step-by-step deductions and explanations. The Summarization Agent then distills these steps into a structured caption without direct access to video or audio, ensuring reasoning quality is independent of raw inputs. A Multi-Modal Evaluator Agent assigns a similarity score based on how well the reasoning aligns with the original content, and a Feedback Agent iteratively refines the reasoning process to improve coherence and accuracy. Once the reasoning achieves an optimal evaluation score, it is integrated into the input before being fed into the target model. This explicit reasoning injection significantly enhances the model’s ability to derive accurate, interpretable answers while minimizing errors and hallucinations.
The process begins with the Reasoning Generator Agent, which analyzes the input set and produces step-by-step reasoning alongside an explanation for each step. Following this, the Summarization Agent interacts with the reasoning steps and generates a detailed caption crafted solely from the reasoning steps without any direct knowledge of the video or audio. This ensures that the caption’s quality and accuracy are entirely dependent on the correctness of the generated reasoning.
Next, a Multi-Modal Evaluator Agent assesses the alignment between the generated caption and the original video-audio content, assigning a similarity score between 1 and 10. A score of 1 indicates no alignment, while a 10 signifies perfect correspondence. Based on this evaluation, a Feedback Agent iteratively refines the reasoning steps by guiding the Reasoning Generator Agent to enhance its output by generating more coherent reasoning steps, aiming to maximize the evaluation score. This iterative loop continues until the reasoning quality surpasses a predefined threshold.
Once the evaluation score pertaining to the reasoning steps reaches an optimal level, the reasoning information obtained at that step is integrated with the original audio, video, and question before being fed into the target model. By incorporating structured reasoning through distillation, \ourapproach significantly improves the model’s reasoning and overall performance.

\section{\ourbenchmark Statistics}
\label{data statistics}

\subsection{Data Distribution}
\label{appendix_data_size}
\begin{figure*}[t]
    \centering
    \begin{subfigure}{0.65\textwidth}
        \centering
        \includegraphics[width=\linewidth]{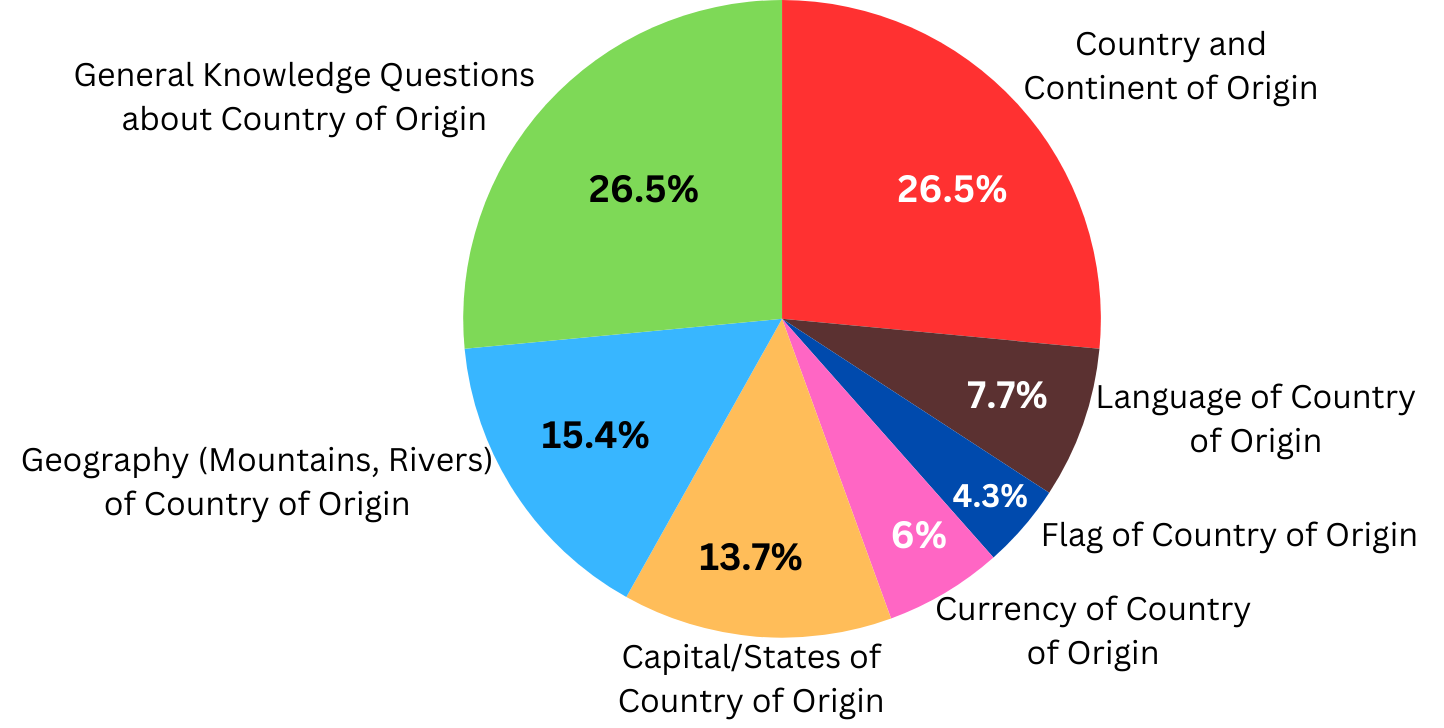}
        \caption{Distribution of \ourtask task.}
        \label{fig:pie-chart-1}
    \end{subfigure}
    
    \vspace{10pt} % Adjust spacing between the charts
    
    \begin{subfigure}{0.52\textwidth}
        \centering
        \includegraphics[width=\linewidth]{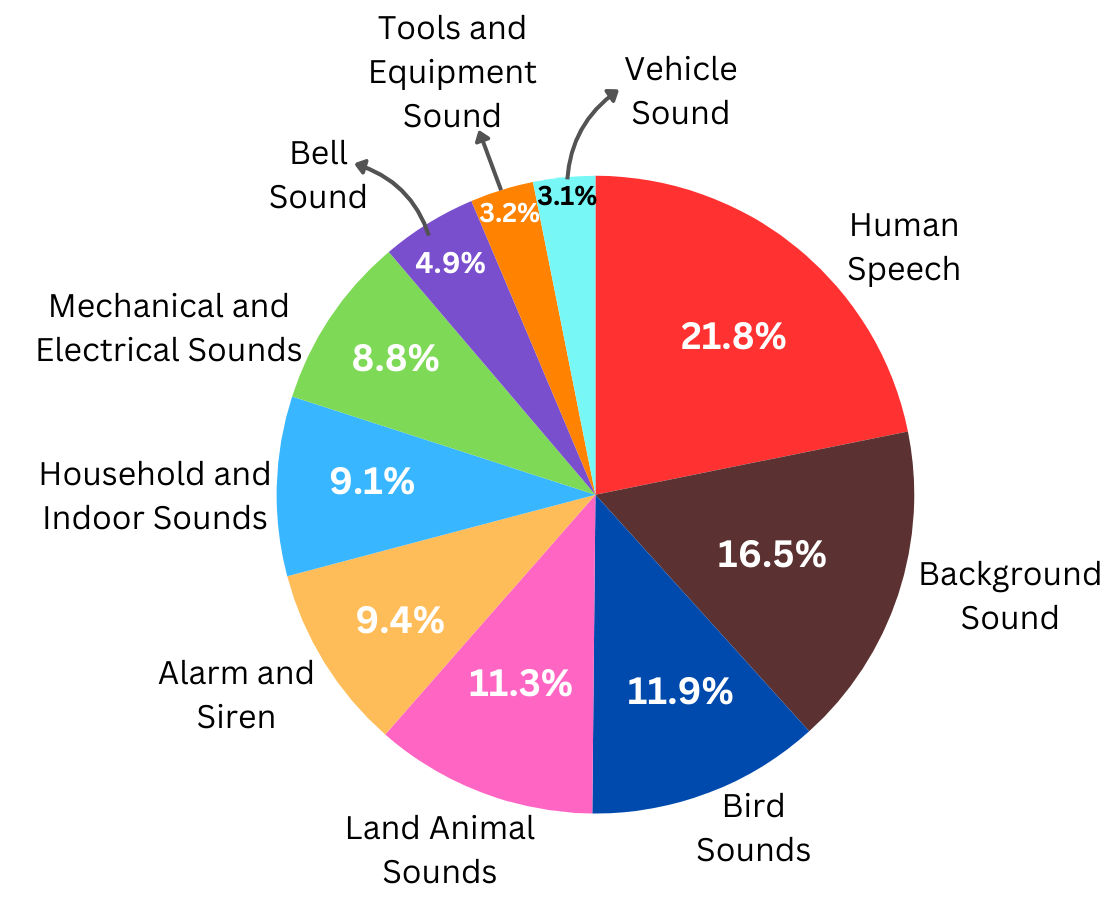}
        \caption{Distribution of AV-Compositional Understanding task.}
        \label{fig:pie-chart-2}
    \end{subfigure}
    
    \caption{\textbf{Distribution of \ourtask and AV-Compositional Understanding tasks. }\textbf{(a)} The pie chart shows the distribution of samples from our proposed \ourtask task. The collected samples exhibit diverse geographical and cultural characteristics. \textbf{(b)} The pie chart shows the distribution of samples from the AV-Compositional Understanding task. As seen from the pie chart, the data samples are collected from a diverse range of practical audio visual scenarios.}
    \label{fig:combined-pie-charts}
\end{figure*}

\cref{tab:benchmark_statistics} reports different tasks along with various question categories associated with them. For example, QA pairs for \ourtask are collected from diverse categories of scenarios that require geographical and cultural knowledge combined with strong audio-visual reasoning. Similarly, samples for other tasks are also collected from diverse domains that span various categories. \cref{fig:combined-pie-charts} reports data distribution for \ourtask and AV-Compositional understanding.      

\begin{table*}[h]
    \centering
    \renewcommand{\arraystretch}{1.0}
    % \vspace{3pt}
    % \vspace{-0.6em}
    \small
    \resizebox{0.8\linewidth}{!}{
\begin{tabular}{@{}ccccc@{}}
\toprule
Task ID & Question Category    & Task Name & Class & Number \\ \midrule
1      & Country Recognition               & \ourtask     & 17    & 21 \\
2 & Famous Landmark & \ourtask & 18 & 23 
\\  
3 & Popular Dish/Food & \ourtask & 16 &  19
\\
4 & Currency & \ourtask & 12 & 13
\\
5 & Continent & \ourtask & 5 & 17 
\\
6 & Flag Specifics & \ourtask & 10 & 15
\\
7 & Popular Dance Form & \ourtask & N/A & 20
\\
8 & Geographical & \ourtask & N/A & 31
\\
9 & Language & \ourtask & 11 & 13
\\
10 & Commonsense Reasoning & \ourtask, AV-Meme, AV-Dance Match & N/A & 165
\\
11 & Musical Performances & Music-AVQA, \ourtask & N/A & 1014
\\
12 & Dynamic Scene & AVSD & N/A & 931
\\
13 & Meme and Humor & AV-Meme & N/A & 50 
\\
14 & Dance Performances & AV-Dance Match & N/A & 100
\\
15 & Indoor/Kitchen Scenarios & VALOR & N/A & 945
\\
16 & Compositional & AV-Comp & N/A & 968
\\
17 & Miscellaneous  & \ourtask, AVSD, VALOR & N/A & 159
\\
\bottomrule
\label{taskdata}
\end{tabular}
}
\caption{\textbf{Task Statistics. }Table shows detailed task statistics in \ourbenchmark.} \label{tab:benchmark_statistics}
\end{table*}

% \textcolor{red}{ADD PIE CHARTS}

\section{Breakdown results}
\label{Breakdown results}
In this section, we report the performance at a more granular level on \ourbenchmark. We identify samples belonging to certain categories and consider only them for evaluation. 

\subsection{Performance on musical videos}
We report the performance on musical videos category in \cref{tab:appendix_musical_results}. The samples under consideration require the AVLLMs to comprehend fine grained audio visual interactions followed by reasoning them with general knowledge/geo-cultural understanding. Experimental results demonstrate -- best performance is achieved by VITA powered by its strong multimodal understanding. On an average, AV-compositional understanding task achieves most gains due to the reasoning supplement.

\subsection{Performance on commonsense reasoning videos}
\cref{tab:appendix_commonsense_results} reports similar breakdown on commonsense reasoning examples. VITA outperforms other opensource models to achieve significantly improved performance upon treated with reasoning enhanced data generated by \ourapproach. Highest performance gains are observed in \ourtask confirming the requirement of strong practical understanding of AV scenarios for this task.

\begin{table*}[t]
    \centering
    \renewcommand{\arraystretch}{1.2}
    \resizebox{0.9\linewidth}{!}{%
    \begin{tabular}{l|c|c|c|c|c|c|c}
\hline
\rowcolor{gray!20}
\multicolumn{1}{c|}{} &
  \multicolumn{2}{c|}{\textbf{AV-QA}} &
  \multicolumn{1}{c|}{\textbf{AV-Captioning}} &
  \multicolumn{1}{c|}{\multirow{1}{*}{\textbf{AV-Compositional}}} &
  \multicolumn{1}{c|}{\multirow{1}{*}{\textbf{AV-GeoIQ}}} &
  \multicolumn{1}{c|}{\multirow{1}{*}{\textbf{AV-Meme}}} &
  \multicolumn{1}{c}{\multirow{1}{*}{\textbf{DM-Match}}} \\
\cline{2-3}
\rowcolor{gray!20}
\multirow{-2}{*}{\textbf{Models}} &
  \textbf{Music-AVQA} &
  \textbf{AVSD} &
   &
   &
   &
   &
\\
\hline
\multicolumn{8}{c}{\textit{Open-Source Models in ZS}} \\
\hline

\cellcolor{gray!20}NExT-GPT & 53.5 &  52.1 &  62.5 & 27.7 &  25.3 &  17.5 &  26.2\\
% \cellcolor{gray!20}Unified-IO-2 L &  55.1 & 57.9 &  70.1 &  27.2 &  21.5 &  20.0 &  27.5 \\
\cellcolor{gray!20}Unified-IO-2 XL &  53.6 &  52.6 &  76.7 &  29.4 &  23.4 &  23.1 & 28.3 \\
\cellcolor{gray!20}Bay-CAT &  55.7 &  54.2 &  68.2 &  26.5 &  22.8 &  23.3 &  28.7\\
\cellcolor{gray!20}Video-SALMONN &  57.6 &  58.8 &  73.4 &  25.5 &  23.0 &  23.0 &  24.5\\
\cellcolor{gray!20}VITA &  59.2 &  62.3 &  74.6 &  27.4 &  26.6 &  46.4 &  28.8\\
\hline
\multicolumn{8}{c}{\textit{Open-Source Models with \ourapproach}} \\
\hline
\rowcolor{magenta!5}

\rowcolor{magenta!5}
\cellcolor{gray!20}NExT-GPT & 56.8 & 55.3 & 66.5 & 30.1 & 29.2 & 22.0 & 30.5\\
% \rowcolor{magenta!5}
% \cellcolor{gray!20}Unified-IO-2 L & 61.9 & 62.0 & 74.6 & 32.4 & 36.5 & 35.0 & 33.5\\
\rowcolor{magenta!5}
\cellcolor{gray!20}Unified-IO-2 XL & 56.3 & 57.7 & 79.6 & 32.6 & 28.5 & 27.2 & 33.0\\
\rowcolor{magenta!5}
\cellcolor{gray!20}Bay-CAT & 57.6 & 59.1 & 73.2 & 29.6 & 27.0 & 26.0 & 32.5\\
\rowcolor{magenta!5}
\cellcolor{gray!20}Video-SALMONN & 61.8 & 62.6 & 76.8 & 29.1 & 28.6 & 28.0 & 29.0\\
\rowcolor{magenta!5}
\cellcolor{gray!20}VITA & 61.4 & 65.3 & 78.3 & 32.5 & 30.7 & 49.2 & 33.9\\
  \hline
\end{tabular}%
}
% \vspace{0.05in}
\caption{\textbf{Breakdown results on musical videos.} Performance comparison of various models before and after applying \ourapproach.}
\label{tab:appendix_musical_results}
% \vspace{-7mm}
\end{table*}

\begin{table*}[t]
    \centering
    \renewcommand{\arraystretch}{1.2}
    \resizebox{0.9\linewidth}{!}{%
    \begin{tabular}{l|c|c|c|c|c|c|c}
\hline
\rowcolor{gray!20}
\multicolumn{1}{c|}{} &
  \multicolumn{2}{c|}{\textbf{AV-QA}} &
  \multicolumn{1}{c|}{\textbf{AV-Captioning}} &
  \multicolumn{1}{c|}{\multirow{1}{*}{\textbf{AV-Compositional}}} &
  \multicolumn{1}{c|}{\multirow{1}{*}{\textbf{AV-GeoIQ}}} &
  \multicolumn{1}{c|}{\multirow{1}{*}{\textbf{AV-Meme}}} &
  \multicolumn{1}{c}{\multirow{1}{*}{\textbf{DM-Match}}} \\
\cline{2-3}
\rowcolor{gray!20}
\multirow{-2}{*}{\textbf{Models}} &
  \textbf{Music-AVQA} &
  \textbf{AVSD} &
   &
   &
   &
   &
\\
\hline
\multicolumn{8}{c}{\textit{Open-Source Models in ZS}} \\
\hline
\cellcolor{gray!20}NExT-GPT & 51.2 &  50.3 &  59.6 & 25.7 &  22.7 &  16.9 &  24.7\\
\cellcolor{gray!20}Unified-IO-2 XL &  50.4 &  51.7 &  73.2 &  28.0 &  22.2 &  22.0 & 25.3 \\
\cellcolor{gray!20}Bay-CAT &  51.7 &  52.2 &  66.4 &  24.9 &  20.3 &  21.1 &  25.2\\
\cellcolor{gray!20}Video-SALMONN &  53.7 &  52.2 &  70.1 &  22.7 &  21.3 &  20.2 &  21.9\\
\cellcolor{gray!20}VITA &  55.7 &  59.7 &  71.2 &  24.0 &  22.3 &  43.5 &  26.5\\
\hline
\multicolumn{8}{c}{\textit{Open-Source Models with \ourapproach}} \\
\hline
\rowcolor{magenta!5}

\rowcolor{magenta!5}
\cellcolor{gray!20}NExT-GPT & 55.2 & 54.8 & 63.1 & 29.6 & 26.7 & 21.0 & 28.3\\
\rowcolor{magenta!5}
\cellcolor{gray!20}Unified-IO-2 XL & 54.3 & 55.2 & 76.8 & 32.1 & 27.4 & 26.3 & 29.5\\
\rowcolor{magenta!5}
\cellcolor{gray!20}Bay-CAT & 55.6 & 56.1 & 70.2 & 28.6 & 25.8 & 25.4 & 29.5\\
\rowcolor{magenta!5}
\cellcolor{gray!20}Video-SALMONN & 58.8 & 57.6 & 74.8 & 27.1 & 26.6 & 25.0 & 26.2\\
\rowcolor{magenta!5}
\cellcolor{gray!20}VITA & 60.4 & 64.7 & 74.7 & 29.5 & 27.7 & 48.1 & 31.2\\
  \hline
\end{tabular}%
}
% \vspace{0.05in}
\caption{\textbf{Breakdown results on commonsense reasoning videos}. Table shows the performance comparison of various models before and after applying \ourapproach specifically on commensense reasoning related videos.}
\label{tab:appendix_commonsense_results}
% \vspace{-7mm}
\end{table*}

% \section{Comparison with other test-time scaling methods}
% \label{Comparison with other test-time scaling methods}
% To further validate the effectiveness of AURELIA, we present a comparison of AURELIA with an existing test-time scaling method i.e. best of N search. We present the results in Table \ref{tab:comparison-testtime-methods} with five target models across 6 tasks. We observe the AURELIA outperforms the best of N search method, showcasing its effectiveness in instilling reasoning inside the AVLLMs.

% \section{Latency comparison}
% The addition of reasoning distillation at test-time via AURELIA can increase the inference time of the target AVLLMs. Table \ref{} shows the latency comparison of models with and without AURELIA. Although the latency increases with AURELIA, the performance is relatively better. Thus, there is a trade-off between compute time and performance. However, for the sake of reliable predictions, we can overlook the time constraints in the current state.

\section{Results on other benchmarks}
\label{other benchmark results}
We compare the performance of Video-SALMONN and Unified-IO-2 on VideoMME and report them in \cref{videomme_benchmark_results_video_salmonn} and \cref{videomme_benchmark_results_unifiedio}. As can be clearly seen, our synthetic reasoning data augmentation pipeline is generalizable to other benchmarks. Employing reasoning enhanced annotations generated by \ourapproach boosts the performance in all the models. Instilling strong reasoning capabilities improves the average performance significantly.

\begin{table*}
\centering

\begin{adjustbox}{max width=\linewidth}
\begin{tabular}{crlllllll} 
\toprule
\multirow{3}{*}{\textbf{Subset}} & \multicolumn{1}{c}{\multirow{3}{*}{\textbf{Modality}}} & \multicolumn{7}{c}{\textbf{Category }}                                                                                                                                                                                                                                                                                                                                                \\ 
\cmidrule{3-9}
                                 & \multicolumn{1}{c}{}                               & \multicolumn{1}{c}{\multirow{2}{*}{\textit{Knowledge~}}} & \multicolumn{1}{c}{\multirow{2}{*}{\textit{Film \& Television}}} & \multicolumn{1}{c}{\textit{Sports}}      & \multicolumn{1}{c}{\textit{Artistic~}}   & \multicolumn{1}{c}{\textit{Life}}   & \multicolumn{1}{c}{\multirow{2}{*}{\textit{Multilingual }}} & \multicolumn{1}{c}{\multirow{2}{*}{Overall}}  \\
                                 & \multicolumn{1}{c}{}                               & \multicolumn{1}{c}{}                                     & \multicolumn{1}{c}{}                                                             & \multicolumn{1}{c}{\textit{Competition}} & \multicolumn{1}{c}{\textit{Performance}} & \multicolumn{1}{c}{\textit{Record}} & \multicolumn{1}{c}{}                                        & \multicolumn{1}{c}{}                          \\ 
\midrule
\multirow{2}{*}{Short} 
 & ZS & 78.6  & 84.2  & 75.1  & 82.9  &  82.0 & 83.6  & 81.2  \\
  & + \ourapproach & 82.1 & 88.3 & 78.4 & 85.7 & 85.2 & 86.4 & \textbf{84.8} \\
\midrule
\multirow{2}{*}{Medium} 
 & ZS & 77.3 & 80.7  & 69.0  &  80.6 & 72.6  & 96.8  & 78.5  \\
 & + \ourapproach & 80.1 & 83.7 & 72.8 & 84.7 & 75.8 & 97.1 & \textbf{82.7} \\
\midrule
\multirow{2}{*}{Long} 
 & ZS & 78.6  & 70.8  & 69.4  & 60.3  & 63.0  & 80.9  & 70.2  \\
 & + \ourapproach & 82.8 & 74.7 & 72.1 & 64.4 & 66.6 &  83.8 & \textbf{73.7}  \\
\midrule
\multirow{2}{*}{Overall} 
 & ZS & 77.5  & 79.6  & 71.7  &  75.8 & 75.9  &  85.7 &   77.1\\
  & + \ourapproach & 80.5 & 82.7 & 74.9 & 78.4 & 78.7 & 89.0 & \textbf{80.0} \\
\bottomrule
\end{tabular}
\end{adjustbox}
\caption{\textbf{Performance of Video SALMONN across Video-MME}. The evaluation is done on audio-visual inputs.}
\label{videomme_benchmark_results_video_salmonn}
\end{table*}
 %%%%%%%%%%%%% for Unified IO2

\begin{table*}[!h]
\centering

\begin{adjustbox}{max width=\linewidth}
\begin{tabular}{crlllllll} 
\toprule
\multirow{3}{*}{\textbf{Subset}} & \multicolumn{1}{c}{\multirow{3}{*}{\textbf{Modality}}} & \multicolumn{7}{c}{\textbf{Category }}                                                                                                                                                                                                                                                                                                                                                \\ 
\cmidrule{3-9}
                                 & \multicolumn{1}{c}{}                               & \multicolumn{1}{c}{\multirow{2}{*}{\textit{Knowledge~}}} & \multicolumn{1}{c}{\multirow{2}{*}{\textit{Film \& Television}}} & \multicolumn{1}{c}{\textit{Sports}}      & \multicolumn{1}{c}{\textit{Artistic~}}   & \multicolumn{1}{c}{\textit{Life}}   & \multicolumn{1}{c}{\multirow{2}{*}{\textit{Multilingual }}} & \multicolumn{1}{c}{\multirow{2}{*}{Overall}}  \\
                                 & \multicolumn{1}{c}{}                               & \multicolumn{1}{c}{}                                     & \multicolumn{1}{c}{}                                                             & \multicolumn{1}{c}{\textit{Competition}} & \multicolumn{1}{c}{\textit{Performance}} & \multicolumn{1}{c}{\textit{Record}} & \multicolumn{1}{c}{}                                        & \multicolumn{1}{c}{}                          \\ 
\midrule
\multirow{2}{*}{Short} 
 & ZS & 76.2  & 82.8  & 73.9  & 80.7  & 79.9  & 81.9  & 79.0  \\
  & + \ourapproach & 78.9 & 84.1 & 74.9 & 82.9 & 81.0 & 83.1 & \textbf{81.1} \\
\midrule
\multirow{2}{*}{Medium} 
 & ZS & 75.6 & 78.9  & 67.9  & 78.1  & 70.7  & 94.4  & 76.4  \\
 & + \ourapproach & 78.9 & 81.8 & 70.4 & 82.8 & 73.8 & 95.3 & \textbf{80.5} \\
\midrule
\multirow{2}{*}{Long} 
 & ZS & 76.7  & 68.8  & 67.3  & 58.5  & 61.9  & 78.0  &  68.2 \\
 & + \ourapproach & 80.8 & 72.6 & 70.4 & 62.6 & 64.6 & 81.7 & \textbf{71.6}  \\
\midrule
\multirow{2}{*}{Overall} 
 & ZS &  75.7 &  77.7 & 68.9  & 73.4  &  73.8 &  82.8 &  75.8 \\
  & + \ourapproach & 79.5 & 80.4 & 72.6 & 76.5 & 76,8 & 87.4 & \textbf{77.6} \\
\bottomrule
\end{tabular}
\end{adjustbox}
\caption{\textbf{Performance of Unified-IO-2 across Video-MME}. The evaluation is done on audio-visual inputs.}
\label{videomme_benchmark_results_unifiedio}
\end{table*}

\section{Qualitative Results}
\label{appendix_qual_results}
\begin{figure*}
    \centering
    \begin{subfigure}{\textwidth}
        \centering
        \includegraphics[width=0.9\textwidth]{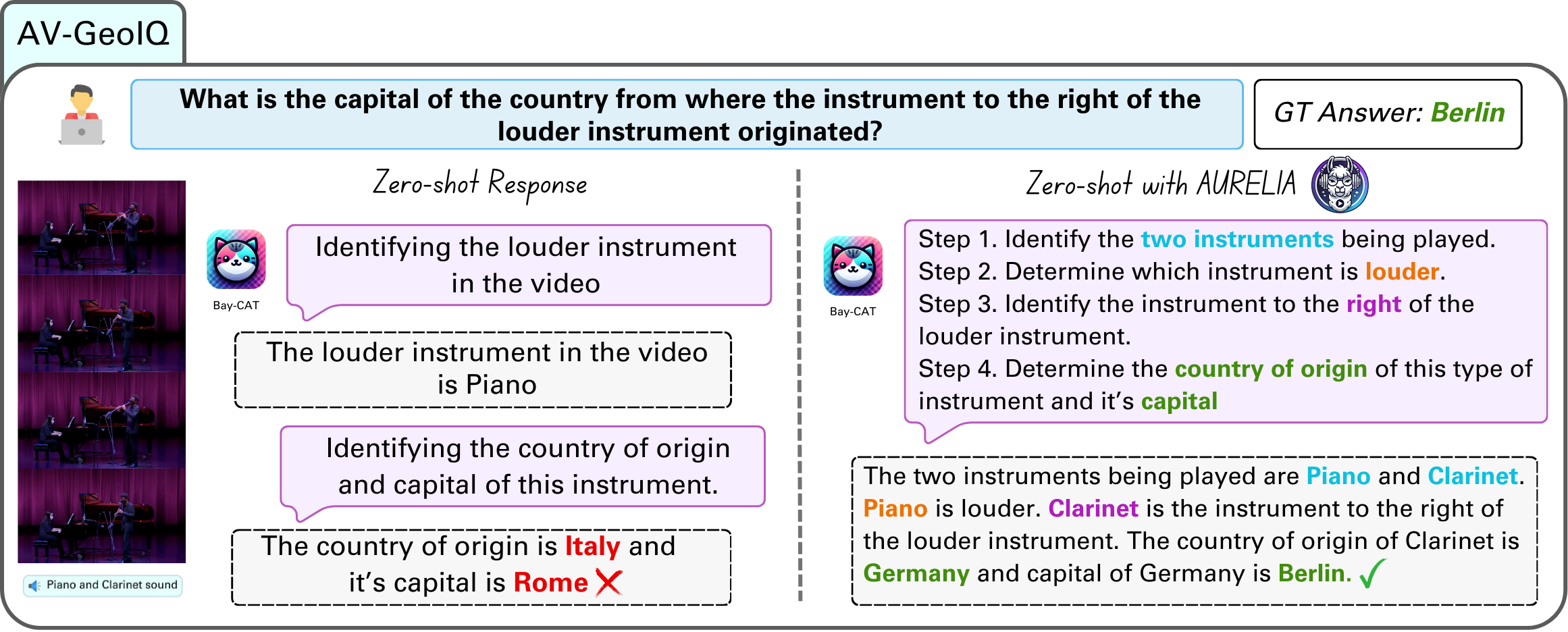}
        \caption{}
        \label{fig:qual-geo-1}
    \end{subfigure}
    
    \begin{subfigure}{\textwidth}
        \centering
        \includegraphics[width=0.9\textwidth]{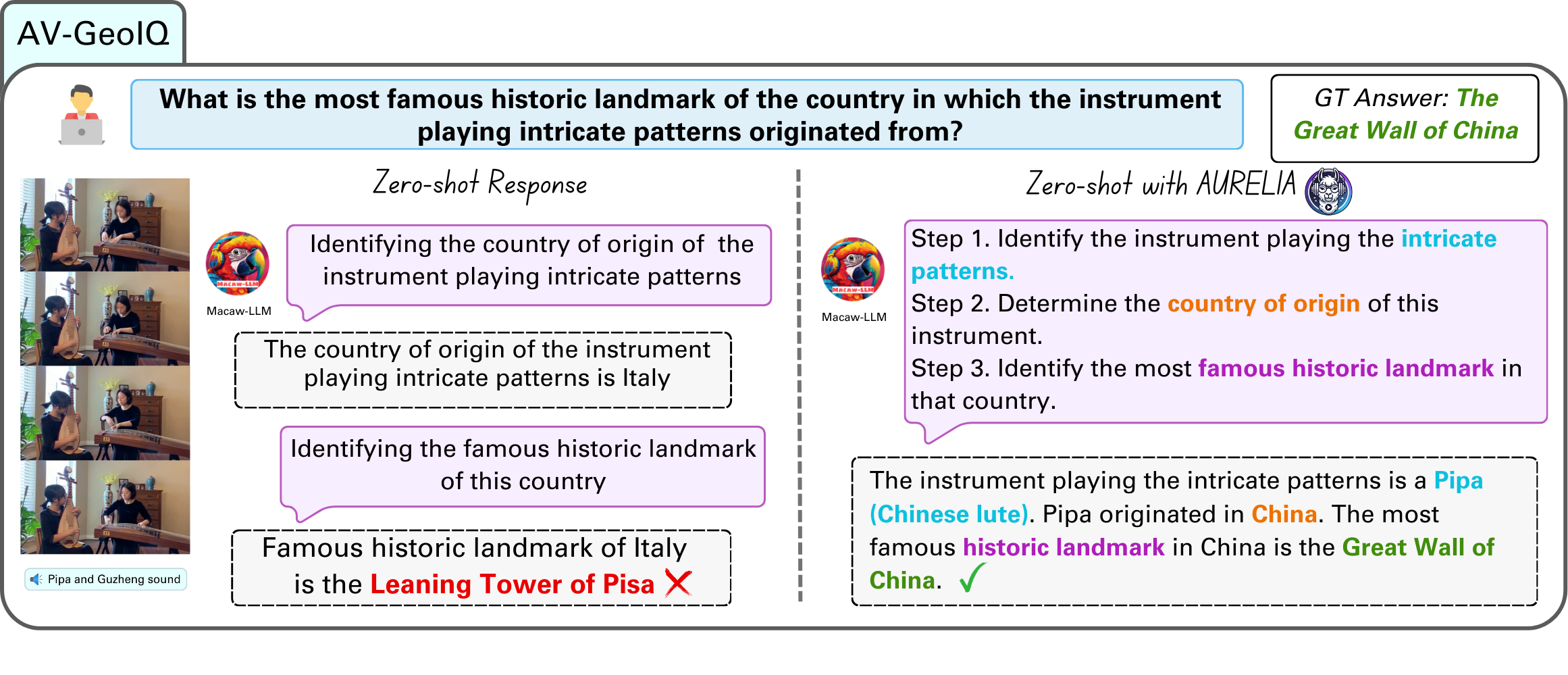}
        \caption{}
        \label{fig:qual-geo-2}
    \end{subfigure}
    
    \begin{subfigure}{\textwidth}
        \centering
        \includegraphics[width=0.9\textwidth]{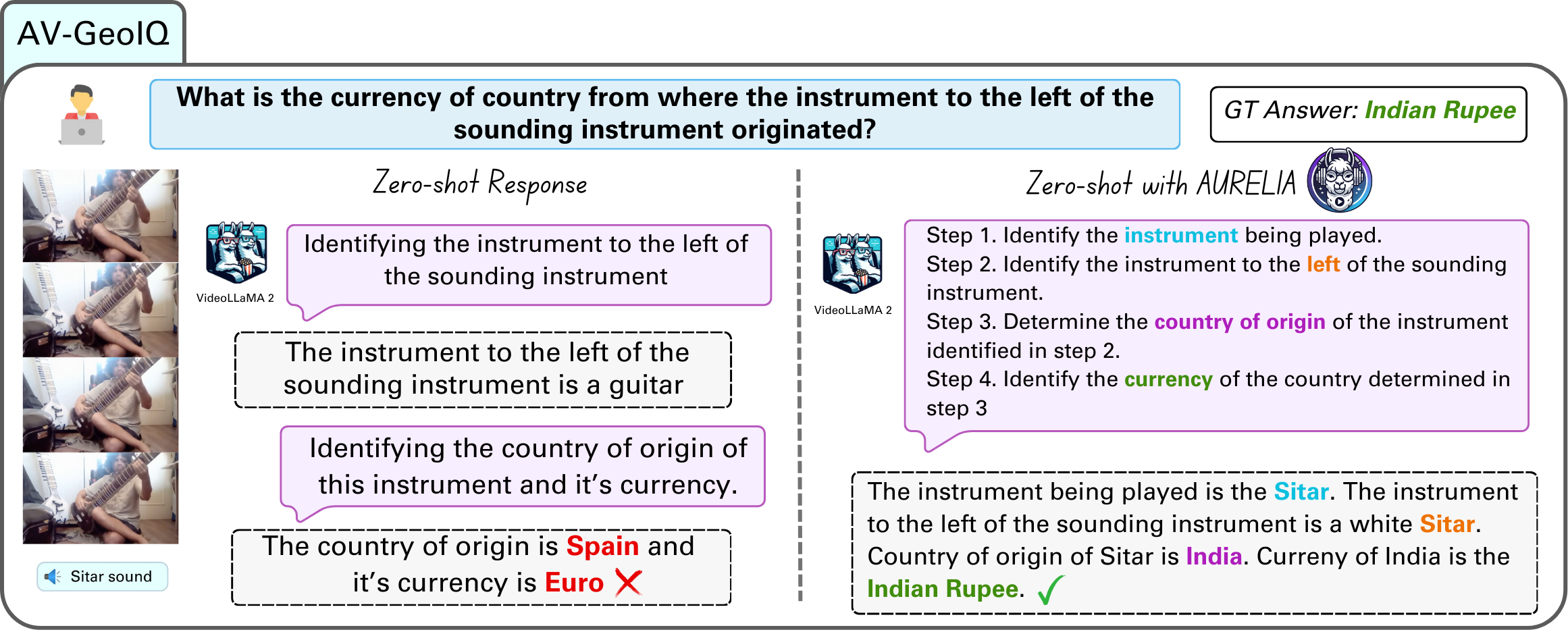}
        \caption{}
        \label{fig:qual-geo-3}
    \end{subfigure}

    \caption{Qualitative visualization of \ourapproach's reasoning distillation across AV-GeoIQ task. } 
    \label{fig:qual-geo-combined}
\end{figure*}

\cref{fig:qual-geo-combined} - \cref{fig:qual-comp-combined} demonstrate several qualitative examples for each task. For \ourtask we design questions which require the model to reason at multiple levels and go through a series of derived steps to be able to come up with the correct response. As seen from these examples, injecting reasoning annotations into the AVLLMs significantly improves the performance in various audio-visual scenarios which require critical multimodal comprehension. Similar improvements can be observed for other tasks as well. \ourapproach equips the models with a series of critical reasoning sequences which enables better decision making through step by step reasoning. Powered by reasoning annotated data significant improvements can be observed in AV-compositional understanding, AV-Meme understanding and AV-Dance matching tasks.  

\section{Key Observations}
\label{discussion on failure cases}
This section highlights key insights into the performance of AVLLMs when injected with reasoning data generated by \ourapproach.
% \begin{figure}
%     \centering
%     \includegraphics[width=0.5\textwidth]{ICCV2025-Author-Kit-Feb/figures/failure_cases.png}
%     \caption{Examples of Failure Cases.}
%     \label{fig:qualitative_example_failure}
% \end{figure}

\noindent
\textbf{Open-ended evaluations. }We observe that AVLLMs injected with the reasoning data generated by \ourapproach, in addition to being effective on AV samples under close ended MCQ setting, are also effective in case of open-ended answers. The former evaluation has a predefined set of options out of which only one option is correct while latter is relatively harder to answer as it is not bounded by word vocabulary. We find that employing our reasoning augmented data also improves the open-ended evaluation of existing AVLLMs.

\noindent
\textbf{Emphasis on one modality. }It is observed that existing AVLLMs occasionally prioritizes one modality over the other, introducing biases in its decision-making process. Since \ourapproach works on the synergy of AV input through the interaction of multiple agents, in such cases, our approach can mitigate the bias induced due to the model's focus on one modality by providing additional cues about the other modality through reasoning steps. However, we also notice occasionally (such as in Fig. 4 (left) of main paper), reasoning distillation becomes less effective in such extreme cases, as the model remains biased towards the dominant modality, neglecting the valuable information from the other.
% As illustrated in Fig. 4 (left) of main paper, \ourapproach fails to integrate the audio modality, leading to an incorrect output. Specifically, referring to the failure case example (left): 
In this specific example, the AVLLM incorrectly assumes the dog is silent, even when audio information is present. We hypothesize that the error in such cases can propagate through the reasoning stages due to model being biased in initial step itself, ultimately resulting in a flawed conclusion.

\noindent
\textbf{Suboptimal Comprehension. }\ourapproach systematically distills the reasoning information in the AVLMMs to advance their AV comprehension capability. Leveraging strong multi-agent LLMs, \ourapproach has an advanced comprehension of intricate AV relationships, which can help mitigate the weak reasoning comprehension in AVLLMs. Even though based on strong closed-source LLMs, \ourapproach can also incur errors sometimes in AV comprehension. Since \ourapproach relies on a synergy of multi-modal agents, making any misunderstanding of audio-video input could be detrimental to the entire reasoning pipeline. Fig. 4 (right) of main paper illustrates such a case, where \ourapproach struggles to grasp the interplay between video and audio.
\section{Future Work}
\label{future_work}
% \ourapproach introduces a novel approach to distilling reasoning information into the target model in a zero-shot setting, eliminating the need for explicit training. 
Currently, the multi-agent framework of \ourapproach leverages a combination of closed-source LLMs as agents. A promising future direction would be to replace these proprietary models with open-source alternatives, enhancing accessibility and transparency. Additionally, another avenue for improvement lies in integrating reasoning directly into the training or instruction-tuning phase, rather than generating it dynamically at inference time. This would enable AVLLM to inherently develop step-by-step reasoning capabilities, allowing it to derive answers more naturally and effectively.

\section{Societal Impact}
\label{societal impact}
In this work, we perform an extensive analysis of reasoning capabilities of existing AVLLMs. Our study reveals that models lack sufficient audio-visual comprehension skills and most often fail to address scenarios that require common-sense reasoning. We believe our work can be useful to the community, and our findings can reveal the potential threats associated with deploying these models in real-time or accuracy-critical setups. We employ existing public datasets and in some cases, collect samples to curate the benchmark. We don't use any personal/human subject data without consent during data preparation and experiments.

\begin{figure*}
    \centering
    \begin{subfigure}{0.9\textwidth}
        \centering
        \includegraphics[width=\textwidth]{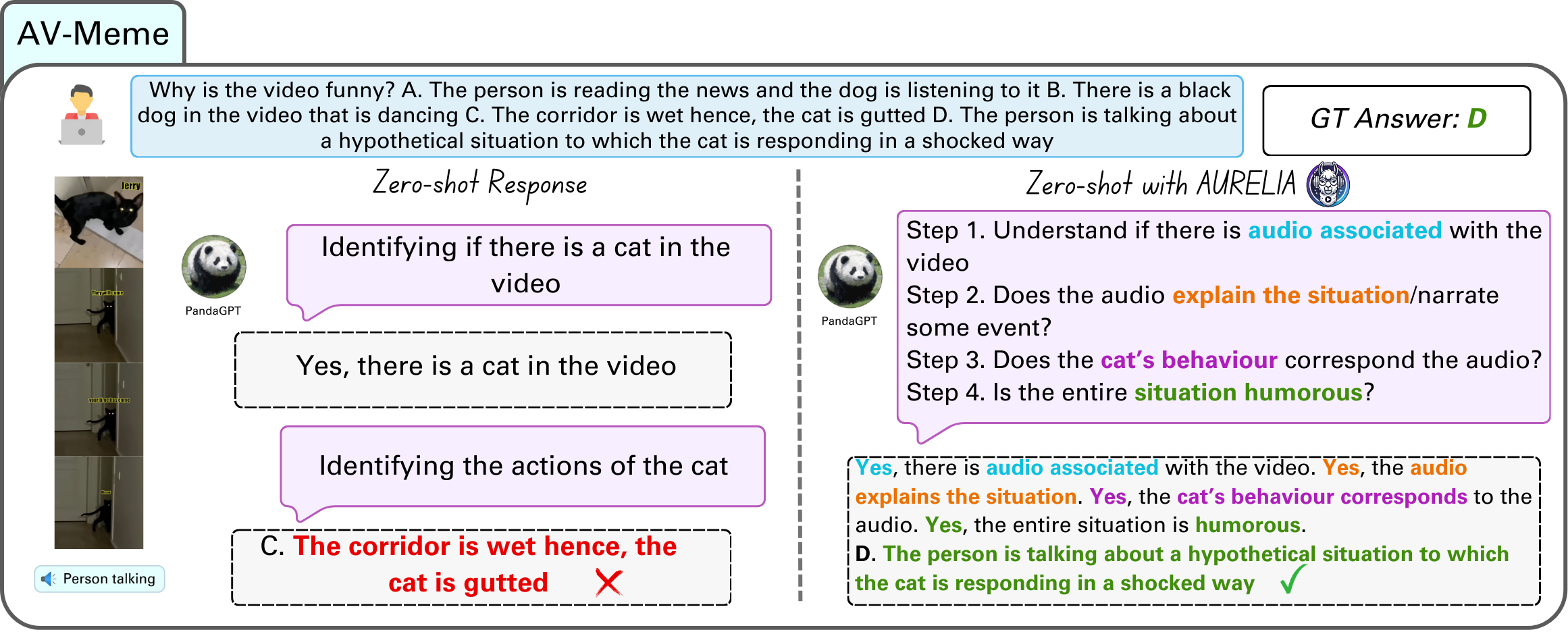}
        \caption{}
        \label{fig:qual-meme-1}
    \end{subfigure}
    
    \begin{subfigure}{0.9\textwidth}
        \centering
        \includegraphics[width=\textwidth]{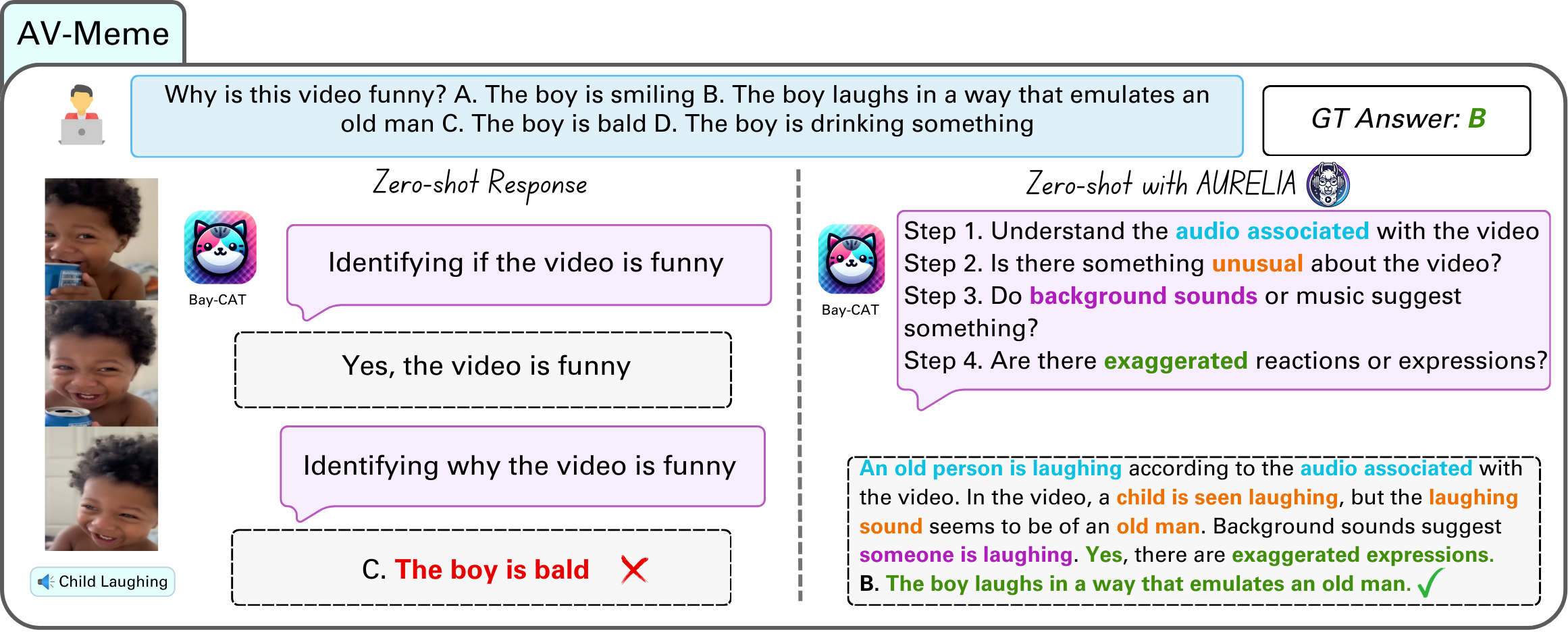}
        \caption{}
        \label{fig:qual-meme-2}
    \end{subfigure}
    
    \begin{subfigure}{0.9\textwidth}
        \centering
        \includegraphics[width=\textwidth]{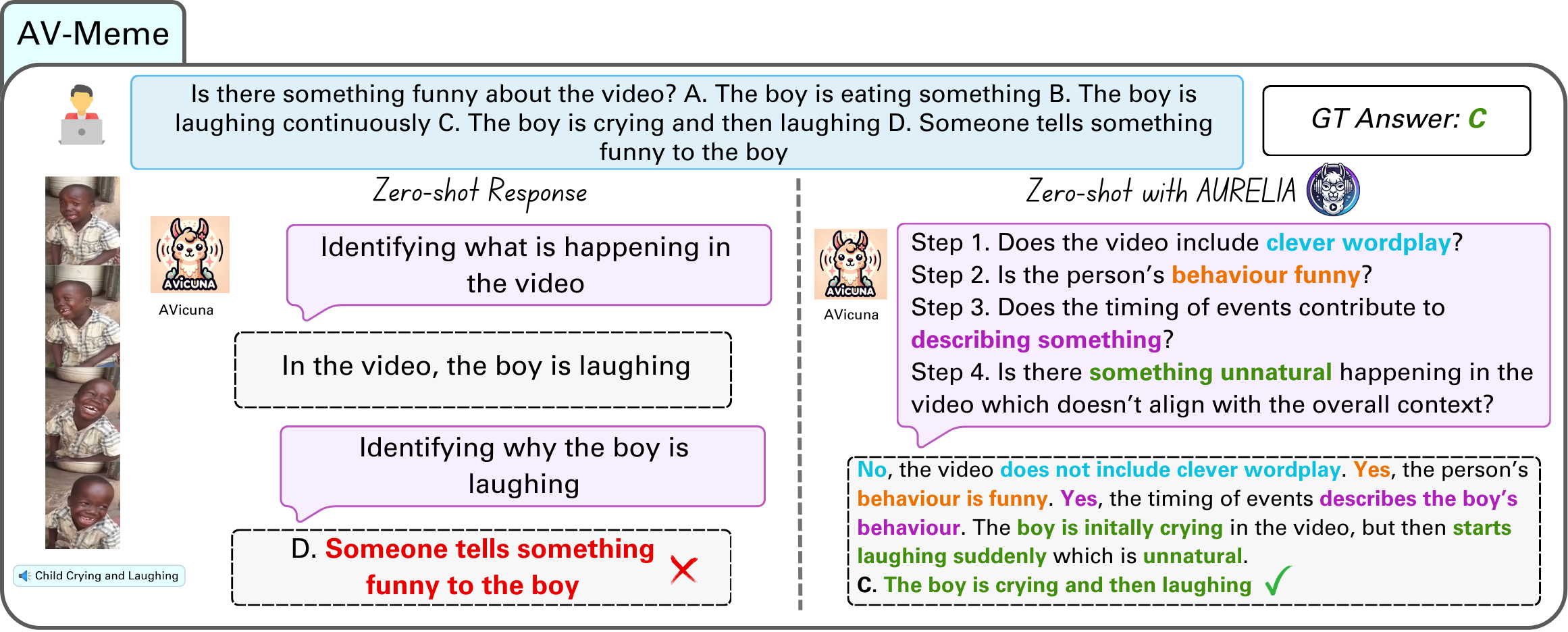}
        \caption{}
        \label{fig:qual-meme-3}
    \end{subfigure}

    \caption{Qualitative visualization of \ourapproach's reasoning distillation across AV-Meme task.} 
    \label{fig:qual-meme-combined}
\end{figure*}

\begin{figure*}
    \centering
    \begin{subfigure}{0.9\textwidth}
        \centering
        \includegraphics[width=\textwidth]{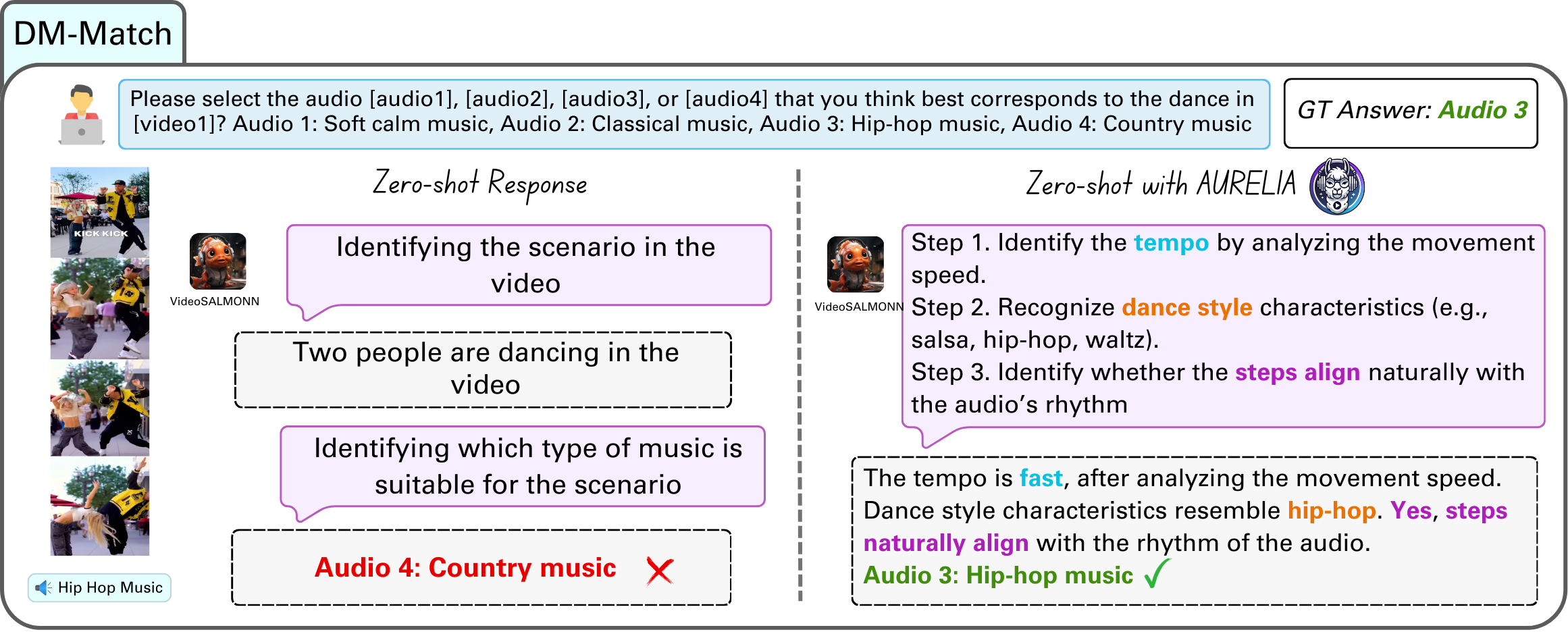}
        \caption{}
        \label{fig:qual-dance-1}
    \end{subfigure}
    
    \begin{subfigure}{0.9\textwidth}
        \centering
        \includegraphics[width=\textwidth]{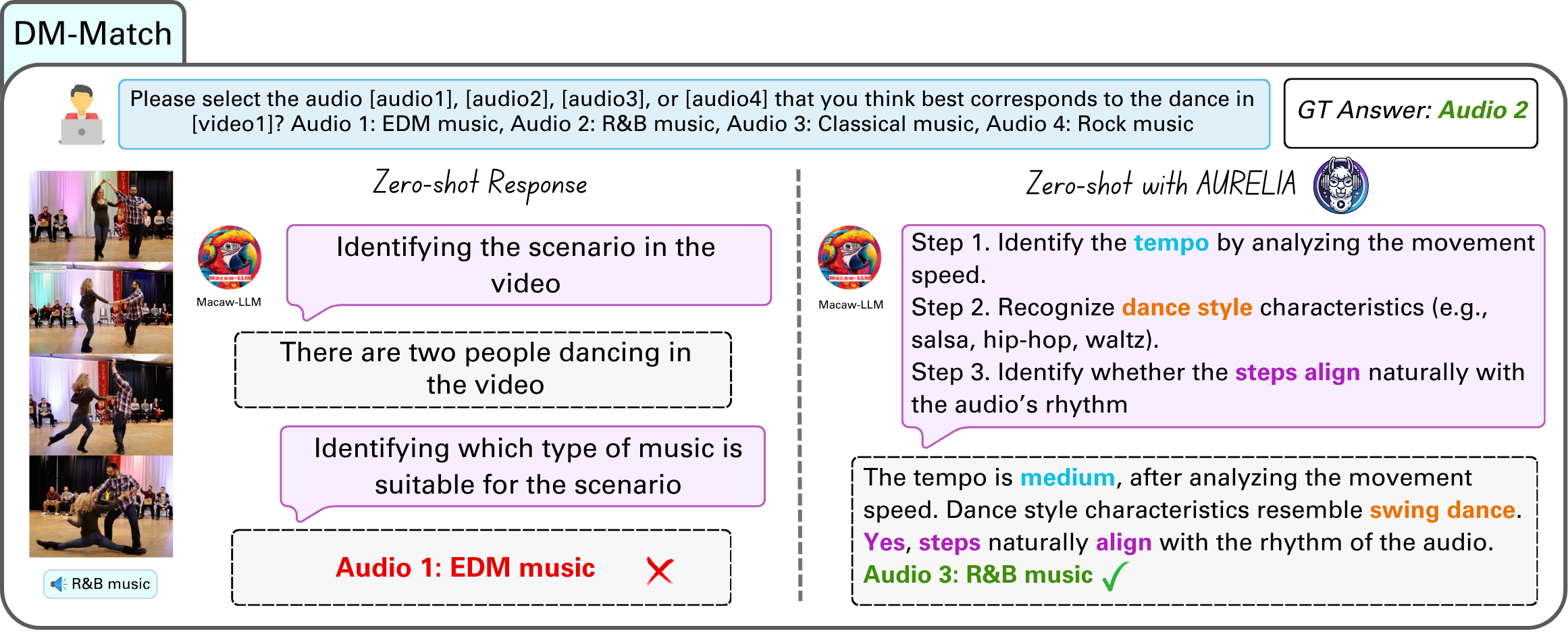}
        \caption{}
        \label{fig:qual-dance-2}
    \end{subfigure}
    
    \begin{subfigure}{0.9\textwidth}
        \centering
        \includegraphics[width=\textwidth]{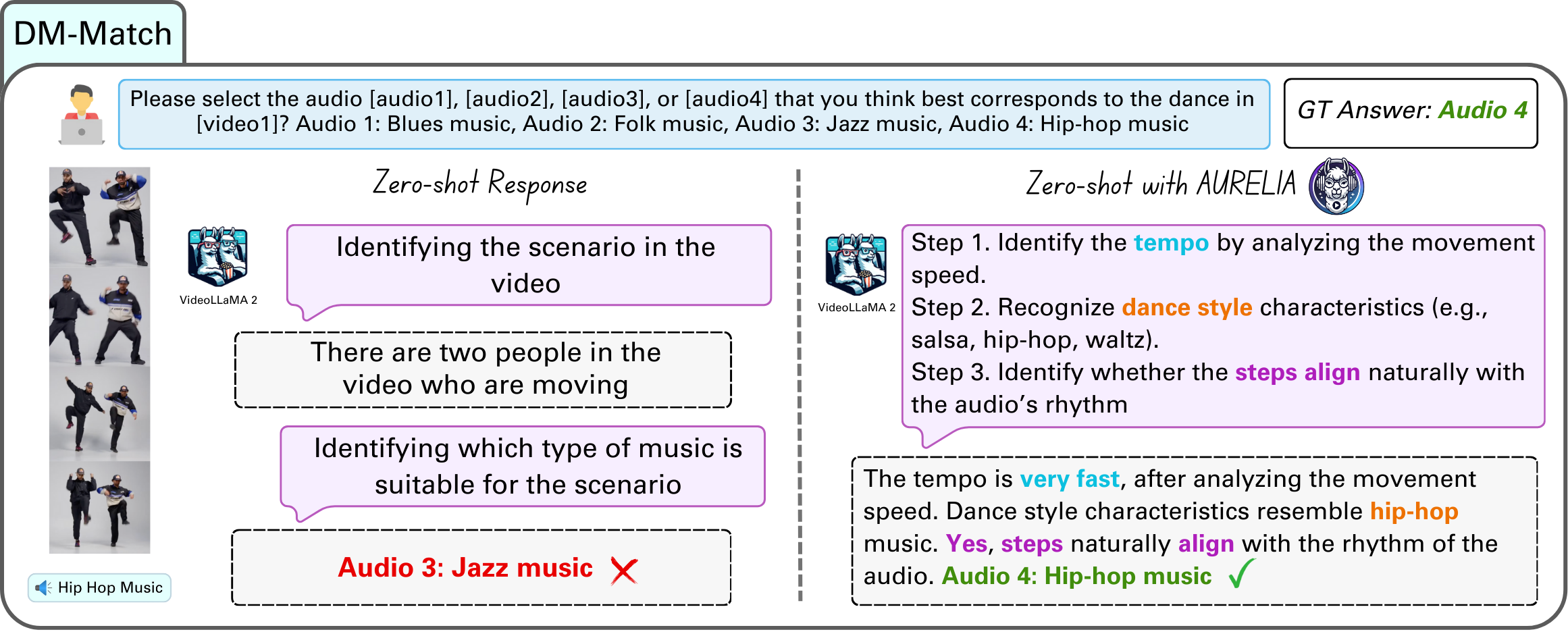}
        \caption{}
        \label{fig:qual-dance-3}
    \end{subfigure}

    \caption{Qualitative visualization of \ourapproach's reasoning distillation across DM-Match task.} 
    \label{fig:qual-dance-combined}
\end{figure*}

\begin{figure*}
    \centering
    \begin{subfigure}{0.9\textwidth}
        \centering
        \includegraphics[width=\textwidth]{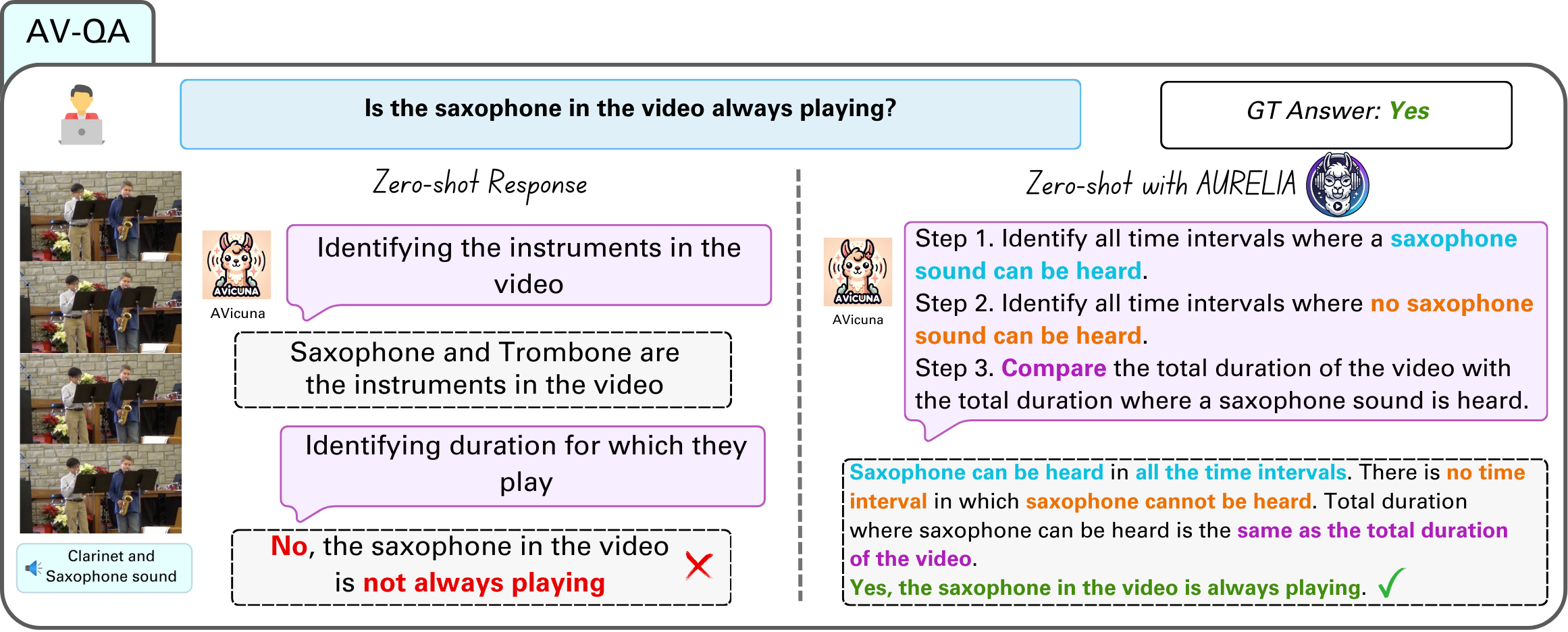}
        \caption{}
        \label{fig:qual-avqa-1}
    \end{subfigure}
    
    \begin{subfigure}{0.9\textwidth}
        \centering
        \includegraphics[width=\textwidth]{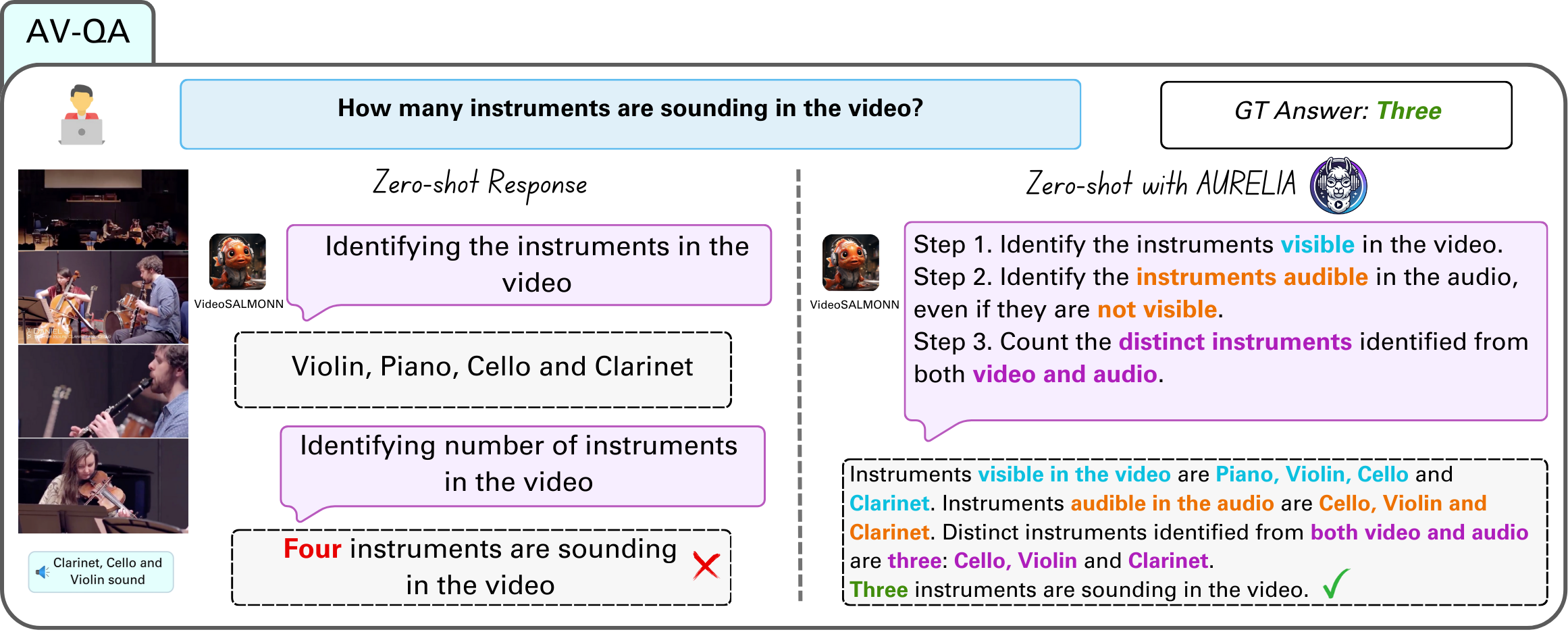}
        \caption{}
        \label{fig:qual-avqa-2}
    \end{subfigure}
    
    \begin{subfigure}{0.9\textwidth}
        \centering
        \includegraphics[width=\textwidth]{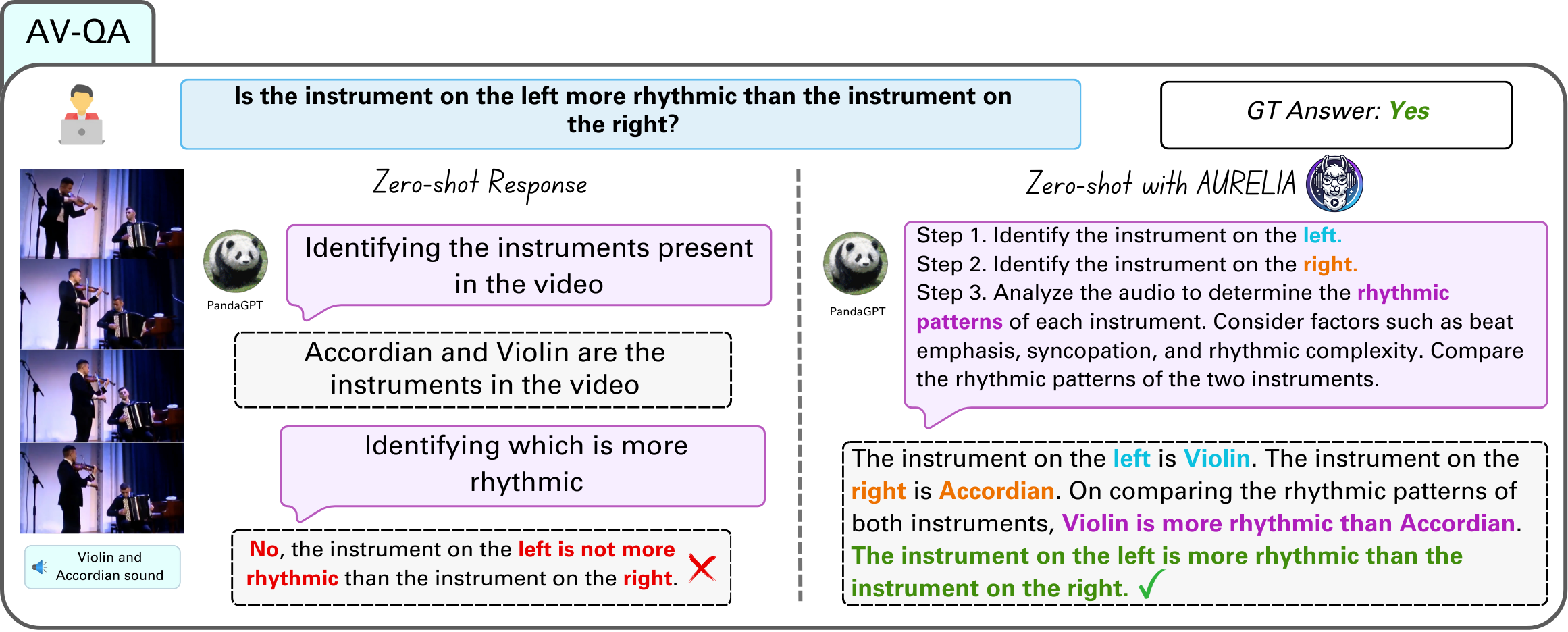}
        \caption{}
        \label{fig:qual-avqa-3}
    \end{subfigure}

    \caption{Qualitative visualization of \ourapproach's reasoning distillation across AV-QA task.} 
    \label{fig:qual-avqa-combined}
\end{figure*}

\begin{figure*}
    \centering
    \begin{subfigure}{0.9\textwidth}
        \centering
        \includegraphics[width=\textwidth]{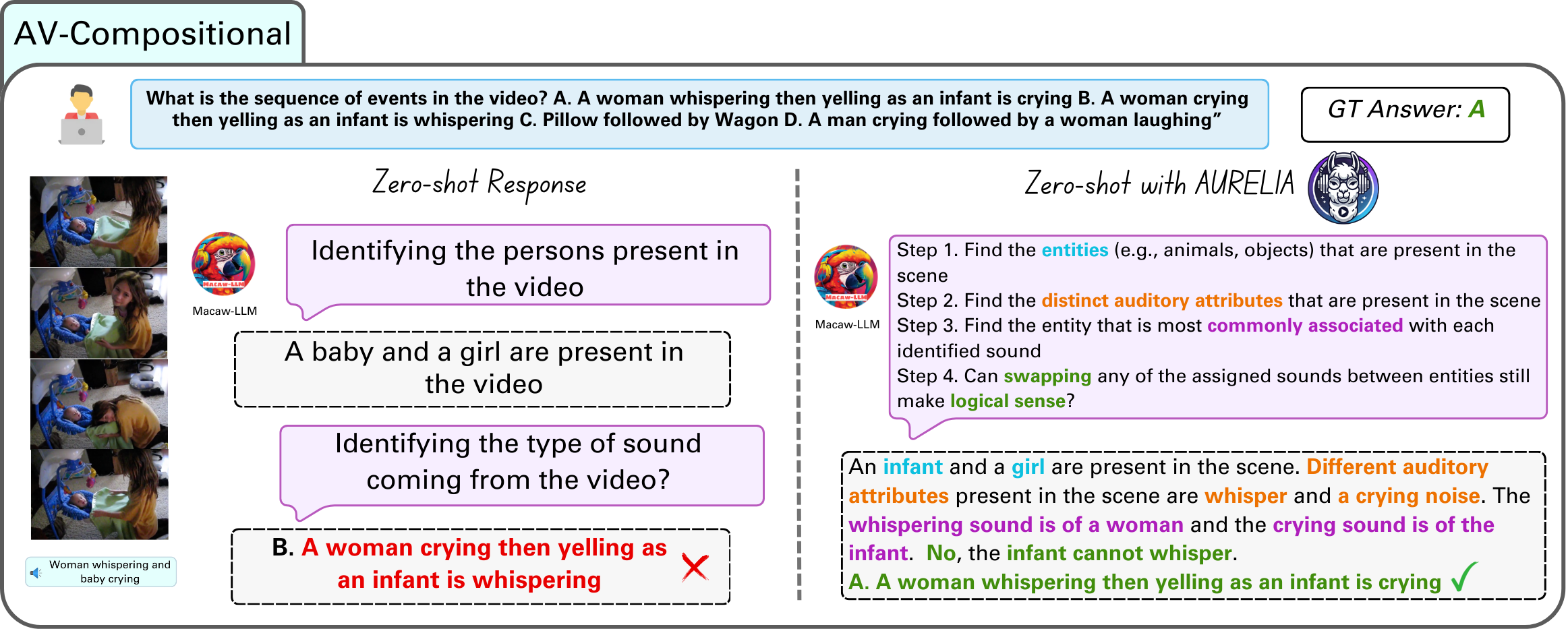}
        \caption{}
        \label{fig:qual-comp-1}
    \end{subfigure}
    
    \begin{subfigure}{0.9\textwidth}
        \centering
        \includegraphics[width=\textwidth]{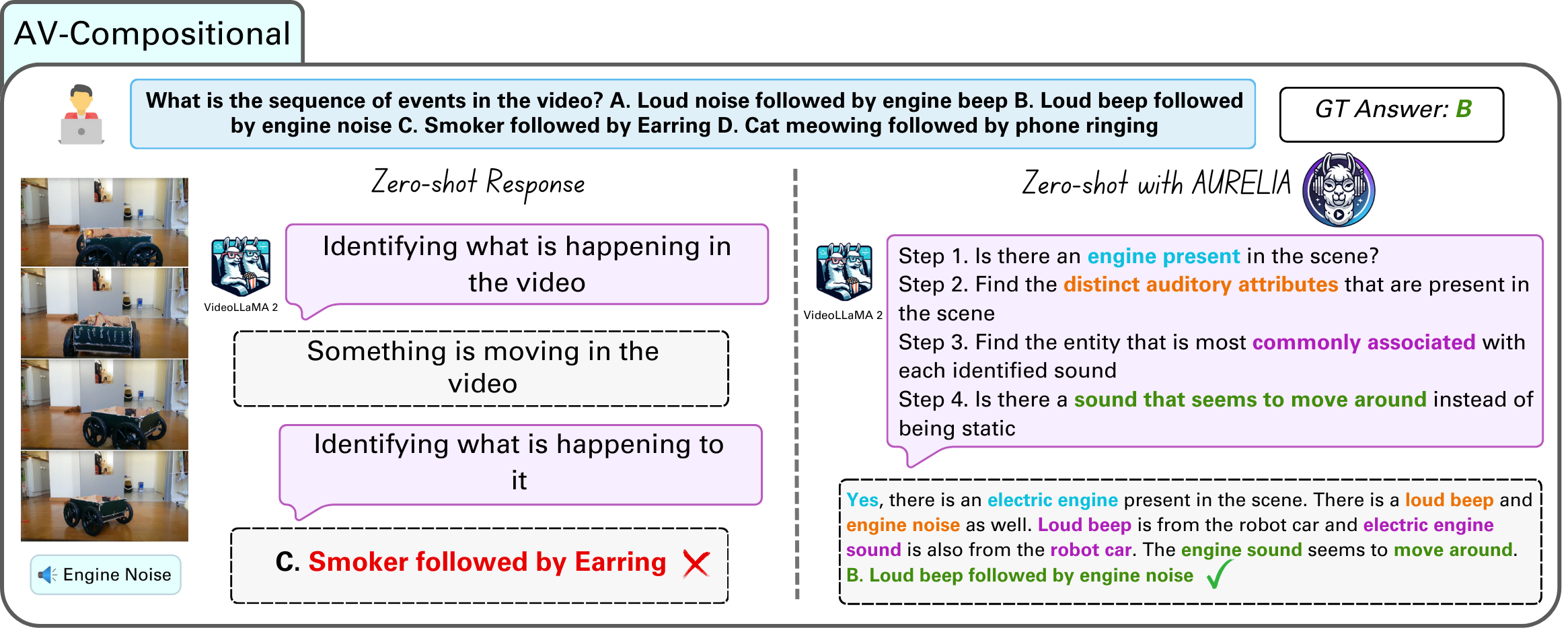}
        \caption{}
        \label{fig:qual-comp-2}
    \end{subfigure}
    
    \begin{subfigure}{0.9\textwidth}
        \centering
        \includegraphics[width=\textwidth]{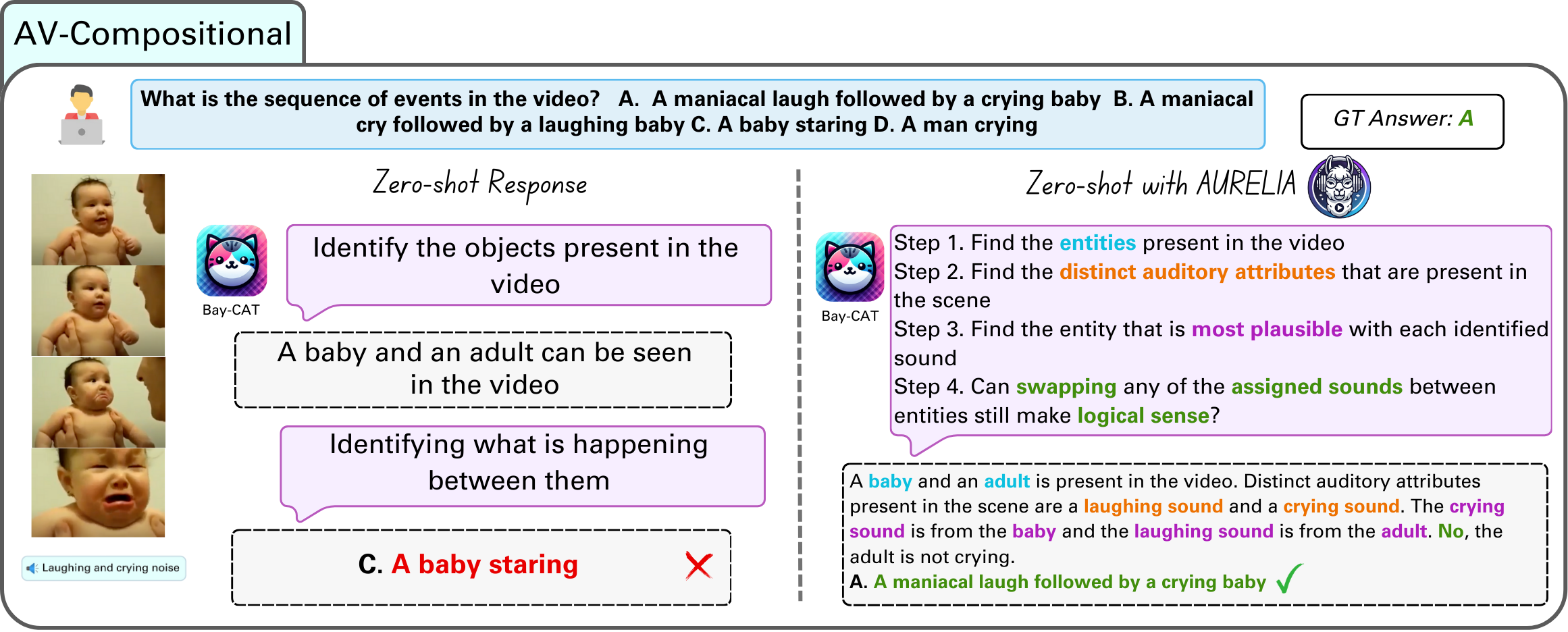}
        \caption{}
        \label{fig:qual-comp-3}
    \end{subfigure}
    \caption{Qualitative visualization of \ourapproach's reasoning distillation across AV-Compositional task.} 
    \label{fig:qual-comp-combined}
\end{figure*}

\begin{figure*}
    \centering
    \begin{subfigure}{0.9\textwidth}
        \centering
        \includegraphics[width=\textwidth]{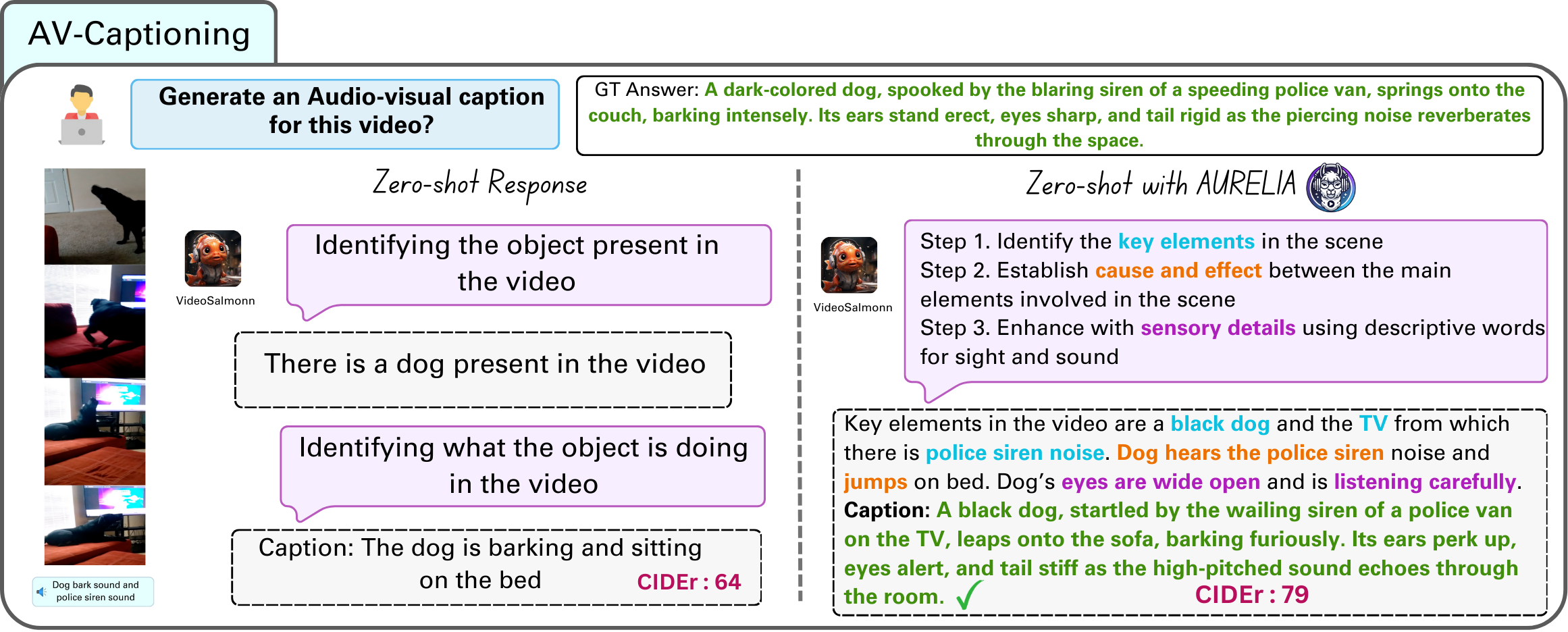}
        \caption{}
        \label{fig:qual-cap-1}
    \end{subfigure}
    
    \begin{subfigure}{0.9\textwidth}
        \centering
        \includegraphics[width=\textwidth]{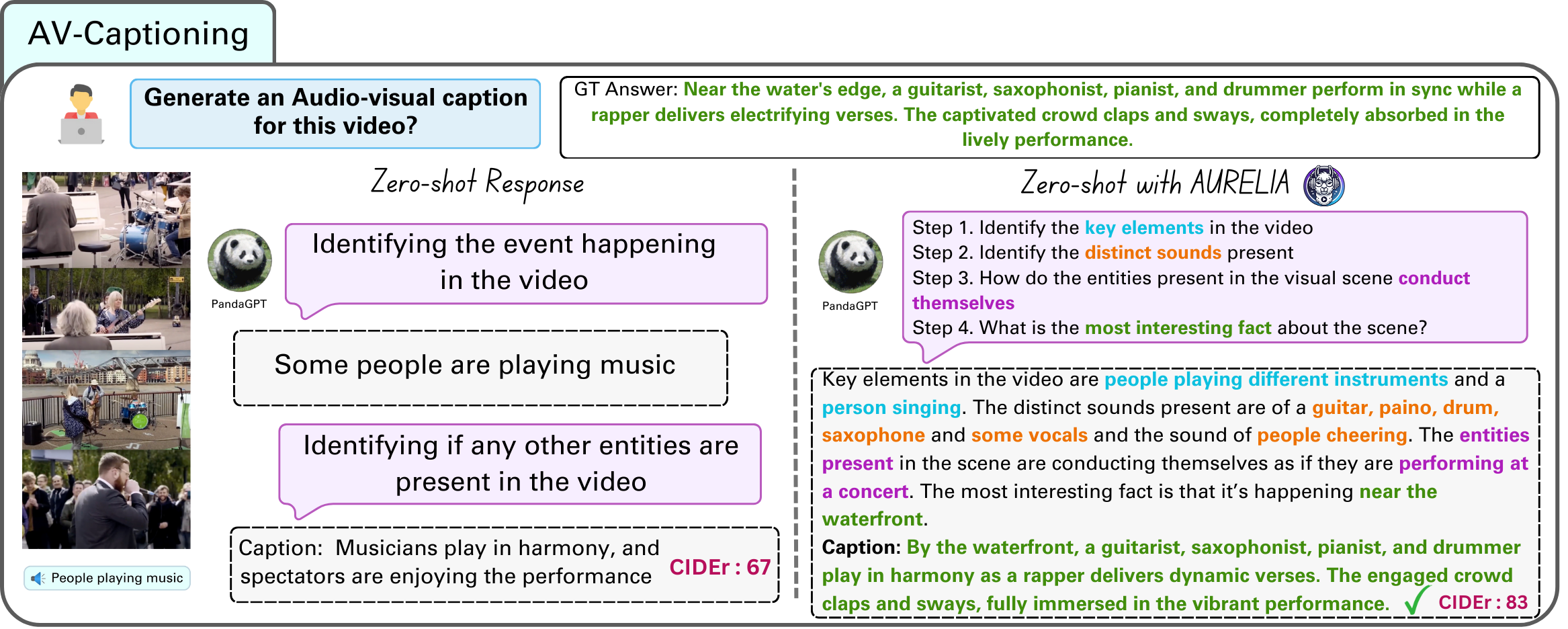}
        \caption{}
        \label{fig:qual-cap-2}
    \end{subfigure}

    \caption{Qualitative visualization of \ourapproach's reasoning distillation across AV-Captioning task.} 
    \label{fig:qual-comp-combined}
\end{figure*}

\clearpage
\raggedbottom
\clearpage

% {
%     \small
%     \bibliographystyle{ieeenat_fullname}
%     \bibliography{main}
% }

\end{document}